\documentclass[aps,prb,amsmath,amssymb,reprint,author-year,author-numerical,floatfix,superscriptaddress]{revtex4-2}
\usepackage{graphicx}
\graphicspath{{./images/}}
\usepackage{dcolumn}
\usepackage{bm}
\usepackage{mathtools}
\usepackage{amssymb}
\expandafter\let\csname equation*\endcsname\relax
\expandafter\let\csname endequation*\endcsname\relax
\usepackage{amsmath}
\usepackage{amsbsy}
\usepackage{amsthm}
\usepackage{booktabs}
\usepackage{verbatim}
\usepackage{array}
\usepackage{braket}
\usepackage{url}
\usepackage{float}
\usepackage{enumitem}
\usepackage{xcolor}
\usepackage[colorlinks = true,
            linkcolor = blue,
            urlcolor  = blue,
            citecolor = blue,
            anchorcolor = blue]{hyperref}
\usepackage[title]{appendix}  
\AtBeginDocument{\renewcommand{\natexlab}[1]{#1}}%

\begin{document}

\preprint{AIP/123-QED}
\title[MANSOORI]{Electron transport in a 1.6~nm-thick	double-gated (100) silicon nanosheet: \\
                 A theoretical study accounting for phonon confinement and remote-phonon scattering}
\author{Shoaib Mansoori}
\affiliation{Department of Materials Science and Engineering, The University of Texas at Dallas\\
             800 W. Campbell Rd., Richardson, TX 75080} 
\author{Bimin Cai}
\affiliation{Department of Materials Science and Engineering, The University of Texas at Dallas\\
             800 W. Campbell Rd., Richardson, TX 75080}
\author{Edward Chen}
\affiliation{Corporate Research, Taiwan Semiconductor Manufacturing Company Ltd.\\
             8 168, Park Ave. II, Hsinchu Science Park, Hsinchu 300-75, Taiwan} 
\author{Dallin O. Nielsen}
\affiliation{Department of Materials Science and Engineering, The University of Texas at Dallas\\
             800 W. Campbell Rd., Richardson, TX 75080}    
\author{Massimo V. Fischetti}
\email[email: ]{max.fischetti@utdallas.edu}
\affiliation{Department of Materials Science and Engineering, The University of Texas at Dallas\\
             800 W. Campbell Rd., Richardson, TX 75080} 

\date{\today}

\begin{abstract}
We present the results of our theoretical study of electron transport in an extremely scaled top- and bottom-gated (100) 1.6~nm-thin silicon nanosheet with 
SiO$_2$/HfO$_2$ gate stacks, focusing on how two intrinsic physical processes affect transport: the confinement of phonons and the presence of interface 
hybrid plasmon-phonon excitations (IPPs or `remote phonons'). 
The band structure is calculated using local empirical pseudopotentials; an approximated elastic continuum model is used to consider the confinement of 
acoustic phonons assuming free-standing boundary conditions (FSBCs) and clamped boundary conditions (CBCs) either at the Si/SiO$_2$ interfaces or at the
SiO$_2$/HfO$_2$ interfaces; the dielectric continuum limit is used to deal with the IPPs in these rather complicated dielectric structures. 
We find that the electron mobility is affected significantly by the boundary conditions chosen to deal with
phonon confinement. The more realistic assumption of phonons clamped at the SiO$_2$/HfO$_2$ interfaces and optical phonons at the Si/SiO$_2$ interfaces
results in a room temperature mobility much smaller than what is obtained using the common assumption of bulk phonons in the elastic, high-temperature 
approximation. We also find that, as a result of the complicated structure of the primed subbands, the high-field saturated velocity is significantly lower 
than its bulk value, as it had been measured in the past in the case of Si inversion layers but never explained theoretically. 
Finally, we find that IPP scattering does depress the low-field mobility but to a small extent, thanks to the presence of the interfacial SiO$_2$ layers and 
to the proximity of the metal gates. Moreover, by keeping electrons `cooler', IPP scattering results in a higher saturated velocity. Therefore, the presence of 
high-$\kappa$ materials in the gate-insulator stacks should not affect negatively the performance of field effect transistors based on Si nanosheets.          
\end{abstract}

\maketitle

\section{Introduction}
\label{sec:intro} As the semiconductor industry continues to extend Moore's law by scaling transistors, gate-all-around (GAA) nanosheet transistors have emerged 
as the preferred architecture and have been adopted in production at the 2--3~nm node~\cite{Loubet_2017,Park_2022,Yeap_2024,Agrawal_2024}. These devices 
employ channels consisting of Si nanosheets (or nanoribbons) with thickness as low as 3 unit cells ($\approx$~1.5~nm)~\cite{Agrawal_2024} and $\approx$~10~nm wide.
These ultra-thin Si layers exhibit degraded electronic-transport properties: Quantum confinement modifies the electron dispersion (subband structure), the confined
wavefunctions lead to enhanced scattering with phonons, the phonon spectrum changes, and scattering with interface roughness becomes stronger. For these reasons,
interest has shifted to alternative channel materials, such as graphene nanoribbons~\cite{Thornhill_2008}, silicene and germanene,~\cite{Houssa_2011,Molle_2018}, 
phosphorene~\cite{Liu_2014}, and van der Waals materials~\cite{Das_2015}, especially transition metal dichalcogenides~\cite{Daus_2021}. Yet, despite their 
advantages, these two-dimensional (2D) materials have not yet demonstrated the combination of mobility, interface quality, and process compatibility 
needed for production~\cite{Liu_2018,Shen_2021,Gopalan_2022,Baikadi_2024,Mansoori_2025}, and silicon remains the channel material of choice.\\

\begin{figure}[tb]
\centerline
{\hbox{
\includegraphics[width=8.60cm]{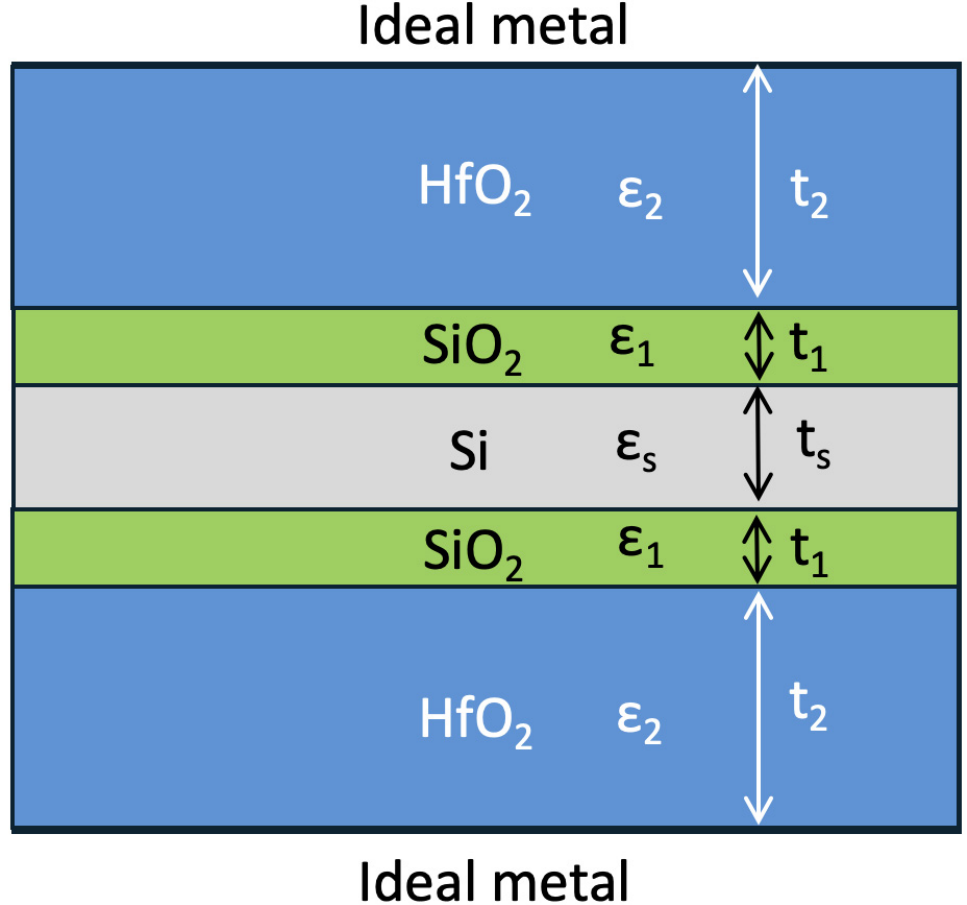}
}}
\caption{Idealized structure of a double-gate field-effect transistor (FET) with a silicon (Si) nanosheet channel, surrounded by SiO$_2$ and HfO$_2$
         dielectric layers. The structure includes the labeled dielectric permittivities ($\epsilon_1$, $\epsilon_2$) and layer thicknesses ($t_1$, $t_2$,
         $t_{\rm s}$). The specific dimensions considered are $t_1 \equiv t_{\rm SiO_2} =$ 0.6~nm, $t_2 \equiv t_{\rm HfO_2} =$1.2~nm, and 
         silicon nanosheet thickness $t_{\rm s} =$ 1.6~nm. The top and bottom metal gates are considered ideal equipotential metals.}
    \label{fig:si_geometry}
\end{figure}

Given the interest on extremely thin Si nano-sheets/ribbons, the purpose of this work is to study how two major {\em intrinsic} scattering processes control
their electronic-transport properties, emphasizing their importance: {\it i}) The geometry-induced confinement, not only of the electrons but also of the 
phonons and {\it ii)} the coupling of the electrons to the interface excitations resulting from the coupling of the plasmons of the two-dimensional Si electron 
gas (2DEG) and the optical phonons of the insulators. This latter process is commonly referred to as 
`remote-phonon  scattering'~\cite{Mahan_1972,Hess_1979,Moore_1980,Fischetti_2001}, but we prefer to call it more appropriately `scattering with interface 
plasmon/phonon (IPP) excitations'~\cite{Gopalan_2022}. A comprehensive investigation of the dependence of these processes on temperature, electron sheet density,
layer thickness, and gate-insulator stacks is beyond the scope of the present work. Rather, we consider only one specific case and analyze in detail the physical
ingredients that control these scattering processes. Similarly, beyond the scope of this work is a study of the characteristics of devices based on these 
nanosheets that would require tackling additional issues, such as {\em extrinsic} scattering processes ({\it e.g.}, electron scattering with charged defects,
impurities, interface roughness) and `parasitic' effects such as the quantum resistance in the source/channel region. These effects are `extrinsic' in the sense that 
they depend on processing and device design. Therefore, we focus on the room-temperature properties of electron transport in the thinnest nanosheet fabricated 
so far~\cite{Agrawal_2024}, a (100) Si sheet with the thickness of 3 cubic cells (1.63~nm) with a double-gate stack consisting of a 0.6~nm-thick SiO$_2$ layer 
and a 1.2~nm-thick HfO$_2$ film, as shown in Fig.~\ref{fig:si_geometry}.\\

The theoretical study of electron transport in thin Si films has a long history, dating back to the late 1990s (see, for example,
Refs.~\cite{Gamiz_1999,Gamiz_2001a,Esseni_2002,Esseni_2004,Khakifirooz_2004,Barraud_2005,Lucci_2005,Donetti_2006,Donetti_2007,Uchida_2007}). 
However, most of these studies deal with
silicon on insulator (SOI) structures or, when dealing with double-gate structures, ignore scattering with the IPPs and, in many cases, ignore also phonon confinement (with the notable exceptions of Refs.~\cite{Donetti_2006} and \cite{Donetti_2007}). Here, we intend to fill this gap.\\

We have organized the discussion as follows: Section~\ref{sec:Band structure} deals with the calculation of the band structure of Si nanosheets using empirical
pseudopotentials, arguing that in extremely thin sheets the gate bias does not affect it in any significant way. In Sec.~\ref{sec:Phonons}, we discuss an 
empirical model, based on our previous study of electron transport in Si nanowires and graphene nanoribbons~\cite{Bimin_2025}, to approximate the dispersion of 
phonons confined in the sheet. In Sec.~\ref{sec:El-ph} we discuss the electron-phonon interactions and in Sec.~\ref{sec:IPP} we discuss scattering with interface hybrid oxide-optical-phonon/2DEG-plasmon excitations. We present our results for the low-field mobility and high-field transport in Sec.~\ref{sec:Results} and 
draw our conclusions in Sec.~\ref{sec:Conclusions}.
\section{The band structure of 1.6~\lowercase{nm}-thick S\lowercase{i} nanosheets}
\label{sec:Band structure}
The device geometry under consideration is shown in Figure~\ref{fig:si_geometry}. The structure represents an idealized double-gate field-effect transistor (FET) based on a silicon (Si) nanosheet channel, sandwiched between SiO$_2$ and HfO$_2$ dielectric layers. The SiO$_2$ layers serve as inner spacers adjacent to the channel, while the outer HfO$_2$ layers act as high-$\kappa$ gate dielectrics separated by ideal metallic electrodes. The respective layer thicknesses are $t_{\text{SiO}_2} = 0.6$~nm, $t_{\text{HfO}_2} = 1.2$~nm, and a channel thickness $t_{\rm{s}}$ varying between 1.6~nm and 2.2~nm. This multilayer configuration provides a realistic platform for studying confined phonon behavior and electron-phonon interactions in nanosheet-based transistor architectures.\\

The first problem we face is to assess the electronic (and vibrational as well) dimensionality of such structures: Should we treat them as 2D `nanosheets' or 
1D `nanoribbons'? We discuss this issue elsewhere~\cite{Fischetti_2026a}, so here only a general overview of our conclusions. 
In general, electrons and phonons should be considered confined only in structures with dimensions smaller than their decoherence length $\lambda_{\phi}$ 
(that is, their inelastic mean-free-path). From the scattering rates $1/\tau_{\rm th}$ at the thermal energy $k_{\rm B}T$, shown in Fig.~\ref{fig:El-ph_3-cell}
and \ref{fig:IPP_rates}, we can extract an electron decoherence length $\lambda_{\phi} =\upsilon_{\rm g}/\tau_{\rm th} \sim $8~nm (assuming a group velocity 
$\upsilon_{\rm g}=\hbar \pi/(m_{\rm L}t_{\rm s})$, with $m_{\rm L} = 0.91 \ m_{0}$ and $t_{\rm s}$ = 1.6~nm, along the along the confinement direction in the 
ground-state subband) for phonon scattering and even smaller for IPP scattering. Therefore, a sheet-width of 12~nm appears to be too large to induced quantum confinement along its width. We also assume that phonons are confined only by the interfaces, although this case is more subtle, especially for long-wavelength acoustic phonons. This would require a discussion the goes beyond the scope of the present work.\\

We assume that the sheet lies on the $(x,y)$ plane, that transport occurs along the $x$ direction. Thus, the electron dispersion along the channel
will be given by the Schr\"{o}dinger equation:
\begin{equation}
\left [ - \frac{\hbar^2}{2m} \nabla^2 + V^{\rm (lat)}({\bf r}) + V^{\rm (ext)}({\bf r}) \right ] \psi({\bf r}) = E \psi({\bf r})  \ ,
\label{eq: Schroed_nanosheet}
\end{equation}
where $V^{\rm (lat)}$is the lattice potential and  $V^{\rm (ext)}({\bf r})$ is the (external) potential due to the source-drain and gate bias.
For the lattice potential we prefer to employ the local empirical pseudopotentials of Ref.~\cite{Zunger_1993} for both Si and the terminating H atoms,
rather than self-consistent {\it ab initio} 
pseudopotentials since their empirical nature guarantees values for the band gap and effective masses (so, more importantly, energies of the subbands) in 
agreement with experimental data. The price we must pay is being forced to determine empirically the electron-phonon matrix elements. However, as discussed
in Ref.~\cite{Fischetti_2019}, these have been determined with sufficient accuracy, at least in the case of bulk Si.\\

As a matter of notation, we shall write all 3-vectors ${\bf a}$ as $({\bf A},a_z)$; that is, using upper-case symbols to denote the components of the vector on the 
$(x,y)$ plane. The only exception is constituted by the vectors ${\bf G}$ of the reciprocal lattice: to use the notation used almost universally in the literature,
we shall write them as ${\bf G} = ({\bf G}_{\parallel},G_z)$. 
We assume for the nanosheets a thickness $t_{\rm s}$ that is an integer multiple of a cell; that is, $t_{\rm s}=Na_0$, where $a_0$ is the lattice constant. 
Accounting also for the thickness $N_{\rm v}a_0$ of the vacuum padding between layers in adjacent supercells, $L_z=(N+N_{\rm v})a_0$.  We set $N_{\rm v} = 2$ and
define $\Omega$ to be the area of the cell on the $(x,y)$ plane. Finally, we consider the external potential $V^{\rm (ext)}({\bf r})$ to be constant along 
the transport direction, $x$.\\ 

With the assumptions just listed, we must solve the eigenvalue problem:  
\begin{multline}
\sum_{{\bf G}^{\prime}} \left [ \frac{\hbar^2}{2m} \vert {\bf k}+{\bf G} \vert^2 \ \delta_{{\bf G}{\bf G}^{\prime}} + 
                 V^{\rm (lat)}_{{\bf G}-{\bf G}^{\prime}} + V^{\rm (ext)}_{x,{\bf G}-{\bf G}^{\prime}} \right ] u^{\rm (\nu)}_{{\bf k},{\bf G}^{\prime}} \\ =
                                 E^{\rm (\nu)}_{\bf k} u^{\rm (\nu)}_{{\bf k},{\bf G}} \ ,
\label{eq:2D_band_structure}
\end{multline}
where we have used the Bloch form $\psi_{\bf k}^{(\nu)}({\bf r})=\sum_{\bf G} e^{i {\bf G} \cdot {\bf r}} u^{\rm (\nu)}_{{\bf k},{\bf G}}$, the index   
$\nu$ labels the bands or subbands (eigenpairs), $V^{\rm (ext)}_{x,{\bf G}}$ represents the Fourier components of the external potential:
\begin{equation}
V^{\rm (ext)}_{x,{\bf G}} = \delta_{{\bf G}_{\parallel},{\bf 0}} \ \frac{1}{L_z} \int_0^{L_z} {\rm d}z \ V^{\rm (ext)}(x,z) \ e^{-iG_{z}z} \ ,
\label{eq:VextFourier}
\end{equation}

Since the band structure does not depend on the out-of-plane component $k_z$ of ${\bf k}=(\bf K,k_z)$ as long as we consider only electrons with
energies below the vacuum level, we can assume $k_z=0$ and consider only the dependence of the eigenpairs on the in-plane wave vector ${\bf K}$ writing 
$E^{\rm (\nu)}_{\bf K}$ and $u^{\rm (\nu)}_{{\bf K}{\bf G}}$ in the following.\\ 
\begin{figure*}[tb]
\centerline 
{\hbox{
\includegraphics[width=8.6cm]{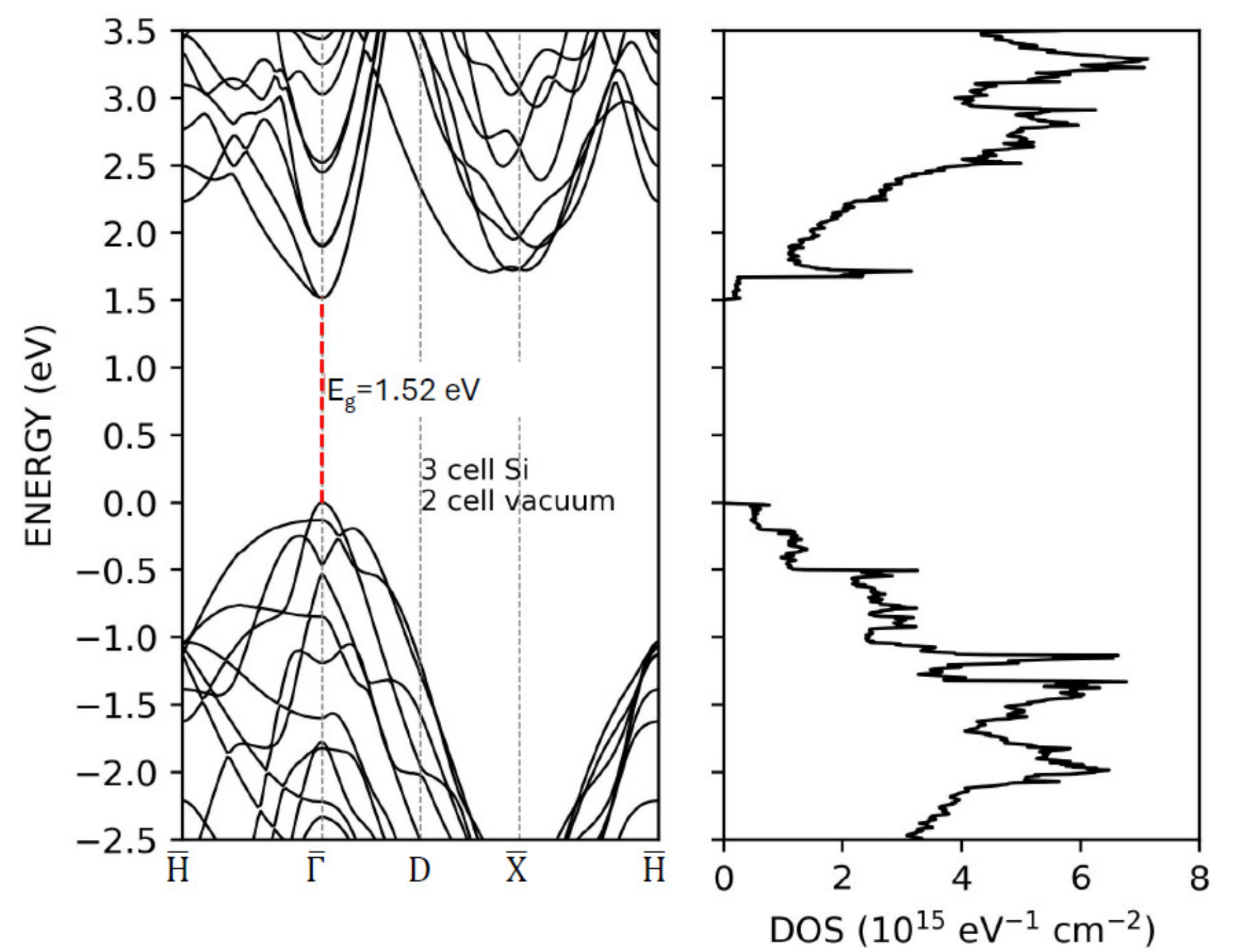}
\includegraphics[width=8.6cm]{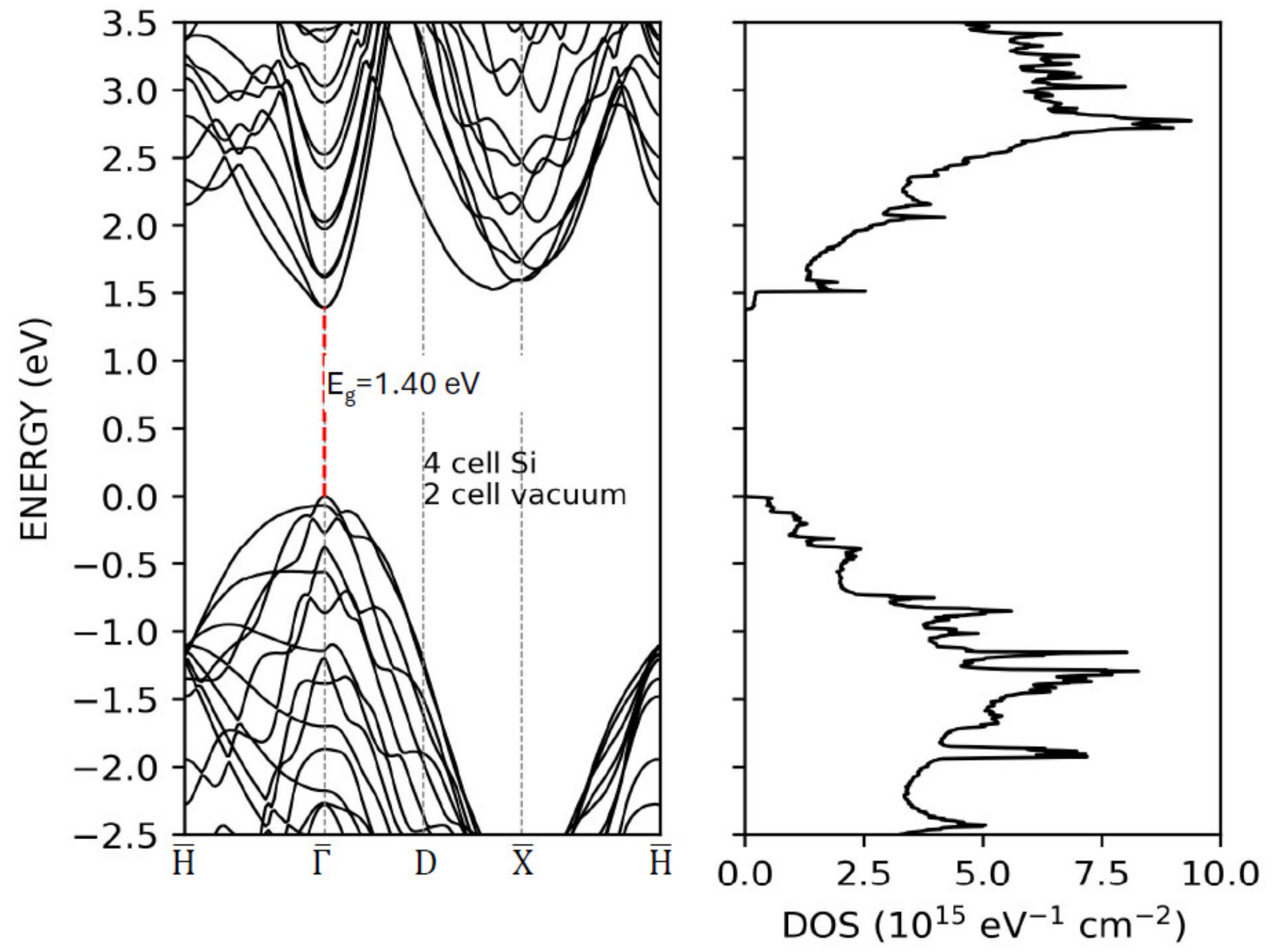}
}}
\caption{Band structure and density of states of the (100) 3-cell (left) and 4-cell (right) H-terminated thick films obtained from empirical 
         pseudopotentials assuming 2 cells of vacuum 'padding'.
         The bandgap of the 3-cell-thick Si nanosheet,$\approx$~1.52~eV, can be viewed as resulting from the ground-state energy of electrons in the conduction band, 
         $E^{0}_{{\overline{\Gamma},c}}=\hbar^2\pi^2/(2m_{\rm L})[1-\alpha \hbar^2\pi^2/(2m_{\rm L})] \approx $~0.143~eV, (assuming a nonparabolicity parameter 
         $\alpha$=0.5/eV and an effective mass $m_{\rm L} \approx$ 0.91~$m_{0}$), from the ground-state energy of holes in the valence band, 
         $E^{0}_{{\overline{\Gamma},v}}=\hbar^2\pi^2/(2m_{\rm h}) \approx$0.201~eV (assuming $m_{\rm h} \approx$ 0.7~$m_{0}$), and from a bulk bandgap of 
         $\approx$~1.20~eV. Similarly, for the 4-cell-thick nanosheet, since now $E^{0}_{{\overline{\Gamma},c}} \approx$~0.069~eV and 
         $E^{0}_{{\overline{\Gamma},v}} \approx$~0.11~eV, we expect a direct bandgap of about 1.38~eV, in rough agreement with the calculated bandgap of 1.40~eV.}    
\label{fig:Bands_3_4-cells}
\end{figure*}
\begin{figure}[tb]
\centerline 
{\vbox{
\hspace*{-0.70cm}\includegraphics[width=8.50cm]{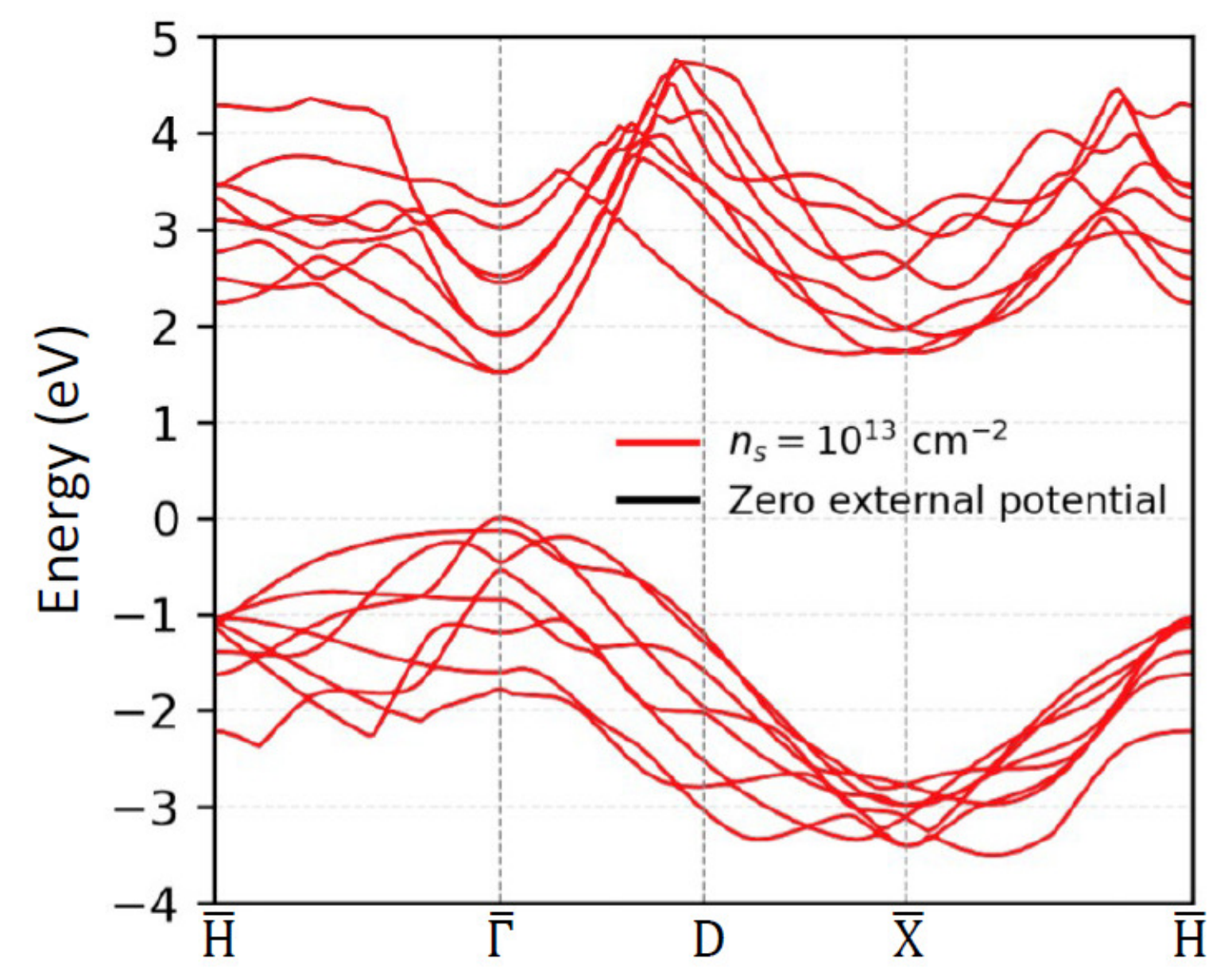}\\
\includegraphics[width=7.75cm]{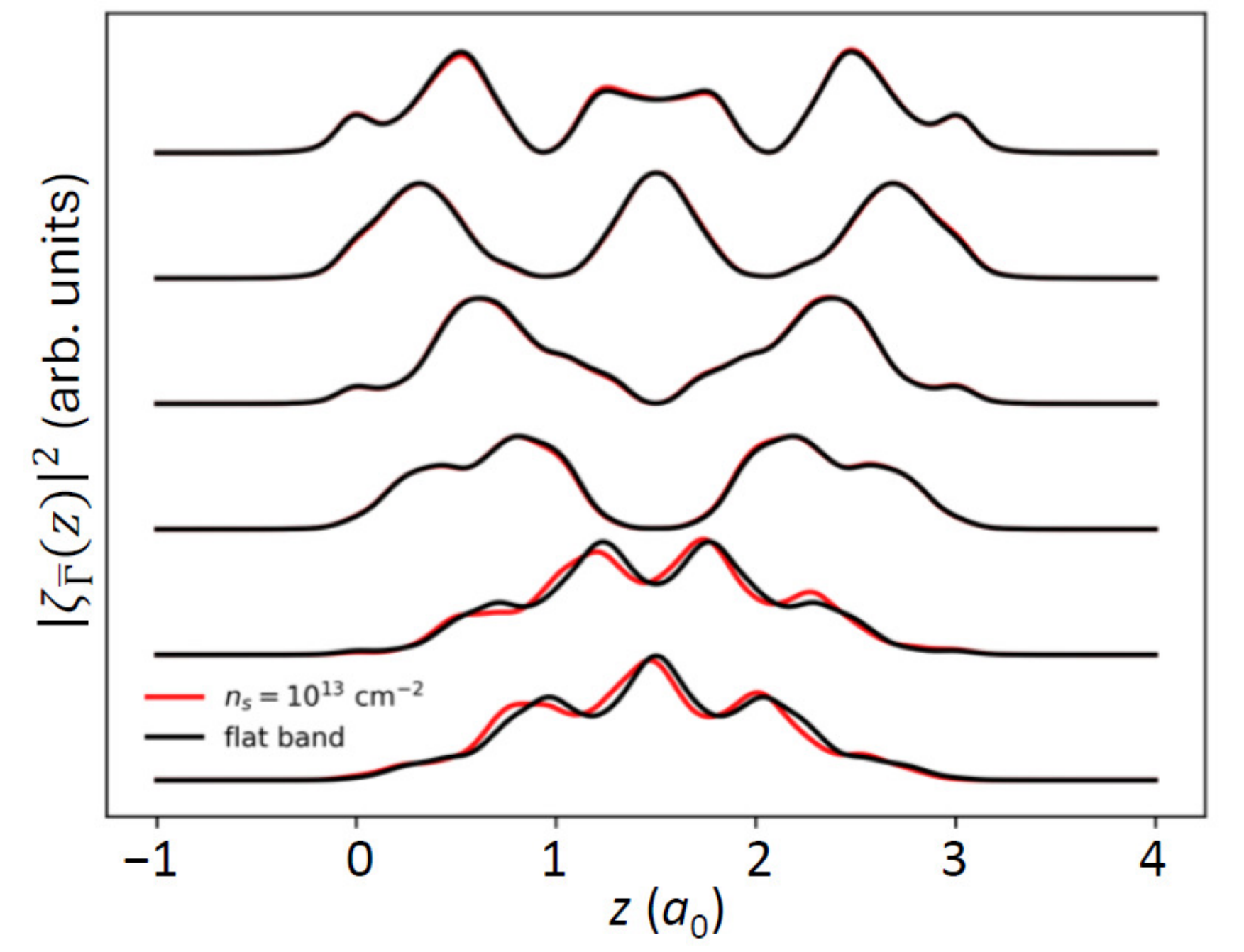}
}}
\caption{{\bf Top}: Band structure of a 3-cell-thick Si nanosheet in the presence of the parabolic potential given by Eq.~(\ref{eq:Vextz})
         (red lines) compared to the band structure at flat bands (black lines). The difference is too small to be visible (the
          band gap at $\overline{\Gamma}$ changes only by about 2~meV, from 1.518~eV to 1.520~eV. {\bf Bottom}: Squared magnitude of the wavefunctions
          at the $\overline{\Gamma}$-point in the six lowest-energy subbands both in the presence of the parabolic potential (red lines) and at flat bands
          (black lines). Note the slight increase of amplitude of the wavefunction in the ground-state subband on the left side and on the right side for the
          first excited state (almost degenerate) wavefunction. The changes of the wavefunctions of higher excited states are too small to be visible.}  
\label{fig:Bands_wavefunctions_1e13}
\end{figure}
 
In principle, the external potential $V^{\rm (ext)}(z)$ should be calculated self-consistently from the electron density
\begin{equation}
\rho_{\rm el}(z) =
      - \frac{2e}{(2 \pi)^2} \int_{0}^{\infty} {\rm d} {\bf K} \ \vert \zeta_{\bf K}^{\rm (\nu)}(z) \vert^2 \ f_{\rm FD}(E^{\rm (\nu)}_{\bf K}) \ ,
\label{eq:el_dens}
\end{equation}
where $f_{\rm FD}(E)=\{1+\exp[(E-E_{\rm F})/(k_{\rm B}T)]\}^{-1}$ is the Fermi-Dirac distribution at temperature $T$ and Fermi energy $E_{\rm F}$. 
The wavefunction $\zeta_{\bf K}^{\rm (\nu)}(z)$, introduced in Chapter 12 of Ref.~\cite{thebook} (see Eq.~12.31) of this reference) is the average 
of the full wavefunction over a unit cell in the $(x,y)$ plane:
\begin{equation}
\zeta_{\bf K}^{(\nu)}(z) = \frac{1}{{\mathcal{L}}_{{\bf K} \nu}^{1/2}} \sum_{G_z} u^{\rm (\nu)}_{{\bf K},{\bf G}_{\parallel}=0,G_{z}} \ e^{iG_{z}z} \ ,
\label{eq:zetaK}
\end{equation}
where ${\mathcal{L}}_{{\bf K} \nu}=L_{z} \sum_{G_z} |u^{\rm (\nu)}_{{\bf K},{\bf G}_{\parallel}=0,G_{z}}|^2$, so that $\zeta_{\bf K}^{(\nu)}$ is normalized to unity.
Obviously, the functions $\zeta_{\bf K}^{(\nu)}(z)$ are real-valued functions, up to a phase, as they represent bound states ({\it i.e.}, 
standing waves) with zero momentum along the $z$ direction.\\

Having stated what should be done in a full self-consistent approach, we note that the nanosheet is so thin that the variations of the potential
induced by the gate bias are negligible, even at the rather large electron density we consider, $\sim 10^{13}$~cm$^{-2}$, representative of the
density present at the source side of a transistor in the 'on' state. This can be shown with a simple example: We approximate   
$V^{\rm (ext)}(z)$ with a parabolic external potential that mimics the potential profile caused by the application of the same gate bias on the bottom and 
top insulators,
\begin{equation}
V^{\rm (ext)}(z) = \left \{ \begin{array}{ll}
    V_{0} \left ( 1 - \frac{4z}{t_{\rm s}} + \frac{4z^2}{t_{\rm s}^2} \right ) & (0 \le z < t_{\rm s}) \\
    0 & (t_{\rm s} \le z < L_z) 
      \end{array} \right. \ ,
\label{eq:Vextz}
\end{equation}
where $L_z = (N + N_{\rm vac})a_0$ is the thickness of the supercell and $t_{\rm s} = Na_0$ is the thickness of the Si layer. 
This potential, for $V_0 < 0$, has a maximum $V(z=L_z/2)=0$ at the center of the sheet and takes the value $V_0=-en_{\rm el}t_{\rm s}/(8\epsilon_{\rm s})$ 
at the edges of the sheet. It is the solution of the Poisson equation ${\rm d}^2 V(z)/{\rm d} z^2=\rho_{\rm el}/\epsilon_{\rm s}$ 
(where $\epsilon_{\rm s}$ is the static dielectric constant of the semiconductor and $n_{\rm el}$ the electron sheet density) assuming 
a $z$-independent, uniform (classical) electron charge density $\rho_{\rm el}=-en_{\rm el}/t_{\rm s}$ in the nanosheet.
Even at the relatively large electron sheet density $n_{\rm el}=10^{13}$~cm$^{-2}$, the potential drop $V_{0}$ is quite small, $\approx$ 31.5~meV. 
Moreover, the bandgap changes by about 2~meV. Therefore, we can neglect the density-dependence of the band structure. 
Figure~\ref{fig:Bands_wavefunctions_1e13} does indeed show that the changes of the band structure and wavefunctions are small enough to be ignored.
This implies that the electron mobility should exhibit a weak dependence on electron density (or gate bias), since quantum confinement is determined
almost exclusively by the geometry of the quantum well rather than by the external potential. Whatever dependence on electron density the mobility exhibits
is due to the effect of free-carrier screening and the dispersion and coupling constant with the IPPs.\\

Regarding numerical details, the eigenvalues $E^{\rm (\nu)}_{\bf K}$ and eigenvectors, $u^{\rm (\nu)}_{{\bf K},{\bf G}}$, of Eq.~(\ref{eq:2D_band_structure}) 
are tabulated for 12 (sub)bands on a mesh of ${\bf K}$-points (with a spacing $\Delta K = 0.025 (2\pi/a)$) in the triangular irreducible wedge of the 
two-dimensional Brillouin (BZ) zone (thus yielding 41$\times$41 ${\bf K}$-points on each quadrant of the Brillouin zone),
together with the gradients $\nabla_{\bf K}E^{\rm (\nu)}_{\bf k}$ that are needed by the Gilat-Raubenheimer algorithm~\cite{Gilat_1966}, modified
for a two dimensional Brillouin zone~\cite{Fischetti_2011a}, used to calculate the scattering rates. 
Additional points outside the BZ are used for interpolation purposes.
\section{Confined phonons in S\lowercase{i} nanosheets}
\label{sec:Phonons}
\subsection{Boundary conditions without accounting for the confinement of SiO$_2$ acoustic phonons}
As discussed at the beginning of the previous section, the small thickness of the nanosheet makes it necessary to account for the confinement of the phonons.
This effect has been extensively studied in the context of silicon thin films or nanomembranes (see, for example,
Refs.~\cite{Torres_2004,Donetti_2006,Donetti_2007,Cuffe_2012,Cuffe_2013,Karamitaheri_2013,Wang_2014,Mansoor_2015,Neogi_2015}).
Here  we follow the same approach employed in our previous work on Si nanowires and graphene nanoribbons~\cite{Bimin_2025}, suitably modified to deal with
the structure of interest here. To emphasize how strongly phonon confinement can affect electron transport, dealing with acoustic phonons, 
we consider three different sets of boundary conditions: We start with the `usual' boundary conditions considered in Ref.~\cite{Donetti_2006} for thin Si layers 
and in Refs.~\cite{Ramayya_2006,Tienda_2013} for Si nanowires, {\it i.e.}, clamped and free-standing boundary conditions at the Si/SiO$_2$ interfaces (CBCs and FSBCs, respectively). We then follow the work by Donetti {\it et al.}~\cite{Donetti_2007} employing a simplified elastic-continuum model to consider the confinement 
of acoustic phonon also in the SiO$_2$ films by assuming clamped boundary conditions at the SiO$_2$/HfO$_2$ interfaces, as dictated by the strong discontinuity of the
elastic properties of SiO$_2$ and HfO$_2$. On the contrary, for optical phonons, given the strong dielectric mismatch and the mismatch of their dispersion at the
Si/SiO$_2$ interfaces, we consider only clamped boundary conditions. 
In the elastic continuum approach that we shall follow, the deviations from reflection symmetry at the atomistic level that we have mentioned before may be neglected
and we shall assume a perfectly symmetric structure.\\

\begin{figure*}[t!]
\centerline 
{\hbox{
\includegraphics[width=14.0cm]{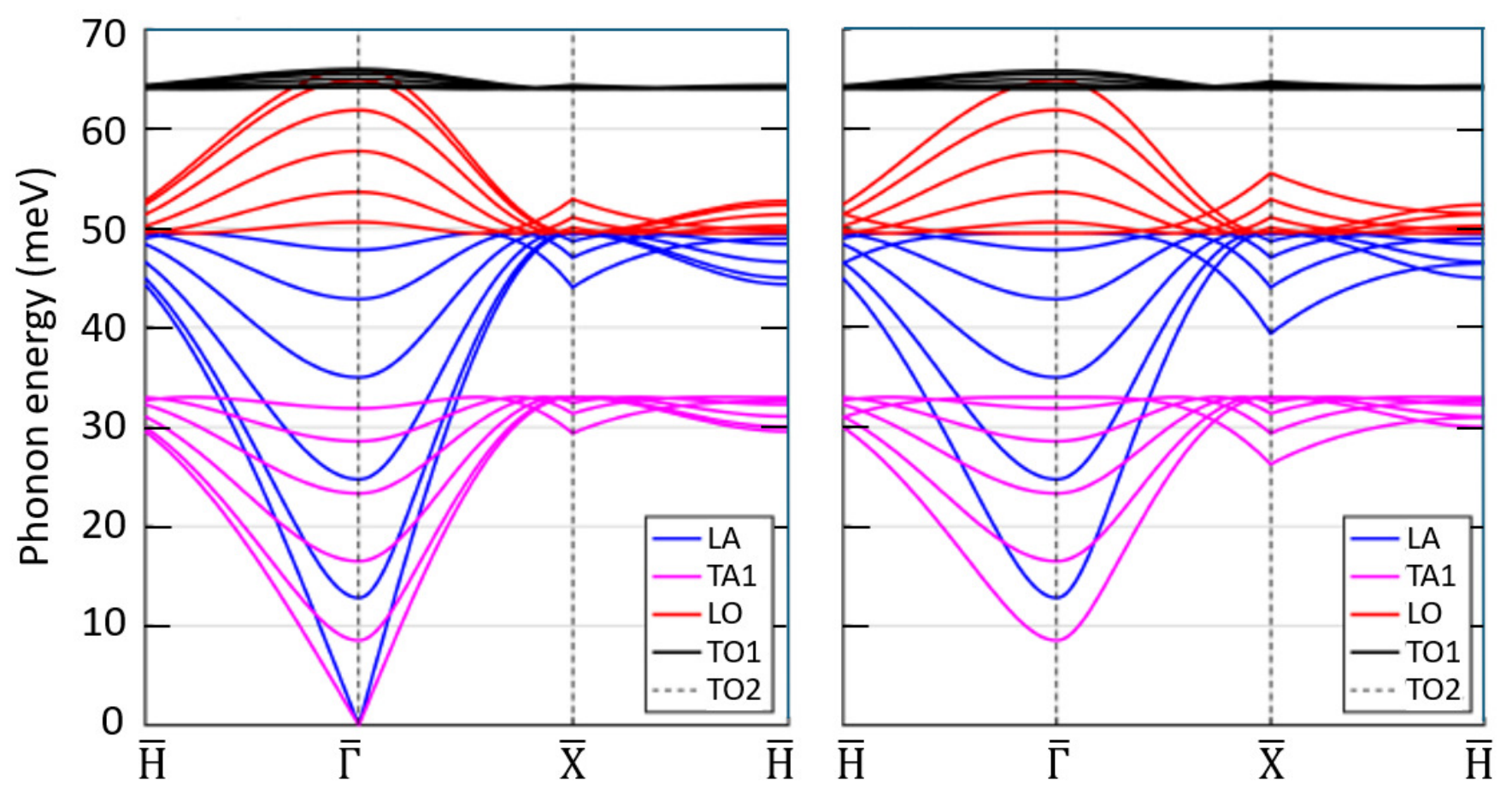}
}}
\caption{Phonon dispersion in a 3-cell thick Si nanosheet with freestanding (left) and clamped (right) boundary conditions, as obtained from the model
         described by Eqs.~(\ref{eq:ac_phon_sheet}) and (\ref{eq:op_phon_sheet}) using the values from Ref.~\cite{Neogi_2015}, also listed in
         Table~\ref{Tab:sheets_parameters}. Although 30 of the 36 branches are shown (not shown are the acoustic shear branches that are decoupled 
         from the electrons), not all branches are visible, since the 6 TO2 (ZO) branches are hidden by the 6 TO1 branches. The labels in the legend, 
         "LA", "TA1", "LO", etc., refer to the bulk modes from which the 2D phonons originate. As stated in the main text,
         bulk TA phonons correspond to shear or flexural modes, whereas the bulk LA phonons correspond to dilatational modes.}  
\label{fig:Phonon_dispersion}
\end{figure*}

As done in Ref.~\cite{Bimin_2025}, for a sheet of thickness $t_{\rm s} = Na_0$, we model the dispersion of the confined phonon starting from the bulk dispersion
but `quantizing' the out-of-plane component, $q_z$, phonon wave vector and `folding' the dispersion into the 2D Brillouin zone obtaining various acoustic and
optical branches with dispersion:
\begin{equation}
\omega^{\rm (LA/TA)}_m(Q) = \omega^{\rm (LA/TA)}_{\rm BZ} \sin \left ( \frac{a_0}{4} q_{m} \right ) \ ,
\label{eq:ac_phon_sheet}
\end{equation}
\begin{multline}
\omega^{\rm (LO/TO)}_m(Q) = \frac{\omega^{\rm (LO/TO)}_{\Gamma}+\omega^{\rm (LO/TO)}_{\rm BZ}}{2} + \\
                       \frac{\omega^{\rm (LO/TO)}_{\Gamma}-\omega^{\rm (LO/TO)}_{\rm BZ}}{2} \cos \left ( \frac{a_0}{2} q_{m} \right )  
\label{eq:op_phon_sheet}
\end{multline}
where ${\bf q}_m = \left ( {\bf Q}, q_{z,m} \right )$ and $q_{z,m} = m\pi/t_{\rm s}$. The integer index $m$ ranges from $0$ to $2N-1$ when
assuming free-standing boundary conditions at the two interfaces of the layer (FSBCs), from $1$ to $2N$ when assuming clamped boundary conditions (CBCs). 
For the phonon frequencies at the $\overline{\Gamma}$ point and at the edge of the Brillouin zone appearing in Eqs.~(\ref{eq:ac_phon_sheet}) and
(\ref{eq:op_phon_sheet}), we adopt the values used before in Ref.~\cite{Bimin_2025}, listed in Table~\ref{Tab:sheets_parameters} below. For acoustic 
phonons, these parameters imply a longitudinal and transverse sound velocity  $c_{\rm L} \approx 10.04 \times 10^5$~cm/s and 
$c_{\rm T} \approx 6.84 \times 10^5$~cm/s. The former is almost identical 
to the experimental value measured in bulk Si (in the range of 9.8 to 10.3$\times 10^5$ cm/s), the latter about 20\% larger than the experimental value, 
$\approx 5.39 \times 10^5$~cm/s. In Fig.~\ref{fig:Phonon_dispersion} we show the phonon spectrum obtained using this simple model: Keeping in mind the different thickness of the sheet, the overall features resemble, at least qualitatively, the molecular dynamic dispersion obtained by 
Neogi {\it et al.}~\cite{Neogi_2015} for a 3~nm-thick film (see Fig.~3(b) of this reference).\\

The coupling between electrons and the confined phonons can be treated using the elastic continuum approximation that is strictly valid only at long
wavelength. In this limit, the coupling with acoustic phonons is proportional to the divergence $\nabla \cdot {\bf u}({\bf R},z;t)$ of the ionic displacement 
field
\begin{equation}
{\bf u}({\bf R},z;t) = \sum_{{\bf Q}m} {\bf w}_{m}(z) e^{i {\bf Q} \cdot {\bf R}} e^{i \omega_{m}(Q) t} \ . 
\nonumber
\end{equation}
Therefore,
\begin{multline}
\nabla \cdot {\bf u}({\bf R},z;t) = \\  \sum_{{\bf Q}m} 
   \left [ i {\bf Q} \cdot {\bf w}_{\parallel m}(z) + \frac{\partial w_{z,m}(z)}{\partial z} \right ]
      e^{i {\bf Q} \cdot {\bf R}} e^{i \omega_{m}(Q) t} \ \\
      \equiv \sum_{{\bf Q}m} 
   \left [ i Q {\mathcal{K}}_{m}(z) + q_{z,m} \mathcal{G}_{m}(z) \right ]
      e^{i {\bf Q} \cdot {\bf R}} e^{i \omega_{m}(Q) t} \ ,   
\label{eq:shape_el_cont}
\end{multline} 
having defined the `shape functions' $\mathcal{G}_{m}(z)=q_{z,m}^{-1} \ \partial w_{z,m}(z)/\partial z$ and ${\mathcal{K}}_{m}(z)=w_{x,m}(z)$. 
In defining this latter function we have assumed that, without loss of generality, that ${\bf Q}$ is along the $x$ axis.
Donetti {\it et al.}~\cite{Donetti_2006} show that the equation for $w_y(z)$ is decoupled from the equations from $w_x$ and $w_z$. These are shear modes and Eq.~(\ref{eq:shape_el_cont}) shows that electrons do not couple with these modes. Therefore, they will be ignored throughout.\\

Now: From the analysis by Donetti {\it et al.}~\cite{Donetti_2006}, assuming either FSBCs or CBCs, in the long-wavelength limit 
the functions ${\mathcal{G}}_m$ and ${\mathcal{K}}_m$ are actually related by the relation (see Eq.~(A20) of Ref.~\cite{Bimin_2025}):
\begin{equation}
  iQ {\mathcal{K}}_m(z) + q_{z,m} {\mathcal{G}}_m(z) \ \rightarrow \ [Q^2+q^2_{z,m}]^{1/2} \ \mathcal{G}_m(z) ,  
\end{equation}
so that Eq.~(\ref{eq:shape_el_cont}) becomes simply:
\begin{multline}
\nabla \cdot {\bf u}({\bf R},z;t) =  \sum_{{\bf Q}m} \ [Q^2+q_{z,m}^{2}]^{1/2} {\mathcal{G}}_{m}(z) \ e^{i {\bf Q} \cdot {\bf R}} e^{i \omega_{m}(Q) t} \\ =
                                     \sum_{{\bf Q}m} \ q_m {\mathcal{G}}_{m}(z) \ e^{i {\bf Q} \cdot {\bf R}} e^{i \omega_{m}(Q) t} \ ,
\label{eq:shape_el_cont_2}
\end{multline}
where, again, ${\bf q}_m=({\bf Q},q_{z,m})$ is the `total' phonon wave vector. 
Following again Refs.~\cite{Donetti_2006} and \cite{Bimin_2025}, we can approximate these `shape functions' as follows: 
\begin{equation}
\mathcal{K}_m(z) = \left \{
\begin{array}{ll}
   \left ( \frac{2}{t_{\rm s}} \right )^{1/2} \sin \left ( \frac{m\pi}{t_{\rm s}} z \right )   & \mbox{for CBCs} \\
   -\left ( \frac{2}{t_{\rm s}} \right )^{1/2} \cos \left ( \frac{m\pi}{t_{\rm s}} z \right )   & \mbox{for FSBCs}
\end{array} \right. 
\label{eq:shape_sheet_K}
\end{equation}
\begin{equation}
\mathcal{G}_m(z) = \left \{
\begin{array}{ll}
   \left ( \frac{2}{t_{\rm s}} \right )^{1/2} \cos \left ( \frac{m\pi}{t_{\rm s}} z \right )   & \mbox{for CBCs} \\
   \left ( \frac{2}{t_{\rm s}} \right )^{1/2} \sin \left ( \frac{m\pi}{t_{\rm s}} z \right )   & \mbox{for FSBCs.}
\end{array} \right. 
\label{eq:shape_sheet_G}
\end{equation}
For optical phonons, instead, the coupling to electrons is given by the ionic displacement field, not by its divergence, so we shall assume that  
it is proportional to $\mathcal{K}_m(z)$.\\

This assumption is motivated by the results of Ridley {\it et al.}~\cite{Ridley_2000a}. Using the elastic-continuum model, they have derived for the acoustic and optical ionic displacement fields, ${\bf u}_{\rm ac}={\bf u}_{\rm A}+{\bf u}_{\rm B}$ and ${\bf u}_{\rm op}={\bf u}_{\rm A}-{\bf u}_{\rm B}$ 
(where ${\bf u}_{A/B}$ is the displacement field of the sublattice A/B) equations that for a homogeneous lattice with two identical ions in the unit cell take
the form:
\begin{multline}
-\omega^2 {\bf u}_{\rm ac} = c_{\rm L}^2 \nabla \nabla \cdot {\bf u}_{\rm ac} - c_{\rm L}^2 \nabla \times \nabla \times {\bf u}_{\rm ac} \\ =
                             c_{\rm T}^2 \nabla^2 {\bf u}_{\rm ac} + (c_{\rm L}^2 - c_{\rm T}^2 ) \nabla \nabla \cdot {\bf u}_{\rm ac}
\label{eq:acoustic_u}
\end{multline}
\begin{multline}
-(\omega^2 - f/\rho){\bf u}_{\rm op} = c_{\rm L}^2 \nabla \nabla \cdot {\bf u}_{\rm op} - c_{\rm T}^2 \nabla \times \nabla \times {\bf u}_{\rm op} \\ =
                             c_{\rm T}^2 \nabla^2 {\bf u}_{\rm op} + (c_{\rm L}^2 - c_{\rm T}^2 ) \nabla \nabla \cdot {\bf u}_{\rm op}, 
\label{eq:optical_u}
\end{multline} 
where $c_{\rm L}$ and  $c_{\rm T}$ are the longitudinal and transverse sound velocities, $f$ is the inter-sublattice force (spring) constant
(so that $\omega_{\rm op} = (f/\rho)^{1/2}$ is the optical phonon frequency), 
$\rho$ is the mass density, and we have assumed a time dependence of the form ${\bf u}_{\rm ac/op} e^{i\omega t}$. These expressions show that the optical and 
acoustic displacement fields have the same spatial dependence. However, a position-independent ${\bf u}_{\rm ac}$ represents a uniform translation of the 
lattice that does not result in any perturbation. On the contrary, a position independent ${\bf u}_{\rm op}$ represent a uniform shift of sublattice A 
with respect to sublattice B. This does result in a deformation-potential perturbation. Finally, note that for $m=0$, when assuming FSBCs the shape function 
$\mathcal{G}_m$ should be taken as $t_{\rm s}^{-1/2}$, whereas $\mathcal{K}_0=0$ when assuming CBCs. Overall, in the case of FSBCs, the ionic-displacement field
obtained using molecular dynamics simulations of freestanding thin Si films at 200~K, as shown in Fig.~9 of Ref.~\cite{Fu_2020}, is consistent with the form of 
$\mathcal{K}_m(z)$ given in Eq.~(\ref{eq:shape_sheet_K}) with $m=1$.\\
 
Overall, modes originating from bulk longitudinal phonons (LA) correspond to dilatational modes, whereas modes originating from bulk transverse phonons correspond 
either to shear waves (TA1) or flexural modes (TA2 or ZA). In total, we have $12N$ phonons, equally divided ($4N$ each) among dilatational, shear, and flexural
modes. This is what we expect, since the supercell contains $4N$ atoms, yielding indeed $12N$ modes. \\

Note that we do account for the coupling of electrons with the flexural (ZA) phonons. Indeed, the atomic structure of the nanosheets we consider 
is not  mirror-symmetric ($\sigma_{\rm h}$-symmetric). Therefore, the first-order electron/ZA-phonon interaction is not forbidden. This interaction does indeed   
result in vanishing intra-subband matrix elements, but results in non-zero matrix elements for inter-subband transitions between electronic states of opposite
parity. A more important problem arises from their (in principle) parabolic dispersion that gives rise to a divergent scattering rate~\cite{Fischetti_2016}
This problem is absent in our model, since the model-dispersion given by Eq.~(\ref{eq:ac_phon_sheet}) does not exhibit a parabolic behavior. Physically, such 
a renormalization of the dispersion may be expected from the `clamping' caused by the interaction of the nanosheet with the gate insulators
(see the brief discussion in Appendix A of Ref.~\cite{Bimin_2025}).
\subsection{Boundary conditions accounting for confinement of SiO$_2$ acoustic phonons}
\label{sec:outer_CBCs}
As shown below in Table~\ref{Tab:phonon-limited-mobility} in Sec.~\ref{sec:Results}, 
assuming the two extreme cases of FSBCs and CBCs for the confined Si phonons results in a vastly different mobility. This uncertainty makes any prediction impossible (or, rather, suspicious). Any way to reduce this uncertainty must rely on a better selection of boundary conditions. Given the strong dielectric stiffness of SiO$_2$ (its dominant optical mode has an energy of about 138~meV, much larger than the frequency of optical phonons in Si, about 60~meV), we may consider CBCs at the Si/SiO$_2$ interface quite realistic. Not so for acoustic phonons. Indeed, assuming acoustic CBCs at the 
Si/SiO$_2$ interfaces seems inappropriate since, as shown in Table~\ref{Tab:elastic_parameters}, the Yong modulus (so, the `stiffness') of Si is close to that of
SiO$_2$; therefore, acoustic vibrations may extend from Si into SiO$_2$ without any significant reflection. Confining acoustic phonons within the Si nanosheet 
assuming CBCs results in a large zero-point energy for the lowest-energy branch, as shown in Fig.~\ref{fig:Phonon_dispersion},
approximately 14~meV for the dilatational modes and 9~meV for the flexural phonons. This lack of low-energy phonons results in reduced scattering for low-energy
electrons and, probably, in an unrealistically large low-field electron mobility, approximately 1,060 cm$^2$/(Vs). On the other hand,
FSBCs do not seem to be appropriate either: A look at the parameters listed in Table~\ref{Tab:elastic_parameters} shows that that HfO$_2$ is very stiff, mainly 
because Hf is a heavy element with an atomic weight, 174.49~amu, that is much larger than Si (28.08~amu) or O (15.999 amu). Therefore, acoustic ionic vibrations are 
likely to be reflected at the SiO$_2$/HfO$_2$ interfaces and it is reasonable to consider CBCs at the SiO$_2$/HfO$_2$ interfaces, while still relying on the
elastic continuum approximation. This amounts to considering also the confinement of acoustic phonons in the SiO$_2$ layers, as has been done before by 
Donetti {\it al.}~\cite{Donetti_2006,Donetti_2007}. The expected effect is a reduction of the zero-point energy of the dilatational and flexural phonons, 
approximately to 4~meV and 3~meV, respectively. As shown in Table~\ref{Tab:phonon-limited-mobility}, this reduces the mobility by a factor of 4.
\setlength\extrarowheight{-5pt}
\begin{table*}[tb]
\centering
\caption{Elastic/mechanical parameters of Si, SiO$_2$ and HfO$_2$.}
\label{Tab:elastic_parameters}
\begin{tabular}{lccccc}
\hline\\
\bf{Quantity}                &        \bf{Symbol}     &     \bf{SiO$_2$}  &      \bf{HfO$_2$}          &   \bf{Si}    & \bf{Units}   \\
\hline\\   
Mass density                 &          $\rho$        &         2.3          &      9.8-10.8           &     2.33     & gr/cm$^{3}$   \\
Longitudinal sound velocity  &       $c_{\rm L}$      &         5.6          &   5.8-6.5$^{\rm (a)}$   &     9.8      & $10^5$ cm/s   \\
Transverse sound velocity    &       $c_{\rm T}$      &         3.4          &   2.9-3.4$^{\rm (a)}$   &     5.4      & $10^5$ cm/s   \\
Young modulus                &  $Y=\rho c_{\rm s}^2$  &       26.6-72.1      &      113-414            &    67-221    & GPa           \\
Young modulus (exp)          &         $Y$            &  10-100$^{\rm (b)}$  &   163-320$^{\rm (c)}$   &   150-180$^{\rm (d)}$    & GPa           \\
\hline
\end{tabular}
\begin{enumerate}[label=(\alph*),itemsep=-0.1cm]
\item Ref.~\cite{Qi_2025}
\item Ref.~\cite{Kudelka_2020}
\item Ref.~\cite{Dole_2006}
\item Ref.~\cite{Hopcroft_2010}
\end{enumerate}
\end{table*}
Therefore, for acoustic phonons, following the approach employed by Donetti {\it et al.}~\cite{Donetti_2006,Donetti_2007}, we may assume that the ionic
displacement on the $(x,y)$ plane for dilatational modes or along the $z$ direction for the flexural modes, including the SiO$_2$ films, has the form:
\begin{equation}
w_{m}(z) =
\left \{
\begin{array}{ll}
A_{m} \sin [ q_{z,m} (z-z_0) ]        & (0 \le z < t_{\rm s}) \\ \\
B_{m} \sin [ \beta_{m} (z+t_1) ]      & (t_1 \le z < 0 \\
                                      & \mbox{ and } t_{\rm s} \le z < t_{\rm s} + t_1 )
\end{array}
\right.
\label{eq:newCP_1}
\end{equation}
with $q_{z,m}=m\pi/(t_{\rm s}+2 z_0)$ and $\beta_{m}=m\pi/(t_{\rm s}+2 t_1)$. The quantities $A_m$, $B_m$, and $z_0$ must be determined imposing the
normalization of this new `shape function' and the continuity of $w_m(z)$ and of the stress tensor at one of the Si/SiO$_2$ interfaces (continuity at the other 
interface is ensured by the assumed symmetry).\\

The continuity of the stress tensor at the two Si/SiO$_2$ interfaces implies: 
\begin{multline}
\begin{array}{l}
T_{xz} = \\ 
      \rho_{\rm SiO_2} c_{\rm T}^{\rm (SiO_{2}) 2} \left ( \frac{{\rm d}w_{x}}{{\rm d}z} +iQ w_z \right ) = 
      \rho_{\rm S} c_{\rm T}^{\rm (Si) 2} \left ( \frac{{\rm d}w_{x}}{{\rm d}z} +iQ w_z \right )   \\ \\
T_{zz} = \\
      \rho_{\rm SiO_2} \left [ c_{\rm L}^{\rm (SiO_{2}) 2} \frac{{\rm d}w_{x}}{{\rm d}z} +iQ(c_{\rm L}^{\rm (SiO_{2}) 2}-c_{\rm T}^{\rm (SiO_{2}) 2}) w_x \right ] = \\
      \rho_{\rm Si} \left [c_{\rm L}^{\rm (Si) 2} \frac{{\rm d}w_{x}}{{\rm d}z} +iQ(c_{\rm L}^{\rm (Si) 2}-c_{\rm T}^{\rm (Si) 2}) w_x \right ] \ .   
\end{array}
\label{eq:newCP_2a}
\end{multline}
Here, we are not interested in obtaining the `correct' phonon dispersion, since we employ, instead, the folded bulk dispersion, Eq.~(\ref{eq:ac_phon_sheet})). On
the contrary, we are interested in finding in the quantized transverse phonon wave vector, $q_{z,m}$, and the amplitude of the ionic displacement field 
$w_m(z)$ (actually, its divergence). Neither of these quantities are likely to depend significantly on the in-plane phonon wave vector ${\bf Q}$. Therefore, we
may consider the small-$Q$ limit of these expressions. Moreover, for dilatational waves we expect $w_z = - \sigma w_x < |w_x|$ and 
$w_x = - \sigma w_z < |w_z|$ for flexural waves, since the Poisson ratio $\sigma$ is about 0.28 for Si and 0.17 for SiO$_2$. Thus, we can consider only
the continuity of the larger component of the displacement at long wavelengths: 
\begin{equation}
\begin{array}{cc}
T_{xz} \approx \rho_{\rm SiO_2} c_{\rm T}^{\rm (SiO_{2}) 2} \frac{{\rm d}w_{x}}{{\rm d}z} =
         \rho_{\rm S} c_{\rm T}^{\rm (Si) 2} \frac{{\rm d}w_{x}}{{\rm d}z}    
                                                             & \mbox{(dilatational)} \\ \\
T_{zz} \approx \rho_{\rm SiO_2} c_{\rm L}^{\rm (SiO_{2}) 2} \frac{{\rm d}w_{x}}{{\rm d}z} =
          \rho_{\rm Si} c_{\rm L}^{\rm (Si) 2} \frac{{\rm d}w_{x}}{{\rm d}z}   
                                                             & \mbox{(flexural)} \ ,
\end{array}
\label{eq:newCP_2}
\end{equation}
Using these equations and accounting also for the continuity of $w_m(z)$ at $z=0$, we obtain for $B_m$:
\begin{equation}
B_m = A_m \frac{\sin(q_{z,m} z_0)}{\sin(\beta_m t_1)} \ ,
\label{eq:newCP_3}
\end{equation}
whereas, from Eq.~(\ref{eq:newCP_2}), $z_0$ is given by the smallest root of the transcendental equation:
\begin{multline}
\frac{1}{\rho_{\rm Si}    c^{\rm (Si)    2}_{\rm L/T}} (t_{\rm s}+2 z_0) \tan \left ( \frac{m \pi}{t_{\rm s}+2 z_0} z_0 \right ) = \\
\frac{1}{\rho_{\rm SiO_2} c^{\rm (SiO_2) 2}_{\rm L/T}} (t_{\rm s}+2 t_1) \tan \left ( \frac{m \pi}{t_{\rm s}+2 t_1} t_1 \right ) \ .
\label{eq:newCP_4}
\end{multline}

In the limits of a small ratio 
$\rho_{\rm Si}c^{\rm (Si)2}_{\rm L/T}/(\rho_{\rm SiO_2} c^{\rm (SiO_2) 2}_{\rm L/T}) = Y_{\rm Si}/Y_{\rm SiO_2}$ and/or large $t_{\rm s}$, $z_0$ approaches the value 
$t_1 (Y_{\rm Si}/Y_{\rm SiO_2}) \approx  t_1 (c^{\rm (Si)}_{\rm L}/c^{(\rm SiO_2)}_{\rm L})^2 \approx  2.5 t_1$ for the dilatational modes, $\approx  3.1 t_1$ 
for the flexural modes. In general, for $m$ in the range (1, 16) (see below why this range matters), the value of the smallest root $z_0$ of
Eq.~(\ref{eq:newCP_4}) varies between 49\% and 91\% of this value. Since we are ignoring a small `penetration' of the displacement field inside the HfO$_2$ 
layers (an effect that would increase the value of $z_0$), as a rough approximation, we may assume the value $t_1 (c^{\rm (Si)}_{\rm L/T}/c^{\rm (SiO_2)}_{\rm L/T})^2$
for all values of $m$.\\

As a consequence of this discussion, optical phonons may be treated assuming CBCs at the Si/SiO$_2$ interfaces, as described above. 
On the contrary, the acoustic-phonon shape function given by Eq.~(\ref{eq:shape_sheet_G}) is replaced by:
\vspace*{-0.2cm}
\begin{multline}
\mathcal{G}_m(z) = \left ( \frac{2}{{\mathcal{N}}_m} \right )^{1/2} \\ \times  \left \{
\begin{array}{ll}
  \cos [ q_{z,m} (z-z_0) ]        & (0 \le z < t_{\rm s}) \\
  \cos [ \beta_{m} (z+t_1) ]      & (t_1 \le z < 0 \mbox{ and } t_{\rm s} \le z < t_{\rm s} + t_1 )
\end{array} \right. , 
\label{eq:newCP_5}
\end{multline}
where the normalization constant ${\mathcal N}_{m}$ is given by the rather cumbersome expression
\vspace*{-0.2cm}
\begin{multline}
{\mathcal{N}}_m = t_{\rm s} + \frac{\sin(q_{z,m} t_{\rm s})}{q_{z,m}} + 2 \frac{\sin^2(q_{z,m}z_0)}{\sin^2(\beta_{m}t_1)} \\ \times
     \left [ t_1 - \frac{1}{\beta_m} \sin (\beta_m t_1) \cos(\beta_m t_1) \right ] \ .
\label{eq:newCP_6}
\end{multline}  
In the same spirit in which we assumed an $m$-independent value for $z_0$, we may approximate this normalization factor with the $m$-independent form:
\vspace*{-0.2cm}
\begin{equation}
{\mathcal{N}}_m \approx t_{\rm s} + 2 \frac{c_{\rm L/T}^{\rm (Si)}}{c_{\rm L/T}^{\rm (SiO_2)}} z_0 \ .
\label{eq:newCP_7}
\end{equation}
This term acts as a booster of the mass density of the nanosheet: Phonons now oscillate with a smaller amplitude, as the mass density accounts fogr both the Si sheet 
and the top and bottom SiO$_2$ gate insulators. This reduces the strength of the electron-phonon interaction (see the factor $\rho_{\rm x}$ in the denominator of
Eq.~(\ref{eq:ep_coupling})). However, this is compensated by the larger number of phonon branches that are confined in the thicker Si/SiO$_2$ system, 
as discussed in the next paragraph. We should recall that the coupling between electrons these acoustic modes can be obtained by following the procedure 
described in Appendix of Ref.~\cite{Bimin_2025}: Equations~(A20) and (A24) of this reference yield for the ${\bf Q}$-component of the Fourier-Bessel transform of
the divergence of the ionic displacement field, $\nabla {\bf u}$, the expression  
$i{\bf Q} \cdot {\bf w}_{\parallel}+{\rm d}w_z/{\rm d}z \rightarrow q_m \mathcal{G}_m(z)$, with ${\bf q}_{m} =({\bf Q}, q_{z,m})$.\\

\begin{figure}[tb]
\centerline{\hbox{
\includegraphics[width=8.60cm]{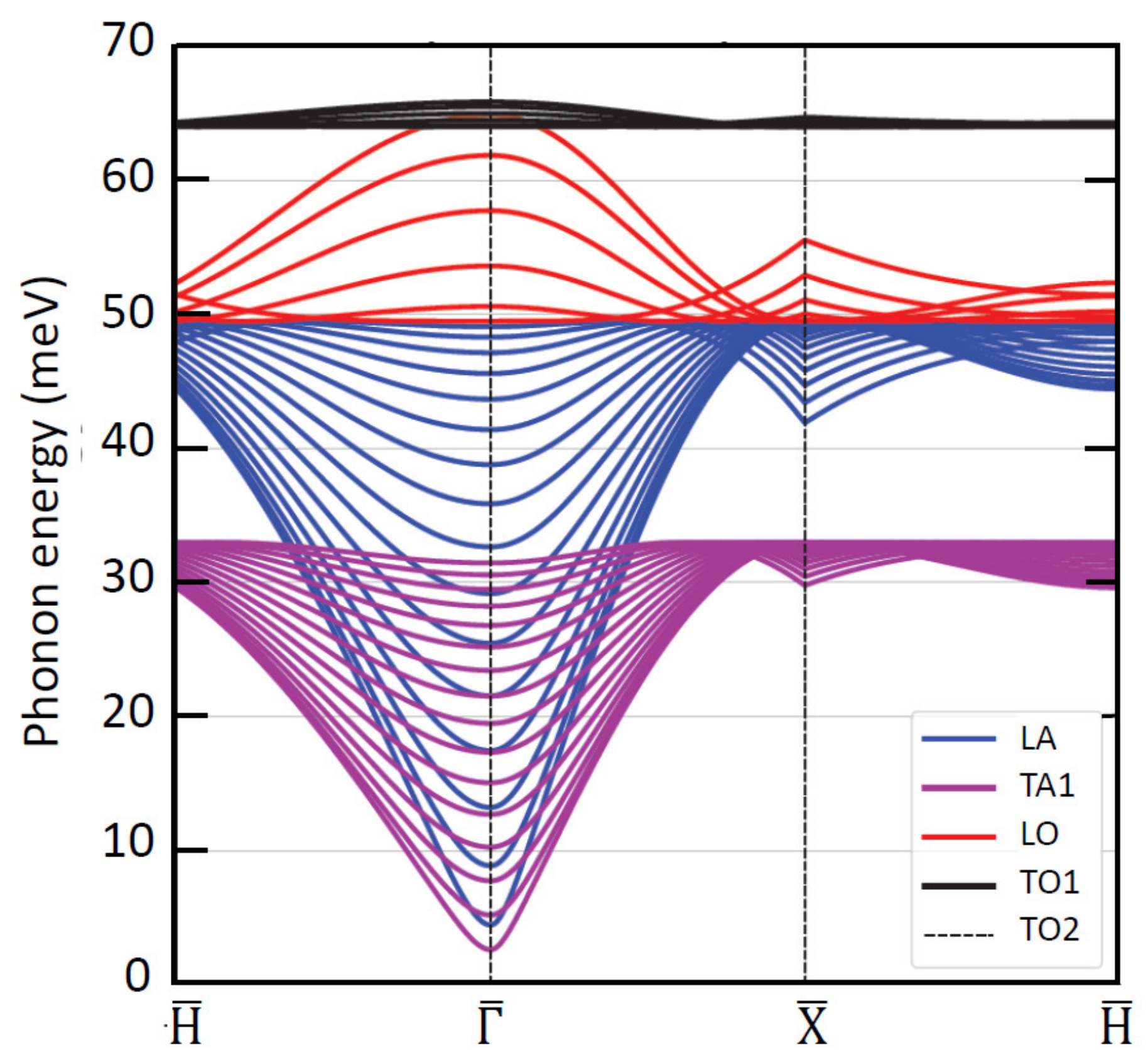}}}
\caption{As in the center frame of Fig.~\ref{fig:Phonon_dispersion}, but considering confinement of the acoustic phonons also in the SiO$_2$ layers assuming 
         clamped boundary conditions at the SiO$_2$/HfO$_2$ interfaces using the elastic continuum approximation given by 
         Eqs.~(\ref{eq:newCP_2})-(\ref{eq:newCP_4}). Comparison with the center frame of Fig.~\ref{fig:Phonon_dispersion} shows a smaller `zero-point' energy 
         of the acoustic phonons as well as a larger number of acoustic branches.}
\label{fig:phonon_dispersion_new}             
\end{figure}

Finally, the dispersion of the acoustic branches is still given by Eqs.~(\ref{eq:ac_phon_sheet}) with unchanged parameters except, of course, for 
$q_{z,m}$ that now is $m\pi/(t_{\rm s}+2 z_0)$. Their identification (dilatational modes emerging from the bulk LA phonons, flexural modes from
the bulk TA phonons) also remains unchanged. However, in addition to the different `shape function', Eq.~(\ref{eq:newCP_5}) instead of 
$\mathcal{K}_m(z)$ given by Eq.~(\ref{eq:shape_sheet_K}), what changes is the number of acoustic modes we should consider. This is now larger: The maximum transverse wave vector in Si, set by the size of the bulk Si primitive cell, remains $2N\pi/t_{\rm s}$. However, since now $q_{z,m}=m\pi/(t_{\rm s}+2z_0)$, the maximum number 
of modes $m$ consistent with the size of the Si cell is roughly $2N(t_{\rm s}+2z_0)/t_{\rm s} \approx$ 16 for both dilatational and flexural modes (assuming 
$z_0 = t_1 [(\rho_{\rm Si}c^{\rm (Si)}_{\rm L/T})/(\rho_{\rm SiO_2} c^{\rm (SiO_2)}_{\rm L/T})]^2$). Considering also the shear modes (that we ignore since they 
do not couple to electrons), this leads to 16$\times$3=48 total acoustic modes. This value is consistent with an atomistic `count': A 0.6~nm-thin SiO$_2$ layer is
slightly thicker than a single primitive cell of $\alpha$-quartz that contains 9 atoms (3 SiO$_2$ groups). Therefore, assuming 10 atoms per SiO$_2$ layer 
and 12 Si atoms in the nanosheet, we have a total of 32 atoms. This indeed results in a total of 32$\times$3=96 phonon branches, 48 of which
are acoustic. Figure~\ref{fig:phonon_dispersion_new} shows the dispersion of the phonons in a 3-cell-thick Si nanosheet obtained assuming these new
boundary conditions.\\

\section{Electron-phonon scattering rates}
\label{sec:El-ph}
Using the model for the dispersion of phonons confined in 2D, as discussed in Sec.~\ref{sec:Phonons} and Fermi's golden rule, 
the rate at which an electron with wave vector ${\bf K}$ in (sub)band $\nu$ emits or absorbs a phonon in branch $(\eta,m)$ can be written as:
\begin{multline}
\hspace*{-0.4cm}
\frac{1}{\tau^{(\eta,m,\pm)}({\bf K},\nu)} = \frac{2 \pi}{\hbar} \sum_{\nu^{\prime}} \ \frac{1}{(2 \pi)^2} \int {\rm d} {\bf K}' 
  \left \vert \braket{{\bf K}^{\prime} \nu^{\prime}|{\widehat{V}}^{(\eta)}_{q_m}|{\bf K} \nu} \right \vert^2 \\ \times
    \delta [ E^{(\nu^{\prime})}({\bf K}^{\prime}) - E^{(\nu)}({\bf K}) \pm \hbar \omega^{(\eta)}_m({\bf K}^{\prime}-{\bf K}) ] \ .
\label{eq:tau_ep_sheet}
\end{multline}
where ${\bf q}_m = ({\bf K}^{\prime}-{\bf K},q_{z,m})$ is the phonon wave vector, $\omega^{(\eta)}_m({\bf Q})$ is the frequency of a phonon of branch $(\eta,m)$ 
and wave vector ${\bf Q}$ (a frequency that we take as isotropic, dependent only of the magnitude $Q$ of the phonon wave vector), 
and the plus/minus sign refers to emission/absorption processes, respectively. The coupling Hamiltonian ${\widehat{V}}^{(\eta)}_{q_m}$ is 
$V^{(\eta)}_{q} {\mathcal{G}}_m(z)$ for acoustic phonons, $V^{(\eta)}_{q}{\mathcal{K}}_m(z)$ for optical phonons, and the
`coupling constant' $V^{(\eta)}_{q}$ is:
\begin{multline}
V^{(\eta)}_{q_m} = \left ( \frac{\hbar}{2 \rho_{\rm x} \omega^{(\eta)}_m(Q)} \right )^{1/2} (DK)({\bf q}_m) \\ \times
     \left \{ \begin{array}{l} \{ 1+N[\omega^{(\eta)}_m(Q)]\}^{1/2} \\ N[\omega^{(\eta)}_m(Q)]^{1/2} \end{array}  \right \} \ ,
\label{eq:ep_coupling}
\end{multline}
where $\rho_{\rm x}$ is the bulk mass-density of the crystal and $N(\omega)$ is the phonon Bose-Einstein distribution function at frequency $\omega$.\\

Regarding the deformation potential $(DK)({\bf q})$, as we shall discuss below, we can set it equal to a constant value $(DK)_{\rm op}$ for optical phonons. 
For acoustic phonons, we set $DK({\bf q}) = q \Delta_{\rm dil/flex}$ for dilatational and flexural acoustic phonons 
(setting $\Delta_{\rm dil} = \Delta_{\rm LA}$ and $\Delta_{\rm flex} = \Delta_{\rm TA}$).\\

Using the cell-averaged wavefunctions defined by Eq.~(\ref{eq:zetaK}), the matrix element can be written as:  
\begin{equation}
\braket{{\bf K}^{\prime} \nu^{\prime}|V^{(\eta)}_{q_m}|{\bf K} \nu} = V^{(\eta)}_{q_m}  
  \int_{0}^{t_{\rm s}} {\rm d}z \ \zeta^{(\nu^{\prime}) \ast}_{{\bf K}'}(z) \ {\mathcal{G}}_m(z) \ \zeta^{(\nu)}_{\bf K}(z) \ ,
\label{eq:Matel_ep_sheet}
\end{equation}
for scattering with acoustic phonons,
\begin{equation}
\braket{{\bf K}^{\prime} \nu^{\prime}|V^{(\eta)}_{q_m}|{\bf K} \nu} = V^{(\eta)}_{q_m}  
  \int_{0}^{t_{\rm s}} {\rm d}z \ \zeta^{(\nu^{\prime}) \ast}_{{\bf K}'}(z) \ {\mathcal{K}}_m(z) \ \zeta^{(\nu)}_{\bf K}(z) \ ,
\label{eq:Matel_ep_sheet_op}
\end{equation}
for scattering with optical phonons.
In discretized form, Eq.~(\ref{eq:tau_ep_sheet}) takes the form:
\begin{multline}
\frac{1}{\tau^{(\eta,m,\pm)}({\bf K},\nu)} = \frac{2 \pi}{\hbar} \sum_{j,\nu^{\prime}}{}^{\prime} \ 
  \left \vert \braket{{\bf K}_j \nu^{\prime}|V^{(\eta)}_{q_{j,m}}|{\bf K} \nu} \right \vert^2 \\ \times
   {\mathcal{D}}^{(\nu^{\prime})}_{j}(E^{(\nu)}_{\bf K}\pm \hbar \omega^{(\eta)}_{m,{\bf K}_j-{\bf K}}) \ ,
\label{eq:tau_ep_sheet_discretized}
\end{multline}
where ${\mathcal{D}}^{(\nu)}_{j}(E)$ is the density of states at energy $E$ in band $\nu$ in the mesh element $j$, the `primed' sum extends only over `energy
conserving' mesh elements $j$ in band $\nu^{\prime}$ ({\it i..e.}, those mesh elements that `contain' the final energy 
$E^{(\nu)}_{\bf K}\pm \hbar \omega^{(\eta)}_m({\bf K}_j-{\bf K})$, or, explicitly, such that 
$E^{(\nu^{\prime})}_{j,{\rm min}} \le E^{(\nu)}_{\bf K}\pm \hbar \omega^{(\eta)}_{m,{\bf K}_j-{\bf K}} \le E^{(\nu^{\prime})}_{j,{\rm max}}$), 
and the phonon wave vector now is given by  ${\bf q}_{j,m} = ({\bf K}-{\bf K}_j,q_{z,m})$.\\

As is usually done in bulk Si, we shall consider nonpolar scattering only with the acoustic dilatational and flexural modes arising from only one of the 
bulk transverse acoustic modes, since the other TA phonon is a shear mode. Coupling with the ZO phonons is, in principle, allowed by symmetry, but we shall lump 
this contribution into a single deformation potential ${DK}_{\rm op}$.\\

{\it Bulk phonons, high-temperature,  elastic approximation.} It is interesting to compare the scattering rates calculated using the model-dispersion for 
confined acoustic phonons with those calculated considering bulk acoustic phonons in the high-temperature elastic approximation. In this limit, the phonon 
dispersion at small $q$ is approximated by $\omega^{\rm LA/TA}(q) \rightarrow c_{\rm L/T} q$ where the sound velocity is 
$c_{\rm L/T}=\omega_{\rm BZ}^{\rm (LA/TA)} a_0/4$. Moreover, for $k_{\rm B}T \gg \hbar \omega^{\rm (LA/TA)}(q)$, the Bose-Einstein factors become 
$N[\omega^{\rm LA/TA}(q)] \approx 1+N[\omega^{\rm (LA/TA)}(q)] \rightarrow k_{\rm B}T/[\hbar \omega^{\rm LA/TA}(q)]$.
Therefore, converting the sum over all confined modes $m$ of branch $\eta$= LA/TA in Eq.~(\ref{eq:tau_ep_sheet_discretized}) to an integral over the
bulk phonon dispersion along the $z$ direction, we have:
\begin{multline}
\frac{1}{\tau^{\rm (LA/TA),\pm}({\bf K},\nu)} = 
   \frac{2 \pi}{\hbar} \frac{\Delta_{\rm LA/TA}^2 k_{\rm B}T}{2 \rho c^2_{\rm L/T}} \\ \times \sum_{j,\nu^{\prime}}{}^{\prime} \ 
   \int \frac{{\rm d} q_z}{2 \pi} \ \left \vert \braket{{\bf K}_j \nu^{\prime}|e^{iq_{z} z}|{\bf K} \nu} \right \vert^2 
        {\mathcal{D}}^{(\nu^{\prime})}_{j}(E^{(\nu)}_{\bf K}) \ .
\label{eq:tau_ep_sheet_discretized_bulk}
\end{multline}
In this case, the overlap integral (often called 'form factor') can be reduced to the simple form:~\cite{Price_1981,Ridley_2000}:
\begin{multline}
 \int \frac{{\rm d} q_z}{2 \pi} \left \vert \braket{{\bf K}_j \nu^{\prime}|e^{iq_{z} z}|{\bf K} \nu} \right \vert^2 = \\
     \int_{0}^{t_{\rm s}} {\rm d} z \left \vert \zeta^{(\nu')}_{{\bf K}_j}(z) \right \vert^2 
                                       \left \vert \zeta^{(\nu)}_{\bf K}(z) \right \vert^2 \ \equiv \mathcal{F}^{(\nu \nu')}_{{\bf K}{\bf K}_j} \ . 
\label{eq:form_factor_F}
\end{multline}
Thus, finally, since the emission and absorption rates are identical (as we have assumed elastic scattering and the high-temperature limit implies that
$N[\omega^{\rm LA/TA}(q)] \approx 1+N[\omega^{\rm (LA/TA)}(q)]$), the total rate, accounting for both emission and absorption, becomes: 
\begin{equation}
\frac{1}{\tau^{\rm (LA/TA)}({\bf K},\nu)} \approx   
    \frac{2 \pi \Delta_{\rm LA/TA}^2 k_{\rm B}T}{\hbar \rho c^2_{\rm L/T}} \sum_{j,\nu'}{}^{\prime} \ \mathcal{F}^{(\nu \nu')}_{{\bf K}{\bf K}_j} 
         {\mathcal{D}}^{(\nu^{\prime})}_{j}(E^{(\nu)}_{\bf K}) \ ,
\label{eq:tau_ep_sheet_discretized_bulk_final}
\end{equation}
where, again, ${\mathcal{D}}^{(\nu)}_{j}(E)$ is the density of states per spin at energy $E$ in the mesh element $j$ and band $\nu$.
A similar procedure can be used to express the scattering rate with bulk (not confined) optical phonons, assumed to be dispersionless with frequency 
given by $\omega^{\rm (LO)}_{\overline{\Gamma}}$ (see Eqs.~(13.101) and (13.106) in Ref.~\cite{thebook}.\\
\setlength\extrarowheight{-5pt}
\begin{table*}[tb]
\centering
\caption{Phonon energies and deformation potentials for Si nanosheets. For completeness, commonly accepted values of the dilatation and uniaxial shear deformation
         potentials, $\Xi_{\rm d}$ and $\Xi_{\rm u}$, respectively, are also given in the last two rows.}
\label{Tab:sheets_parameters}
\begin{tabular}{lccc}
\hline\\
\bf{Quantity}                                        &              \bf{Symbol}                 &                     \bf{Value}              &    \bf{Units}  \\
\hline\\
Dilatational acoustic phonon energy at the zone edge & $\hbar \omega^{\rm (dil;ac)}_{\rm BZ}$   &       49.5$^{\rm (a)}$                      &     meV        \\
Flexural acoustic phonon energy at the zone edge     & $\hbar \omega^{\rm (flex;ac)}_{\rm BZ}$  &       33$^{\rm (a)}$                        &     meV        \\
Optical phonon energy at $\Gamma$ (LO)               & $\hbar \omega^{\rm (LO)}_{\Gamma}$    		&       66$^{\rm (a)}$                        &     meV        \\
Optical phonon energy at the zone edge (LO)          & $\hbar \omega^{\rm (LO)}_{\rm BZ}$    		&       49.5$^{\rm (a)}$                      &     meV        \\
Optical phonon energy at $\Gamma$ (TO1)              & $\hbar \omega^{\rm (TO1)}_{\Gamma}$   		&       66$^{\rm (a)}$                        &     meV        \\
Optical phonon energy at the zone edge (TO1)         & $\hbar \omega^{\rm (TO1)}_{\rm BZ}$   		&       64$^{\rm (a)}$                        &     meV        \\
Optical phonon energy at $\Gamma$ (TO2)              & $\hbar \omega^{\rm (TO2)}_{\Gamma}$   		&       66$^{\rm (a)}$                        &     meV        \\
Optical phonon energy at the zone edge (TO2)         & $\hbar \omega^{\rm (TO2)}_{\rm BZ}$   		&       64$^{\rm (a)}$                        &     meV        \\
Dilatation deformation potential (isotropic)         & $\Delta_{\rm dil}=\Delta_{\rm LA,bulk}$  &       10$^{\rm (b)}$                        & eV$^{\rm (c,d)}$ \\
Flexural deformation potential (isotropic)           & $\Delta_{\rm flex}=\Delta_{\rm TA,bulk}$ &        2$^{\rm (b)}$                        & eV$^{\rm (c,d)}$ \\
Optical deformation potential                        & $(DK)_{\rm op}$                          &           1.75$\times 10^8$$^{\rm (c)}$     &     eV/cm    \\   
Dilatation deformation potential (Herring-Vogt)      & $\Xi_{\rm d}$                            &        1.1$^{\rm (e)}$                      &     eV         \\
Uniaxial shear deformation potential (Herring-Vogt)  & $\Xi_{\rm u}$                            &        10.5$^{\rm (e)}$                     &     eV         \\ 
\hline
\end{tabular}
\begin{enumerate}[label=(\alph*),itemsep=-0.1cm]
\item Ref.~\cite{Neogi_2015}
\item Ref.~\cite{Canali_1975}
\item References~\cite{Fischetti_1988} and \cite{Fischetti_1996b} employ values of 1.2~eV for both $\Delta_{\rm LA,bulk}$ and $\Delta_{\rm TA,bulk}$
    for intraband processes within the lowest-energy conduction band, and 1.7~eV for processes involving higher-energy bands. The values listed here have 
    been corrected to account approximately for spin degeneracy and for the underestimated value of the sound velocity used in those references.
\item It may seem that the values for the acoustic deformation potentials used in Refs.~\cite{Jacoboni_1983} and \cite{Canali_1975} are very different from those
    used in Refs.~\cite{Fischetti_1988} and \cite{Fischetti_1996b}. This is not so. Indeed, the value of $\Delta_{\rm LA,bulk}=9$~eV from Refs.~\cite{Canali_1975}
    and \cite{Jacoboni_1983} was used ignoring scattering with the bulk TA phonons. Therefore, the values for $\Delta_{\rm LA,bulk}$ and $\Delta_{\rm TA,bulk}$ from
    Ref.~\cite{Fischetti_1988}, 2-to-2.5~eV and 1.2-to-1.5~eV, respectively, yield an equivalent 'LA-phonon-only' deformation potential 
    $[\Delta_{\rm LA,bulk}^2+2(c_{\rm L}/c_{\rm T})^2 \Delta_{\rm TA,bulk}^2]^{1/2}$$\approx$~5.8-to-8.1~eV and the scattering rates calculated using these 
    different sets of values may differ by as little as 20\% (using the largest value 2.8~eV) or even less, considering non-parabolic effects. Moreover, 
    taking the angle-averaged Herring-Vogt deformation potentials $\Xi_{\rm d}$ and $\Xi_{\rm u}$ (see Eqs.~(24)-(26) 
    of Ref.~\cite{Fischetti_1996}) yields effective deformation potentials around 2-to-2.5~eV for both LAs and TAs, values that are not too dissimilar from those 
    used in Refs.~\cite{Fischetti_1988} and \cite{Fischetti_1996b}.
\item Refs.~\cite{Fischetti_1996,Li_2021,Yang_2024,Williams_2025}
\end{enumerate}
\end{table*}

The deformation potentials we employ for nonpolar scattering with acoustic and optical phonons, as well as all other parameters used to treat electron-phonon
scattering, are listed in Table~\ref{Tab:sheets_parameters}. The dilatational, shear, and optical deformation potentials are similar to those used in the 
past~\cite{Fischetti_1988,Fischetti_1996b,Gamiz_2001,Donetti_2006,Donetti_2007}. They also yield inter-valley  
(inter-subband) deformation potentials consistent with the literature~\cite{Canali_1975,Jacoboni_1983}. In Ref.~\cite{Gamiz_2001} the effective-mass approximation 
was employed with the inter-valley deformation potentials of Refs.~\cite{Canali_1975} and inter-valley deformation potentials very similar to those employed here 
(that is, $\Delta_{\rm LA}$ = 9~eV and $\Delta_{\rm TA}$ = 1~eV) obtaining an electron mobility in double-gate thin Si layers that has been experimentally
verified~\cite{Uchida_2007}.\\
\begin{figure*}[t!]
\centerline 
{\hbox{
\includegraphics[width=5.70cm]{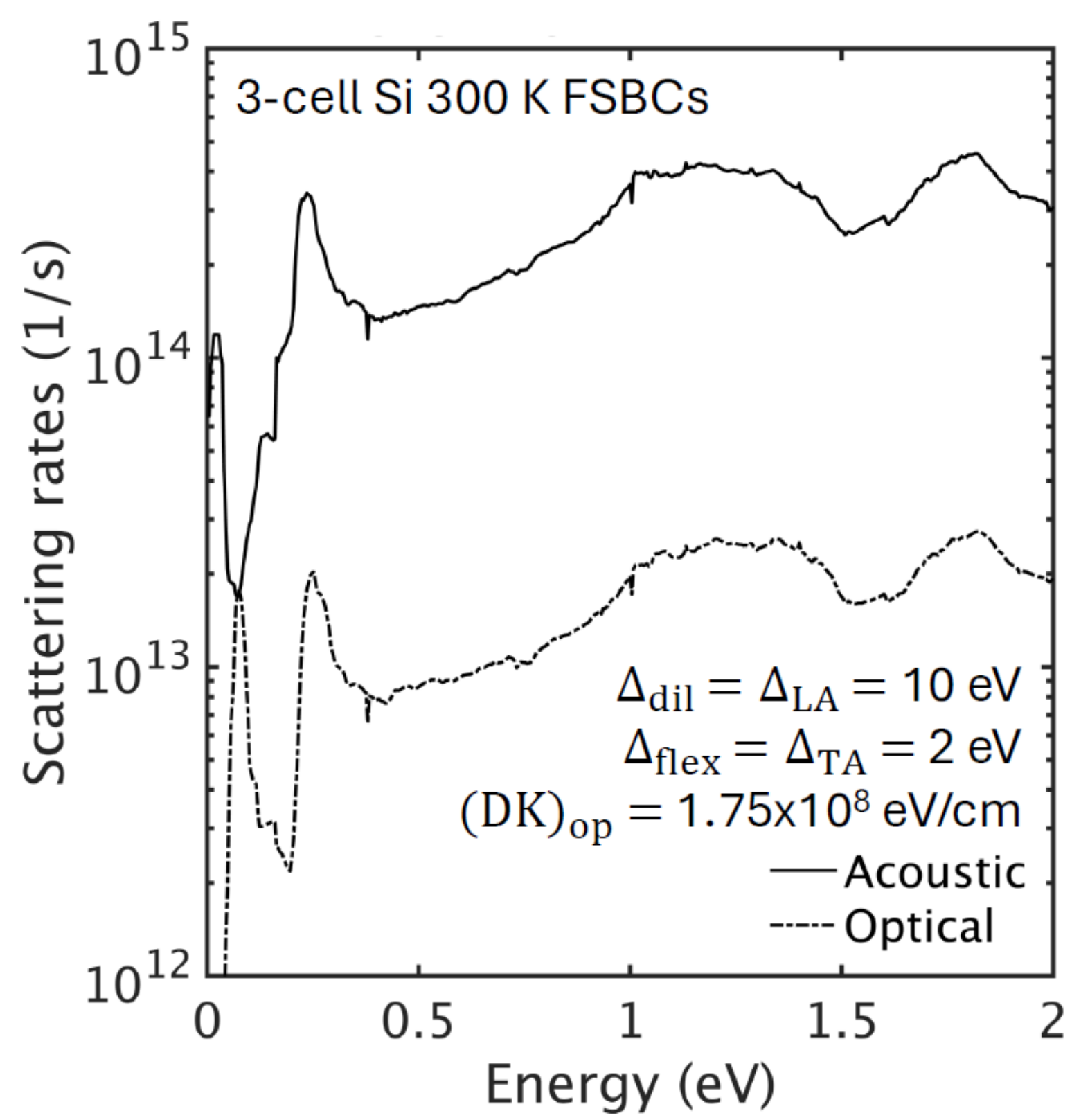}
\includegraphics[width=5.70cm]{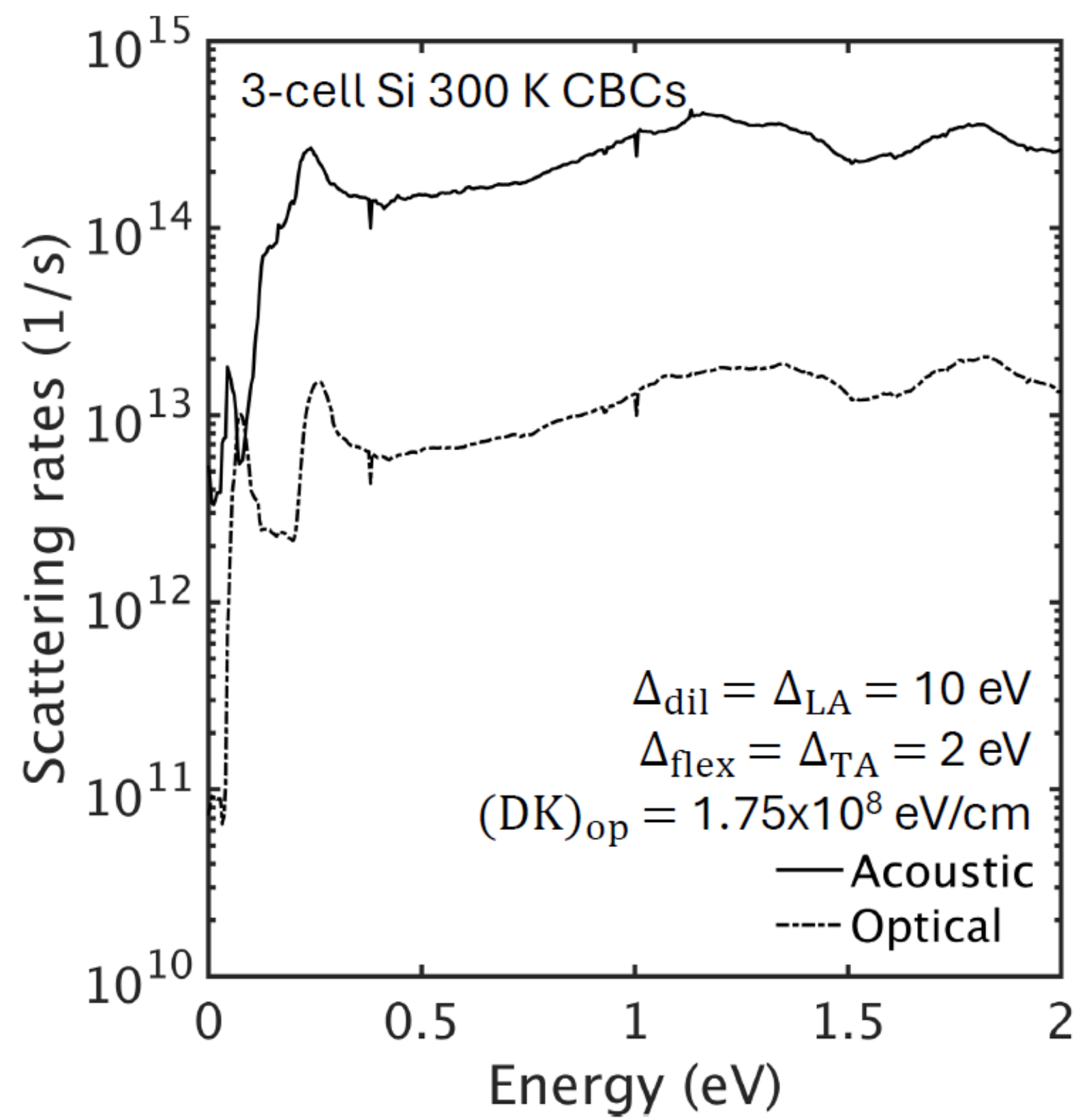}
\includegraphics[width=5.75cm]{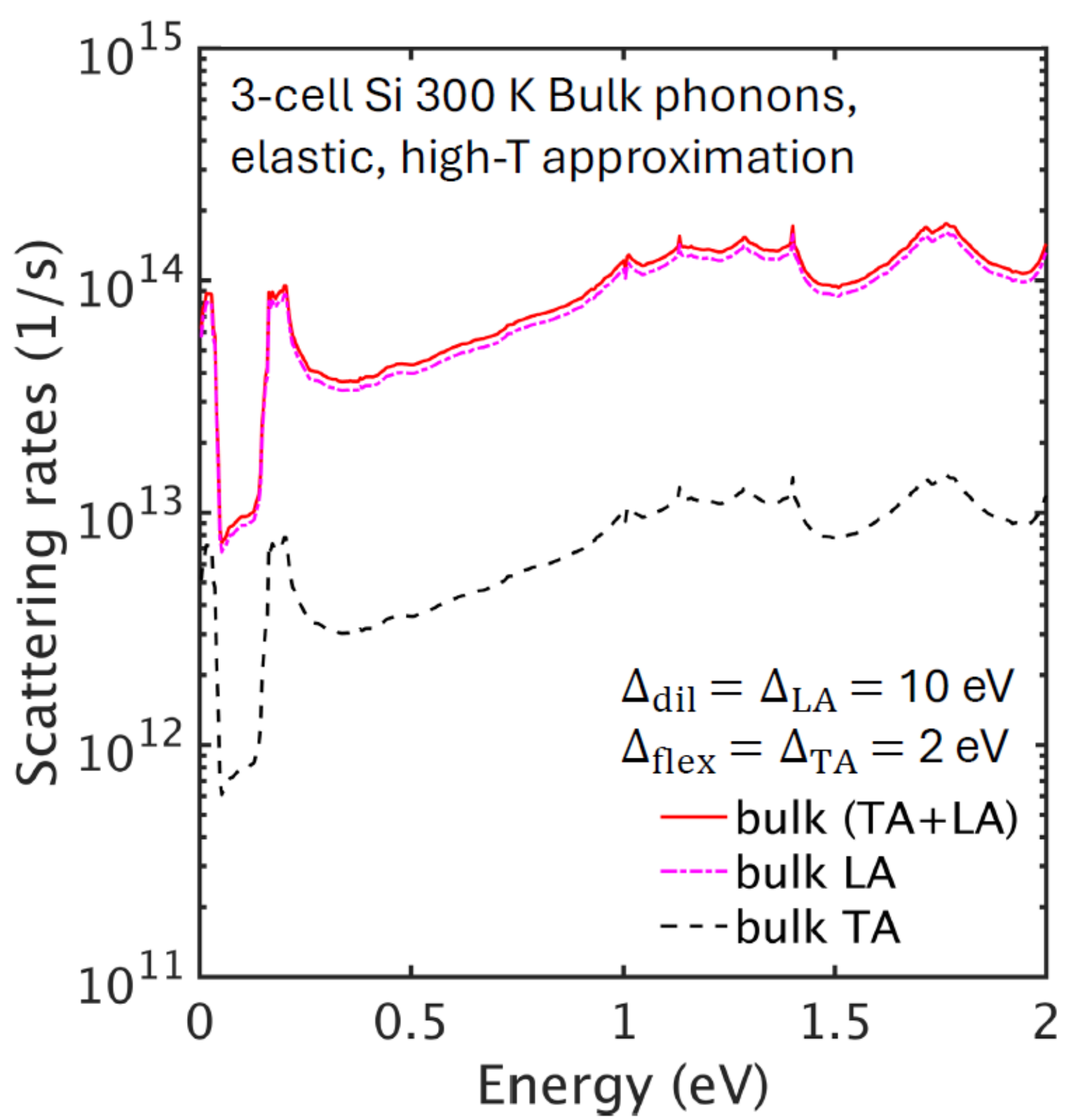}
}}
\caption{{\bf Left:} Unscreened electron-phonon scattering rates at 300~K for a 3-cell-thick Si nanosheet calculated assuming FSBCs.
         {\bf Center:} The same, but assuming CBCs.  
         {\bf Right}: The same, but assuming bulk acoustic phonons and the elastic, high-$T$ approximation often employed in the literature.
         The rates are plotted as a function of electron kinetic energy $E$ measured from the bottom of the conduction band. They have been obtained by
         averaging the total scattering rates $1/\tau({\bf K}, \nu)$ over equienergy shells $E_{(\nu)_{\bf K}}=E$ for any (sub)band $\nu$.          }  
\label{fig:El-ph_3-cell}
\end{figure*}

Figure~\ref{fig:El-ph_3-cell} shows the scattering rates with all acoustic and optical phonon branches, averaged over equienergy shells $E_{(\nu)_{\bf K}}=E$ 
for any (sub)band $\nu$, plotted as a function of $E$ measured from the bottom of the conduction band. The left frame shows the rates calculated assuming 
phonon free-standing boundary conditions, the center frame assuming clamped boundary conditions. The rates calculated assuming bulk phonons and
the elastic, high-$T$ approximation, Eq.~(\ref{eq:tau_ep_sheet_discretized_bulk_final}) (right frame), are similar to those calculated assuming confined 
phonons with FSBCs at low energy (left frame). Indeed, in the low-energy range, the phonon energy is sufficiently small for the the elastic and high-$T$
approximations to be valid. On the contrary, at higher energy, the increased density of states for confined phonons ({\it i.e.}, a smaller sound velocity) 
boosts the rates for confined phonons. On the contrary, scattering with phonons confined by CBCs results in smaller rates throughout the entire energy range
we have considered, since these BCs result in the absence of the lowest-energy phonon branches. Clamped boundary conditions, instead, result in smaller scattering
rates because of the absence of the low-energy acoustic branches. Therefore, we expect a low-field electron mobility higher when assuming CBCs than when assuming
FSBCs, with an intermediate value when assuming bulk phonons; in short, we expect $\mu_{\rm CBC} > \mu_{\rm bulk} > \mu_{\rm FSBC}$. These results are qualitatively
consistent with those obtained by Donetti {\it et al.}~\cite{Donetti_2006} who also found $\mu_{\rm CBC} > \mu_{\rm bulk} > \mu_{\rm FSBC}$ However, when
accounting for phonon confinement in the SiO$_2$ gate insulators~\cite{Donetti_2007}, they obtained a different ordering: 
$ \mu_{\rm bulk} > \mu_{\rm CBC} > \mu_{\rm FSBC}$, presumably since they considered thicker gate insulators ($t_{\rm SiO_2}>1$~nm)that induce a smaller 
zero-point energy of the confined acoustic phonons, thus depressing $\mu_{\rm CBC}$.\\

\begin{figure}[tb]
\centerline 
{\hbox{
\includegraphics[width=8.0cm]{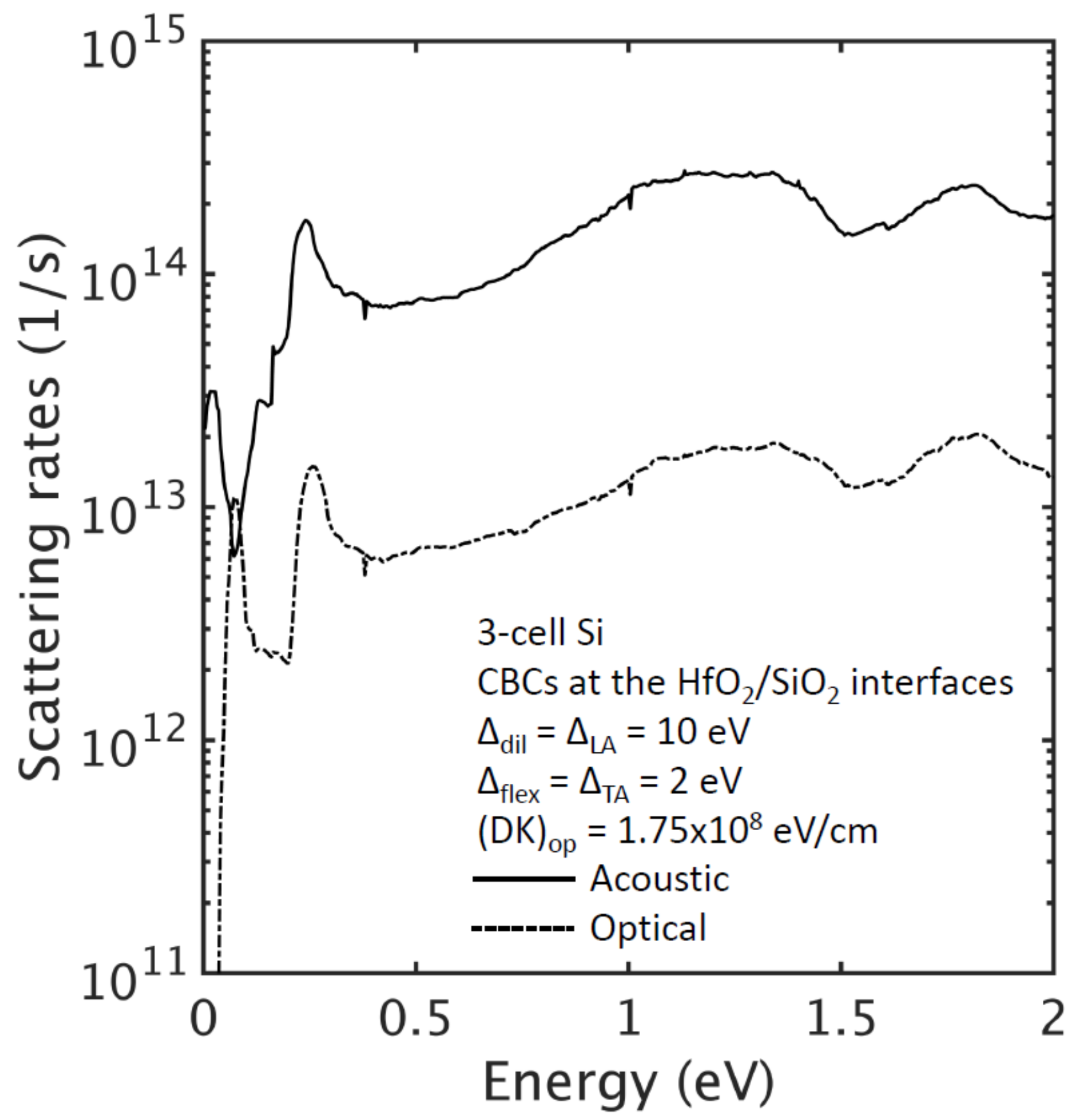}
}}
\caption{As in Fig.~\ref{fig:El-ph_3-cell} but assuming clamped acoustic phonons at the SiO$_2$/HfO$_2$ interfaces, as discussed in Sec. 2.2.}  
\label{fig:El-ph_3-cell_HfO2}
\end{figure}

As we saw before, the boundary conditions considered to obtain the scattering rates shown in the top frames of Fig.~\ref{fig:El-ph_3-cell} 
are not realistic. Therefore we have considered a different model assuming for acoustic phonons clamped boundary conditions at the SiO$_2$/HfO$_2$ interfaces. 
The scattering rates obtained using the model of Sec.~\ref{sec:outer_CBCs} are shown in Fig.~\ref{fig:El-ph_3-cell_HfO2}. Comparing these rates with 
those obtained assuming CBCs at the Si/SiO$_2$ interfaces, we see that they are larger for small electron energies, as a result of the lower `zero-point' 
energy of the acoustic phonons that are confined over a thicker region consisting not of the Si sheet but of the Si/SiO$_2$/SiO$_2$ 'sandwich. 
Table~\ref{Tab:phonon-limited-mobility} below shows that indeed the electron mobility, 1,060~cm$^2$V$^{-1}$s$^{-1}$ when assuming CBCs at the Si/SiO$_2$ interfaces, 
drops to 260~cm$^2$V$^{-1}$s$^{-1}$ when assuming acoustic phonons clamped at the outer interfaces with HfO$_2$.\\

Finally, we must consider how the dielectrics and metal gates that surround the nanosheet screen the electron-phonon interactions. Following the approach 
described in Sec.~II.C of Ref.~\cite{Gopalan_2022}, the screened Hartree components of the electron-phonon matrix elements may be approximately expressed as:  
\begin{multline}
\braket{{\bf K}' \nu^{\prime}|V^{\rm (\eta, screened )}_{q_m}|{\bf K} \nu} \\
      \approx \frac{ \int_{0}^{L_z} {\rm d} z \ \zeta^{(\nu')\ast}_{{\bf K}'}(z) \ 
           G^{\rm (screened)}_{Q,\omega^{(\eta)}_{m,Q}}(z,t_{\rm s}/2) \ \zeta^{(\nu)}_{\bf K}(z) } 
            { 1/(2Q\epsilon_{\rm s}) \  \int_{0}^{L_z} {\rm d} z \ \zeta^{(\nu')\ast}_{{\bf K}'}(z) e^{-Q|z-t_{\rm s}/2|} \ \zeta^{(\nu)}_{\bf K}(z)}                                  
                      \\ \times  \braket{{\bf K}' \nu^{\prime}|V^{(\eta)}_{q_m}|{\bf K} \nu} \ ,
\label{eq:epscreening2}
\end{multline}
where $G^{\rm (screened)}_{Q,\omega}(z,z')$ is the Green's function given by Eq.~(\ref{eq:AGQ}) of Appendix~\ref{AppendixC} with the dielectric 
constants $\epsilon_{\rm s}$, $\epsilon_1=\epsilon_{\rm SiO_2}$, and $\epsilon_2=\epsilon_{\rm HfO_2}$, replaced by the dielectric functions of the 2DEG, 
$\epsilon^{\rm (scalar)}_{2D}(Q,\omega)$ and $\epsilon_{i}(\omega)$ given by Eqs.~(\ref{eq:scalar_0_eps}) and (\ref{eq:epsox}) below
with $\omega = \omega^{(\eta)}_{m,Q}$ and $Q = |{\bf K}-{\bf K}'|$. Therefore, the scattering rates, Eq.~(\ref{eq:tau_ep_sheet_discretized}), 
should be modified by replacing the squared matrix element by the factor:
\begin{equation}
     \left \vert \frac{ \int_{0}^{L_z} {\rm d} z \ \zeta^{(\nu')\ast}_{{\bf K}'}(z) \ 
        G^{\rm (screened)}_{Q,\omega^{(\eta)}_{m,Q}}(z,t_{\rm s}/2) \ \zeta^{(\mu)}_{\bf K}(z) } 
            { 1/(2Q\epsilon_{\rm s}) \  \int_{0}^{L_z} {\rm d} z \ 
                 \zeta^{(\nu')\ast}_{{\bf K}'}(z) \ e^{-Q|z-t_{\rm s}/2|} \ \zeta^{(\nu)}_{\bf K}(z) } \right \vert^2 \ \le \ 1 \ .
\label{eq:tau_ep_sheet_screened}
\end{equation}
As described in Ref.~\cite{Gopalan_2022}, this can be implemented in the Monte Carlo simulation by first considering scattering events with the unscreened rates. 
Once the emission or absorption of a phonon of momentum ${\bf Q}$ and branch $(\eta,m)$ is selected, the rejection technique can be used to accept of reject 
the event with probability given by Eq.~(\ref{eq:tau_ep_sheet_screened}).\\
 
In principle, we should account for the dynamic screening effects expressed by Eq.~(\ref{eq:tau_ep_sheet_screened}). However, the large numerical cost of
calculating and/or tabulating and interpolating these screening corrections for all $({\bf K}]\nu',{\bf K}\nu)$ pairs, $Q$, and $\omega$, forces us to employ 
the approximation commonly used in the literature by screening the electron-phonon interactions statically; that is, by evaluating 
$G^{\rm (screened)}_{Q,\omega^{(\eta)}_{m,Q}}(z,t_{\rm s}/2) \ \zeta^{(\mu)}_{\bf K}(z)$ at $\omega^{(\eta)}_{m,Q}=0$, thus presumably obtaining a lower bound for the scattering rates and an upper bound for the mobility. 
\section{Interface plasmon-phonon scattering}
\label{sec:IPP}
The 'remote-phonon scattering' problem has been studied before and we refer the reader to Ref.~\cite{Gopalan_2022} for a review of the problem, noting only 
that recently Dinar, Macheda, Sohier, and co-workers have extended Hauber's and Fahy's work~\cite{Hauber_2017} to study the problem in the case of 
hBN-encapsulated graphene~\cite{Macheda_2026,Macheda_2026a}.
We shall follow a procedure similar to Ref.~\cite{Gopalan_2022} in the case of transition metal dichalcogenides. The main differences will consist 
in a different Poisson Green's function reflecting the new many occupied subbands and the multi-sheet/gate-stack geometry of interest here. We shall also
assume a symmetric structure, as done above when dealing with phonon confinement. We shall also ignore the effect of the confinement of optical phonons on the
on their static, long-wavelength dielectric constant. This is likely to be a small effect, as shown by Giustino {\it et al.}~\cite{Giustino_2003} and discussed
in Ref.~\cite{Fischetti_2026}.\\

\subsection{Secular equation and IPP dispersion}
We must look for the potential due to interface polarization charges in the absence of external charges; that is, for the
solution of the Laplace equation:
\begin{equation}
   \nabla \cdot \nabla [ \epsilon(z) \Phi({\bf r},t) ] \ = \ 0 
\label{eq:Laplace_Phi}
\end{equation}
obviously with the '`usual' boundary conditions at the interfaces appropriate for  geometry in which we are interested. 
The Fourier transform of $\Phi$ on the $(x,y)$-plane, using translational invariance on this plane, is:
\begin{equation}
      \Phi({\bf r},t) \ = \ \frac{1}{\Omega^{1/2}} \ \sum_{\bf Q} \ \Phi_{Q, \omega}(z) \ e^{i{\bf Q} \cdot ({\bf R}-{\bf R}')} e^{i \omega t} \ .
      \label{eq:Fourier_Phi}
      \end{equation}
The general solution is of the form:
\begin{multline}
      \Phi_{Q,\omega}(z) \ = \\  \left \{
      \begin{array}{ll}
      a_1 e^{Qz} \ + \ a_2 e^{-Qz} & (-t_2-t_1 \leq z \le -t_1) \\
      b_1 e^{Qz} \ + \ b_2 e^{-Qz} & (-t_1 \leq z \le 0) \\
      c_1 e^{Qz} \ + \ c_2 e^{-Qz} & (0 < z \leq t_{\rm s}) \\
      d_1 e^{Qz} \ + \ d_2 e^{-Qz} & (t_{\rm s} < z \leq t_{\rm s}+t_1) \\
      f_1 e^{Qz} \ + \ f_2 e^{-Qz} & (t_{\rm s}+t_1 < z \leq  t_{\rm s}+t_1+t_2) \\
      \end{array}
      \right. \ 
      \label{eq:Phi_IPP_Q}  
\end{multline}
subject to the usual boundary conditions expressing the continuity of $\Phi_{Q, \omega}(z)$ and $\epsilon(z){\rm d}\Phi_{Q, \omega}(z)/{\rm d}z$ at the
four interfaces (the two Si/SiO$_2$ and the two SiO$_2$/HfO$_2$ interfaces) and the vanishing of $\Phi_{Q, \omega}(z)$ at the two HfO$_2$/metal-gate 
interfaces.\\ 
 
Unfortunately, we must face a serious complication. The condition expressing the continuity of the component of the ${\bf D}$-field perpendicular to the 
plane of the interfaces, $\epsilon(z) {\rm d}\Phi_{Q, \omega}(z)/{\rm d}z$, at the two Si/SiO$_2$ interfaces involves the dielectric matrix with elements 
that should be obtained using the full dielectric matrix of a 2DEG, as given in Ref.~\cite{Dahl_1977} and in Appendix~A of Ref.~\cite{Fischetti_2001b}. 
Even considering only the longitudinal response of the 2DEG, this results in a very complicated problem. Indeed, the `correct' procedure would consist 
in starting by finding the potential 
$\Phi_{Q, \omega}(z)$ assuming for the dielectric function of the Si layer the simple dielectric constant $\epsilon_{\rm s}$ that accounts only for the 
response of the valence electrons (that we can safely consider static and wavelength-independent). Now note that the matrix elements
$\Phi^{\rm (screened)}_{Q, \omega, \lambda \lambda'}$ of the potential screened by the free electrons and by the dielectric environment 
are related to the matrix element of an `external' potential, $\Phi_{Q,\omega,\mu \mu'}$: 
\begin{equation}
      \Phi_{Q,\omega,\mu \mu'} =  
           \sum_{\lambda \lambda'} \ [\kappa_{\mu \mu',\lambda \lambda'}(Q,\omega)] \Phi^{\rm (screened)}_{Q, \omega, \lambda \lambda'} \ ,
\label{eq:screened_PhiQ}     
\end{equation}
where $\kappa_{\mu \mu',\lambda \lambda'}(Q,\omega)]$ is the full dielectric matrix of the 2DEG that accounts for the geometry of the system usiung the
appropriate Poisson Green's function. \\
           
A reasonable approximation consists in assuming the scalar dielectric function 
using the Poisson Green's function $G^{\rm (rel,0)}_{Q,\omega}(z,z') = e^{-Q|z-z'|}$ for an isolated Si layer,    
\begin{multline}
      \epsilon^{\rm (scalar,0)}(Q,\omega) = \epsilon_{\rm s} \mbox{Tr} |\kappa^{(0)}_{\mu \mu,\lambda \lambda}(Q,\omega)|  \\
             \approx \epsilon_{\rm s} \left [ 1 - \sum_{\lambda} \frac{\beta^{\rm (2D)}_{\lambda}(Q,\omega)}{Q} \right ] \ ,
\label{eq:scalar_0_eps}
\end{multline}
(having ignored the wavefunction form-factors in the log-wavelength limit)
since the effects of the double-gate, double gate-stack geometry are already included in the expression for $\Phi_{Q, \omega}(z)$ given by
Eq.~(\ref{eq:Phi_IPP_Qhalf_fin}) below. In Eq.~(\ref{eq:scalar_0_eps}) the screening parameter $\beta^{\rm (2D)}_{\lambda}(Q,\omega)$ is defined as
\begin{equation}
      \beta^{\rm (2D)}_{\lambda}(Q,\omega) \approx 
      \beta^{\rm (2D)}_{\rm DH,\lambda}(Q) = - \frac{e^{2} n_{\lambda}}{2 \epsilon_{\rm s} k_{\rm B} T } \ g_{1}(Q \ell_{\lambda}) 
\label{eq:beta_static_nondeg}
\end{equation} 
when the frequency of the perturbation, $\omega$, is smaller then the plasma frequency in subband $\lambda$, $\omega_{\rm P,\lambda}(Q)$,
in the nondegenerate case, and  
\begin{multline}
      \beta^{\rm (2D)}_{\lambda}(Q,\omega) \approx       
      \beta^{\rm (2D)}_{\rm TF,\lambda}(Q) = \frac{e^{2}}{2 \epsilon_{\rm s}} {\mathcal{D}}_{\lambda}(E_{\rm F}) \ g_{1}(Q \ell_{\lambda}) \\ = 
               - \frac{e^{2}}{2 \epsilon_{\rm s}} \frac{g_{\lambda} m_{\rm d,\lambda}} {2 \epsilon_{\rm s} \pi\hbar^{2} } \ g_{1}(Q \ell_{\lambda})                  
\label{eq:beta_static_deg}
\end{multline} 
in the degenerate case. In the opposite case in which $\omega > \omega_{\rm P,\lambda}(Q)$, instead: 
\begin{equation}
\label{chp5a_e44}
      \beta^{\rm (2D)}_{\lambda}({\bf Q},\omega) \simeq  \frac{e^{2} g_{\lambda} n_{\lambda}}{2 \epsilon_{\rm s} m_{\lambda} \omega^{2}} Q^{2} \ \equiv
           \left ( \frac{\omega_{\rm P,\lambda}(Q)}{\omega} \right )^2 Q \ .     
\end{equation}
Therefore, we shall find the IPP scattering potential and the IPP dispersion considering
Eq.~(\ref{eq:Phi_IPP_Qhalf_fin}) and setting $\epsilon_{\rm 2D}(Q,\omega) = \epsilon^{\rm (scalar,0)}(Q,\omega)$ to deal with the boundary conditions.
In these expressions $g_{1}(x)=2 \pi^{1/2}\Phi[x/(4\pi^{1/2})]/x$ where $\Phi(x)$ is the plasma dispersion function~\cite{Fried_1961} and, 
for each subband $\lambda$, $\ell_{\lambda}=[2\pi \hbar^2/(m_{\lambda}k_{\rm B}T)]^{1/2}$ is the thermal wavelength, $g_{\lambda}$, $n_{\lambda}$ and
$m_{\rm d,\lambda}$ are the degeneracy, sheet density, and density-of-states effective mass, ${\mathcal{D}}_{\lambda}(E)$ the density of states
at energy $E$, and the plasma frequency $\omega_{\rm P,\lambda}(Q)$ is implicitly defined by Eq.~(\ref{chp5a_e44}). Note that at small $Q$, 
$g_{1}(Q \ell_{\lambda}) \rightarrow 1$.\\ 
             
For the insulators we assume the usual form of the dielectric function in the long-wavelength limit:       
\begin{multline}
      \epsilon_{i}(\omega ) \ =  \ \epsilon_{i}^{(\infty)} + 
                     [ \epsilon_{i}^{\rm (mid)} - \epsilon_{i}^{(\infty)} ] 
                         \frac{\omega_{{\rm TO}1,i}^{2}}{\omega_{{\rm TO}1,i}^{2} - \omega^{2}} + \\
                     [ \epsilon_{i}^{(0)} - \epsilon_{i}^{(\rm mid)} ] 
                         \frac{\omega_{{\rm TO}2,i}^{2}}{\omega_{{\rm TO}2,i}^{2} - \omega^{2}}  \ ,
\label{eq:epsox}
\end{multline} 
where the index $i \ (=1,2)$ labels the insulators and we have indicated with $\omega_{{\rm TO}1,i}$ and $\omega_{{\rm TO}2,i}$ the high- and low-frequency
optical phonons of the each insulator (assuming, for simplicity, that only two optical modes are present in each insulator). Moreover, $\epsilon^{(0)}$,
$\epsilon^{\rm (mid)}$, and $\epsilon^{(\infty)}$ are the static, intermediate, and optical 
dielectric constants of the insulators.\\

\begin{figure}[tb]
\centerline 
{\hbox{\hspace*{-0.5cm}
\includegraphics[width=8.6cm]{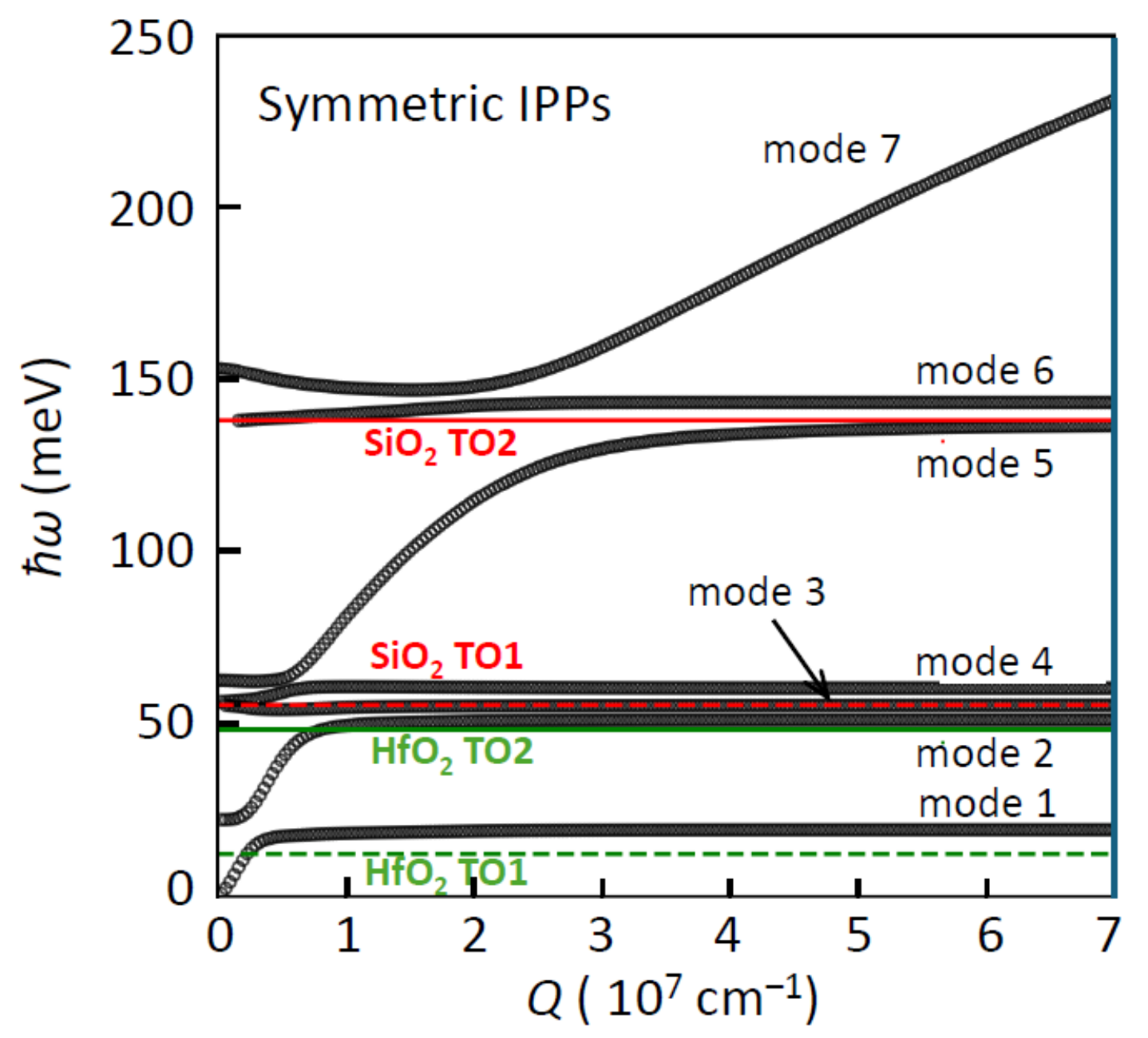}
}}
\caption{Dispersion of the symmetric  hybrid interface plasmon/optical-phonon excitations calculated for an electron 
        sheet density of $1.15 \times 10^{13}$ cm$^{-2}$. Note that the two SiO$_2$ phonons split into two quasi-degenerate hybrid modes at the Si/SiO$_2$ and
        SiO$_2$/HfO$_2$ interfaces. The dispersions shown here have been obtained ignoring Landau damping. For $Q > Q_{\rm LD}$ ($\approx 1.1 \times 10^7$ cm${-1}$ 
        at this electron sheet density), the plasma response of the Si 2DEG is suppressed by Landau damping. The labels identify the bare
        (non hybridized) phonons to which the IPP dispersions tend in the short-wavelength limit.}  
\label{fig:IPP_dispersion}
\end{figure}

Now note that the symmetry of the geometry leads to a very useful simplification: The system of equations
for the coefficients of Eq.~(\ref{eq:Phi_IPP_Q}) should be a system of 14 equations in 14 unknowns, leading to a secular equation of 10$^{\rm th}$ 
degree in $\omega^2$. However, we can split the problem into two linear systems of 7 equations in 7 unknowns that require the solution of an
algebraic equation of 7$^{\rm th}$ degree. Therefore, we can look for solutions of the form:
\begin{multline}
      \Phi^{(\pm)}_{Q,\omega}(z) \ =  \\ \left \{
      \begin{array}{ll}
      a^{(\pm)}_1 e^{Qz} + a^{(\pm)}_2 e^{-Qz}  & (t_{\rm s}/2 \le z \le t_{\rm s}) \\
      b^{(\pm)}_1 e^{Qz} + b^{(\pm)}_2 e^{-Qz}  & (t_{\rm s} < z \le t_{\rm s}+t_1) \\
      c^{(\pm)}_1 e^{Qz} + c^{(\pm)}_2 e^{-Qz}  & (t_{\rm s}+t_1 < z \le t_{\rm s}+t_1+t_2) \\
      \end{array}
      \right. \ .
\label{eq:Phi_IPP_Qhalf}  
\end{multline}
For the symmetric/antisymmetric solutions, the boundary conditions imply:
\begin{widetext}
\begin{subequations}
      \begin{align}
      a^{(\pm)}_1 e^{Qt_{\rm s}/2} \mp a^{(\pm)}_2 e^{-Qt_{\rm s}/2}                       &= 0                                               \label{eq:Fsystema}  \\                                                              
      a^{(-\pm}_1 e^{Qt_{\rm s}}  + a^{(\pm)}_2 e^{-Qt_{\rm s}}                              
                                                       &= b^{(\pm)}_1 e^{Qt_{\rm s}}+b^{(\pm)}_2 e^{-Qt_{\rm s}}                               \label{eq:Fsystemb}  \\
      \epsilon_{\rm s}Q \ [ a^{(-)}_1 e^{Qt_{\rm s}}-a^{(\pm)}_2 e^{-Qt_{\rm s}}]         
                                                       &= \epsilon_{1} Q \  [b^{(\pm)}_1 e^{Qt_{\rm s}}-b^{(\pm)}_2 e^{-Qt_{\rm s}}]           \label{eq:Fsystemc}  \\
      b^{(\pm)}_1 e^{Q(t_{\rm s}+t_1)}+b^{(-)}_2 e^{-Q(t_{\rm s}+t_1)}    &= c^{(\pm)}_1 e^{Q(t_{\rm s}+t_1)}+c^{(\pm)}_2 e^{-Q(t_{\rm s}+t_1)} \label{eq:Fsystemd}  \\
      \epsilon_{1}Q \ [ b^{(\pm)}_1 e^{Q(t_{\rm s}+t_1)}-b^{(\pm)}_2 e^{-Q(t_{\rm s}+t_1)}] 
                                                        &= \epsilon_{2} Q \ [f^{(\pm)} e^{Q(t_{\rm s}+t_1)}-g^{(\pm)} e^{-Q(t_{\rm s}+t_1)}]   \label{eq:Fsysteme}  \\
      c^{(\pm)}_1 e^{Q(t_{\rm s}+t_1+t_2)}+c^{(\pm)}_2  e^{-Q(t_{\rm s}+t_1+t_2)}            
                                                                         &= 0                                                              \label{eq:Fsystemf}   \ .
      \end{align}
\end{subequations}
where, in Eq.~(\ref{eq:Fsystema}), the plus/minus sign refers to the antisymmetric/symmetric modes.\\
      
We can solve these systems by expressing all coefficients in terms of the coefficient $b^{(\pm)}_{1}$ obtaining: 
\begin{subequations}
      \begin{align}
      a^{(\pm)}_1   &= \frac{1}{2\epsilon_{\rm s}} \left [  \left ( \epsilon_{\rm s}+\epsilon_1 \right )  
                         + \Lambda e^{2Qt_1} \left ( \epsilon_{\rm s}-\epsilon_1 \right ) \right ] b^{(\pm)}_1 
                                                                                           \equiv {\mathcal{A}}_{1}(Q,\omega) \ b^{(\pm)}_1  \label{eq:Gsystema}  \\ 
      a^{(\pm)}_2   &= \frac{1}{2\epsilon_{\rm s}} \left [  \left ( \epsilon_{\rm s}-\epsilon_1 \right )  
                         + \Lambda e^{2Qt_1} \left ( \epsilon_{\rm s}+\epsilon_1 \right ) \right ]  e^{2Qt_{\rm s}}  \ b^{(\pm)}_1 
                                                                                           \equiv {\mathcal{A}}_{2}(Q,\omega) \ b^{(\pm)}_1  \label{eq:Gsystemb}  \\                    
      b^{(\pm)}_2   &= e^{2Q(t_{\rm s})+t_1}\Lambda b^{(\pm)}_1                            \equiv {\mathcal{B}}_{2}(Q,\omega) \ b^{(\pm)}_1  \label{eq:Gsystemc}  \\
      c^{(\pm)}_1   &= \frac{1}{2\epsilon_2} \left [ \left ( \epsilon_2+\epsilon_1 \right ) 
                         + \Lambda \left ( \epsilon_2-\epsilon_1 \right ) \right ] b^{(\pm)}_1                            
                                                                                           \equiv {\mathcal{C}}_{1}(Q,\omega) \ b^{(\pm)}_1  \label{eq:Gsystemd}  \\
      c^{(\pm)}_2   &= \frac{1}{2\epsilon_2} e^{2Q(t_{\rm s}+t_1)} \left [ \left ( \epsilon_2-\epsilon_1 \right ) 
                         + \Lambda \left ( \epsilon_2+\epsilon_1 \right ) \right ] b^{(\pm)}_1            
                                                                                           \equiv {\mathcal{C}}_{2}(Q,\omega)\  b^{(\pm)}_1  \label{eq:Gsysteme} \ ,  
      \end{align}
\end{subequations}
\end{widetext}
where
\begin{equation}
      \Lambda = \frac{\epsilon_1 \sinh(Qt_2) + \epsilon_2 \cosh(Qt_2)}{\epsilon_1 \sinh(Qt_2) - \epsilon_2 \cosh(Qt_2)} \ .
\label{eq:Lambda}
\end{equation}
Inserting these expressions for $a^{(\pm)}_1$ and $a^{(\pm)}_2$ into Eqs.~(\ref{eq:Fsystema}) yields an equation that the
coefficients must satisfy. This amounts to requiring the vanishing of the determinant of the coefficients of these homogeneous linear systems, therefore 
we obtain the `secular equation': 
\begin{multline}
      ( \epsilon_{\rm s} + \epsilon_1 ) \mp (\epsilon_{\rm s} - \epsilon_1) e^{Qt_{\rm s}}  \\
             + \Lambda e^{2Qt_1} \left [ ( \epsilon_{\rm s} -\epsilon_1 ) \mp (\epsilon_{\rm s}+\epsilon_1) e^{Qt_{\rm s}} \right ] = 0 \ ,
\label{eq:secular}
\end{multline}      
where the upper `minus' sign refers to the symmetric solutions and the lower `plus' sign refers to the antisymmetric solutions. As an example of the result obtained by solving the secular equation, Eq.~(\ref{eq:secular}), we show in Fig.~\ref{fig:IPP_dispersion} the dispersion of the symmetric IPPs. The anti-symmetric modes
have a very similar behavior.\\
      
We should recall that at sufficiently small wavelength the plasma oscillations of the 2DEG decay into single-particle excitations (Lanadu damping). Although
this has been treated exactly by Hauber and Fahy~\cite{Hauber_2017} and, more recently, by Macheda and Sohier~\cite{Macheda_2026}, here we follow the
simpler approach used before~\cite{Fischetti_2001,Gopalan_2022}: We define a Landau-damping wave vector $Q_{\rm LD}$ given implicitly by the approximate
expression:
\begin{equation}
  Q_{\rm LD} \approx \left [ k_{\rm F}^2 - \frac{2m_{\rm d}\omega_{\rm P}(Q_{\rm LD})^2}{\hbar} \right]^{1/2} + k_{\rm F} \ .
\label{eq:QLD}
\end{equation}
In the spirit of the approximations assumed to express the dielectric response of the 2DEG with Eq.~(\ref{eq:scalar_0_eps}), we have approximated $k_{\rm F}$ 
and the density-of-states effective mass $m_{\rm d}$ with the Fermi wave vector and DoS effective mass of the most-populated 
ground-state unprimed subband, and $\omega_{\rm P}(Q)^2$ with $\sum_{\lambda} \omega_{{\rm P},\lambda}(Q)^2$.    
Then, in the Landau-damping regime, $Q > Q_{\rm LD}$, we effectively neglect the plasmons assuming for $\epsilon_{\rm s}({\bf Q},\omega)$ the
static expressions, Eq.~(\ref{eq:scalar_0_eps}) with $\beta^{\rm (2D)}_\lambda ({\bf Q},\omega)$ given by Eq.~(\ref{eq:beta_static_nondeg}) or
Eq.~(\ref{eq:beta_static_deg}), as appropriate. This reduces the secular equation to a $12^{\rm th}$-degree algebraic equation hat can be reformulated as
two $6^{\rm th}$-degree algebraic equations giving the dispersion of 6 symmetric and 6 antisymmetric modes that represent the hybridization of the
optical phonons of the insulators at the interfaces. The dispersion of these modes can be obtained following a procedure similar to what we have outlined
above. 
\begin{figure}[tb]
\centerline 
{\hbox{
\includegraphics[width=8.6cm]{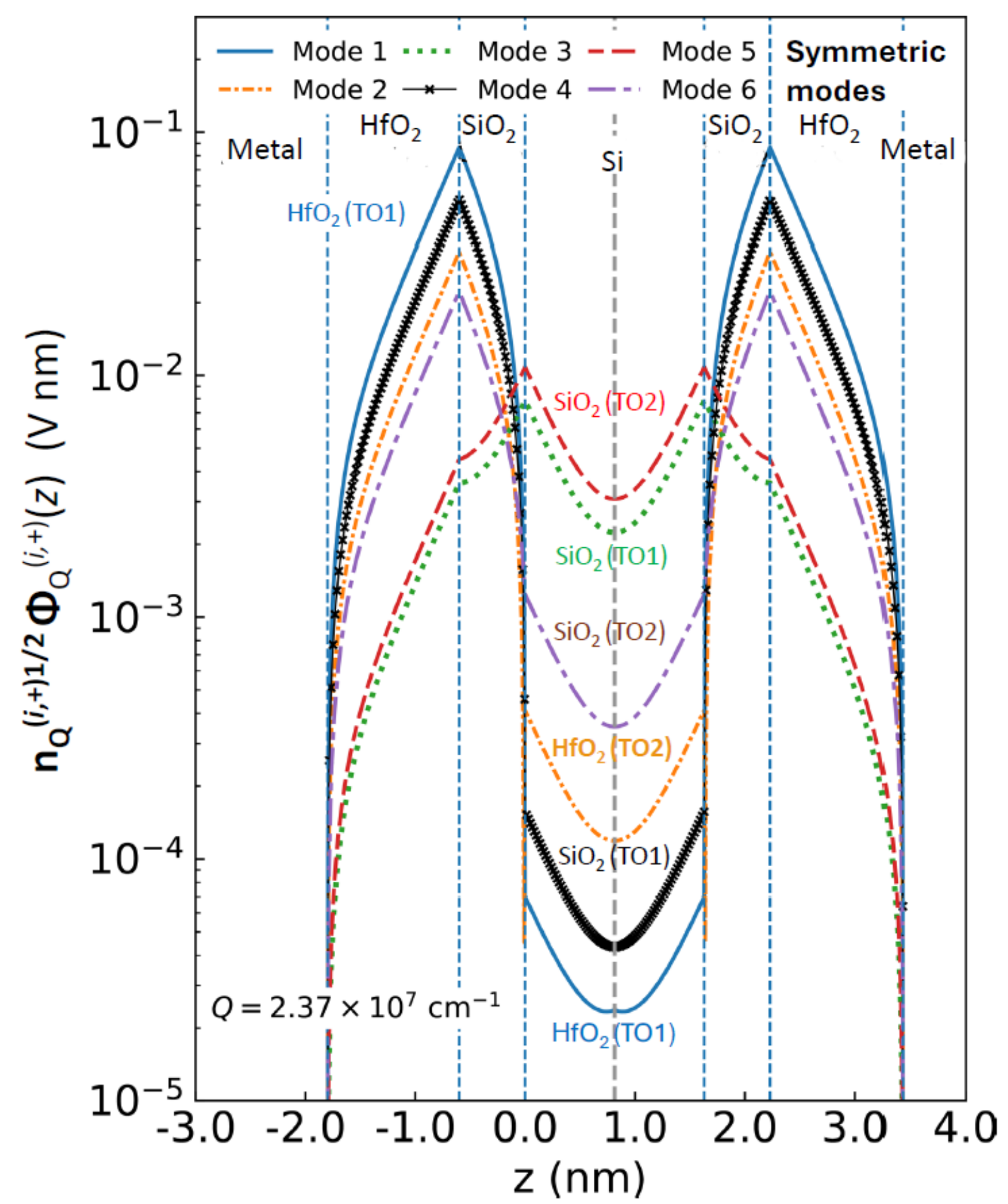}
}}
\caption{Scattering potential $\Phi^{(i+)}_{Q}(z)=\sum_{\alpha}\Phi_{Q,\omega_{Q}^{(i)}}^{(\alpha,i,\pm)}({\bf R =0},z)$ weighted by the square root of the 
         Bose-Einstein occupation $n_{Q}^{(i,+)}$ for each of the symmetric (+) hybrid modes $i$ calculated at $Q=2.37 \times 10^7$/cm (in the Landau-damping region).
         The square of this quantity is approximately proportional to the contribution of each IPP to the total scattering rate. Note how at such a short
         wavelength, the SiO$_2$ interfacial layer depresses the contribution of the low-energy HfO$_2$ mode.}  
\label{fig:IPP_potential}
\end{figure}

\subsection{Scattering potential} 
To determine the multiplicative constant $b^{(\pm)}_{1}$, we follow the quantization procedure described many time before~\cite{Fischetti_2001,Gopalan_2022}
(see Eqs. (28)-(30) of Re,~\cite{Gopalan_2022}), first employed by Stern and Ferrel~\cite{EAStern_1960}:       
For each hybrid mode $i$ we calculate the total energy of the field (time-averaged and including self-energy terms) in terms of $b^{(\pm)}_1$ and set it equal 
to the ground-state energy $\hbar \omega^{(i)}/2$. However, in order to consider separately the contribution $\Pi^{(\alpha)}(\omega)$ due to each bare
phonon $\alpha$ (= TO1, TO2, TO3, TO4, or ZO, for both the top and bottom gate stacks, and the 2DEG plasmon) -- which is the `phonon content discussed
below  --, we consider separately each component $\alpha$ of the time-averaged energy, 
$\langle W^{(\alpha)}_{Q,\omega_{Q}^{(i)}} \rangle$, and set it equal to a fraction $\Pi^{(\alpha)}(\omega_{Q}^{(i)})$ of the oscillator ground-state
energy of the hybrid mode $i$ (Ref.~\cite{Gopalan_2022} shows how to calculate the content $\Pi^{(\alpha)}(\omega_{Q}^{(i)})$ of the bare mode  
$\alpha$ of the hybrid mode $i$). The total energy can be expressed as: 
\begin{multline}
    \frac{2}{\Omega} \ \left \langle W^{(\alpha)}_{Q,\omega_{Q}^{(i)}} \right \rangle \
           = \ \frac{1}{\Omega} \int_{-t_1-t_2}^{t_{\rm s}+t_1+t_2} {\rm d}z \ \rho_{Q,\omega_{Q}^{(i)}}(z) \ \Phi^{(\alpha)}_{Q,\omega_{Q}^{(i)}}(z) \\
                    = \ \frac{1}{2} \ \hbar \omega_{Q}^{(i)} \Pi^{(\alpha)}(\omega_{Q}^{(i)})  \ .
\label{eq:W1}
\end{multline}
where $\Omega$ is the normalization area. 
The electrostatic potential $\Phi^{(\alpha)}_{Q,\omega_{Q}^{(i)}}(z)$ due to the bare phonon mode $\alpha$ 
can be obtained from the Poisson equation considering only the charge density due to this mode, that is (writing $\omega$ for $\omega_{Q}^{(i)}$ for
simplicity): 
\begin{equation} 
      \nabla^{2} \Phi^{(\alpha)}_{Q,\omega}(z) = 
       - \left [ \frac{\rho_{Q,\omega}(z)}{\epsilon_{\rm TOT}^{(\alpha,\rm high)}(\omega)} -
                \frac{\rho_{Q,\omega}(z)}{\epsilon_{\rm TOT}^{(\alpha,\rm low) }(\omega)} \right ] \equiv
                \frac{\rho_{Q,\omega}(z)}{{\overline{\epsilon}}_{\rm TOT}^{(\alpha)}(\omega)} \ ,   
\label{eq:Poisson_alpha} 
\end{equation}  
where $\epsilon_{\rm TOT}^{(\alpha,\rm high)}(\omega)$ and $\epsilon_{\rm TOT}^{(\alpha,\rm low) }(\omega)$ (as defined 
in Refs.~\cite{Fischetti_2001,Gopalan_2022}) are the dielectric 
functions of the system when phonon-mode $\alpha$ does not respond or responds fully, respectively.  
For the calculation of the polarization charge at the interfaces (arising from the discontinuity of the $z$-component of the field ${\bm E}$ at the 
various top and bottom interfaces) we can exploit the symmetry of the system and consider only the two `top' interfaces, multiplying the result by a factor of 2.
Thus, we obtain:
\begin{widetext}
\begin{multline}
      \frac{2}{\Omega} \left \langle W_{Q,\omega} \right \rangle \ = \  \ 2 |b_1|^{2} Q \left \{ \left [ 
         \widetilde{\epsilon}_{\rm 2D}(Q,\omega)[{\mathcal{A}}_{1}(Q,\omega) e^{Qt_{\rm s}}-{\mathcal{A}}_{2}(Q,\omega) e^{-Qt_{\rm s}} ] - \right. \right. \\
           \left. - \widetilde{\epsilon}_{1}(\omega)[e^{Qt_{\rm s}}-{\mathcal{B}}_{2}(Q,\omega) e^{-Qt_{\rm s}}] 
               \right ] [{\mathcal{A}}_{1}(Q,\omega) e^{Qt_{\rm s}}+{\mathcal{A}}_{2}(Q,\omega) e^{-Qt_{\rm s}} ] 
           + \left [ \widetilde{\epsilon}_{1}(\omega)[e^{Q(t_{\rm s}+t_1)}-{\mathcal{B}}_{2}(Q,\omega) e^{-Q(t_{\rm s}+t_1)}]  \right. \\
          \left. \left. - \widetilde{\epsilon}_{2}(\omega)[{\mathcal{C}}_{1}(Q,\omega) e^{Q(t_{\rm s}+t_1)}-{\mathcal{C}}_{2}(Q,\omega) e^{-Q(t_{\rm s}+t_1)}] \right ] 
             [e^{Q(t_{\rm s}+t_1)}+{\mathcal{B}}_{2}(Q,\omega) e^{-Q(t_{\rm s}+t_1)}] \right \}  \\
                \equiv |b_1|^{2} Q  \ \overline{\epsilon}_{\rm TOT}^{(\alpha)}(\omega_{Q}^{(i)}) \ ,
\label{eq:W2}            
\end{multline}
\end{widetext}
Here the last expression defines implicitly the `total dielectric function' of the system, ${\epsilon}_{\rm TOT}^{(\alpha)}(\omega)$ (the
bar indicates the fact that we have isolated the high- and low-frequency response of mode $\alpha$).
In these expressions, the quantities $\widetilde{\epsilon}$ must be evaluated assuming that the mode $\alpha$ does respond fully 
($\epsilon_{\rm TOT}^{(\alpha,\rm low)}(\omega)$) or does not respond at all ($\epsilon_{\rm TOT}^{(\alpha,\rm high)}(\omega)$), as discussed before.\\
   
Finally, we set the time-averaged energy $\langle W^{(\alpha)}_{Q,\omega} \rangle$ equal to the fraction 
$\Pi^{(\alpha)}(\omega)$ of ground-state energy of the of the oscillator (reinserting now explicitly $\omega^{(i)}_{Q}$ for $\omega$),
so that the coefficient $b_{1}$ becomes a function ${\mathcal{B}}^{(\alpha)}_{1}(Q,\omega_{Q}^{(i)})$ of $Q$ and $\omega$ and of the mode $\alpha$:    
\begin{multline}
      \frac{2}{\Omega} \ \left \langle W^{(\alpha)}_{Q,\omega_{Q}^{(i)}} \right \rangle 
          \ = \ \frac{1}{\Omega} \int_{-t_1-t_2}^{t_{\rm s}+t_1+t_2} {\rm d}z \ 
              \rho^{(\alpha)}_{Q,\omega_{Q}^{(i)}}(z) \ \Phi^{(\alpha)}_{Q,\omega_{Q}^{(i)}}(z) \\
                 = \left \vert {\mathcal{B}}^{(\alpha)}_{1}(Q,\omega_{Q}^{(i)}) \right \vert^{2} \ Q \ 
                   \overline{\epsilon}_{\rm TOT}^{(\alpha)}(\omega_{Q}^{(i)})
                    \ = \ \Pi^{(\alpha)}(\omega_{Q}^{(i)}) \ \frac{1}{2} \ \hbar \omega_{Q}^{(i)} \ ,
\label{eq:W1_decomp}
\end{multline}
Thus, we obtain for the scattering amplitude $b_1$ of the potential due to the $\alpha$-phonon-content of hybrid mode $i$ as:
\begin{multline}
      \vert b_1 \vert^{2} \equiv \left \vert {\mathcal{B}}^{(\alpha)}_{1}(Q,\omega_{Q}^{(i)}) \right \vert^{2} \\ 
       = \Pi^{(\alpha)}(\omega_{Q}^{(i)}) \frac{\hbar \omega_{Q}^{(i)}}{2Q} 
             \left \vert \frac{1}{\epsilon_{\rm TOT}^{(\alpha,\rm high)}(\omega_{Q}^{(i)})} 
                                    - \frac{1}{\epsilon_{\rm TOT}^{(\alpha,\rm low)}(\omega_{Q}^{(i)})} \right \vert \ . 
\label{eq:amplitude}
\end{multline}
As stated above, the dielectric functions appearing in this equation represent the response of the system when the bare phonon $\alpha$ 
responds fully ('low') or does not respond at all (`high'), thus isolating its contribution from the response of all other nodes.\\
\begin{figure*}[tb]
\centerline 
{\vbox{
{\hbox{
\includegraphics[width=15.2cm]{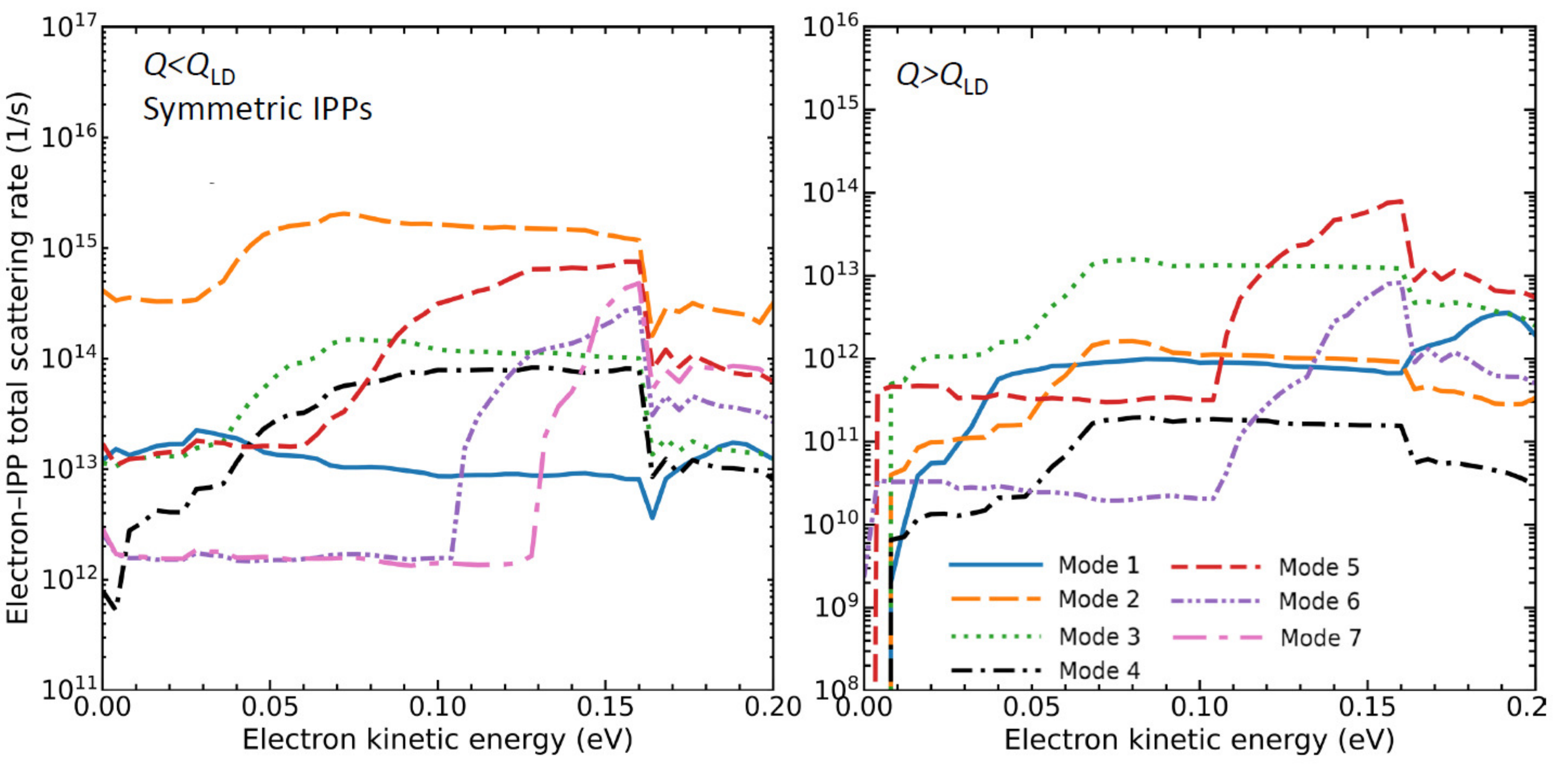}}}
{\hbox{
\includegraphics[width=15.2cm]{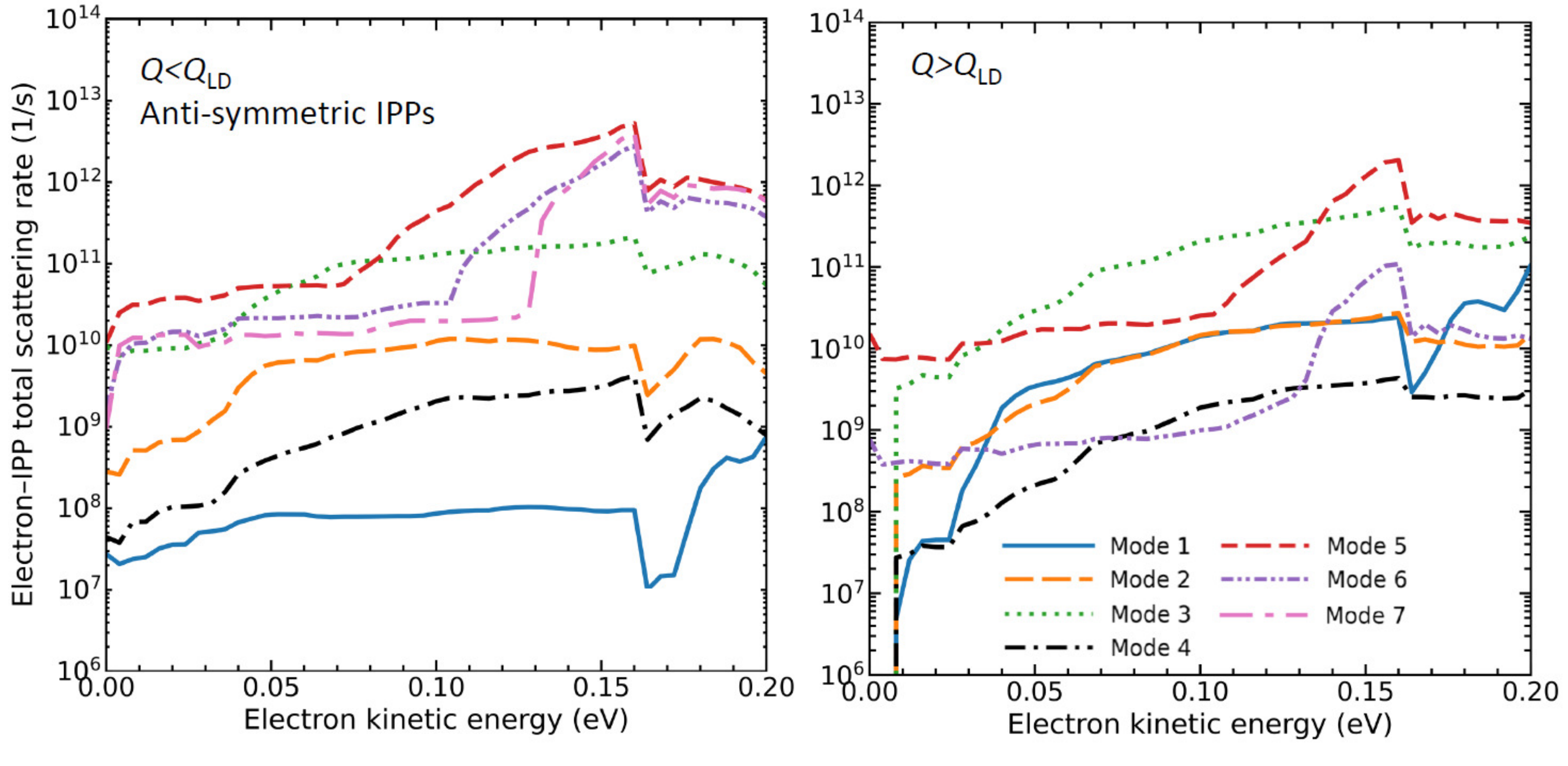}}}
}}
\caption{{\bf Top}: Calculated scattering rates with the symmetric IPPs for all transfer-wave vectors $Q \le Q_{\rm LD}$ (left) and $Q > Q_{\rm LD}$ (right).
         {\bf Bottom}: As in the top panel, but for the antisymmetric IPPs. Note the different scales on the $y$ axis.
}
\label{fig:IPP_rates}
\end{figure*}
\vspace*{-0.25cm}
\subsection{Electron-IPP scattering rates}
Finally, using Fermi's golden rule, and cell-averaged wavefunctions, Eq.~(\ref{eq:zetaK}) employed so far, 
the scattering rate for an electron in band $\mu$ and 2D wave vector ${\bf K}$ to emit or absorb the total phonon
content of a hybrid excitation of mode $(i,\pm)$ with wave vector ${\bf Q}$ and frequency $\omega^{(i,\pm)}_{Q}$ can be written as:
\begin{widetext}
\begin{multline}
      \frac{1}{\tau^{(i,\pm)}_{\mu}({\bf K})} \ \approx \ \frac{2 \pi}{\hbar} \ \sum_{\alpha \mu'} \ \int \frac{{\rm d} {\bf Q}}{(2 \pi)^2} \ 
        \frac{e^{2} \hbar \omega^{(i)}_{Q}}{2 Q} \ 
            \left \vert \phi^{(\alpha,i,\pm)}_{\mu', \mu, {\bf K}+{\bf Q},{\bf K}}(\omega^{(i,\pm)}_{Q}) \right \vert^{2} \\ \times 
                          \left \vert \frac{1}{\epsilon_{\rm TOT}^{(\alpha,\rm high)}(Q,\omega_{Q}^{(i)})} 
                                    - \frac{1}{\epsilon_{\rm TOT}^{(\alpha,\rm low)}(Q,\omega_{Q}^{(i)})} \right \vert 
                       \left \{ \begin{array}{cc} n^{(i)}_{Q} \\ 1+n^{(i)}_{Q} \end{array} \right \} \
                        \ \delta ( E^{(\mu)}_{\bf K} - E^{\nu'}_{{\bf K}'} \pm \hbar \omega^{(i,\pm)}_{Q} ) \ ,
\label{eq:rate5}
\end{multline}
where $n^{(i,\pm)}_{\bf Q}$ is the equilibrium Bose-Einstein occupation of each modem and the upper/lower term refers to absorption/emission of each 
hybrid mode. The matrix elements $\phi^{(\alpha,i,\pm)}_{\mu', \mu, {\bf K}', {\bf K}}(\omega)$  appearing in this expression are:
\begin{equation}
      \phi^{(\alpha,i,\pm)}_{\mu', \mu, {\bf K}', {\bf K}}(\omega) 
        = \ \int_{0}^{L_z} {\rm d}z \ \zeta^{(\mu')\ast}_{{\bf K}'}(z) \ \phi^{(\alpha,i,\pm)}_{Q, \omega^{(i)}_{Q}}(z) \ \zeta^{(\mu)}_{{\bf K}}(z) \ .
\label{eq:rate6}
\end{equation}
From Eqs.~(\ref{eq:Phi_IPP_Qhalf}), (\ref{eq:Gsystema})-(\ref{eq:Gsysteme}), and (\ref{eq:amplitude}), the scattering potential needed to calculate 
the scattering rates with the phonon-content $\alpha$ of the symmetric (+) or anti-symmetric (-) IPP $i$ has the form:
\begin{multline}
      \Phi_{Q,\omega}^{(\alpha,i,\pm)}({\bf R},z) \       
       = e^{i {\bf Q} \cdot {\bf R}} 
        \left  [ \Pi^{(\alpha)}(\omega_{Q}^{(i)}) \frac{\hbar \omega_{Q}^{(i)}}{2Q} 
             \left \vert \frac{1}{\epsilon_{\rm TOT}^{(\alpha,\rm high)}(\omega_{Q}^{(i)})} 
                                    - \frac{1}{\epsilon_{\rm TOT}^{(\alpha,\rm low)}(\omega_{Q}^{(i)})} \right \vert \ \right ]^{1/2}  \\ \times
      \left \{ \begin{array}{ll}
      {\mathcal{A}}_{1}^{(\pm)}(Q,\omega) e^{Qz} + {\mathcal{A}}^{(\pm)}_{2}(Q,\omega) e^{-Qz}  & (t_{\rm s}/2 < z \leq t_{\rm s}) \\
                                          e^{Qz} + {\mathcal{B}}^{(\pm)}_{2}(Q,\omega) e^{-Qz}  & (t_{\rm s} < z \leq t_{\rm s}+t_1) \\
      {\mathcal{C}}_{1}^{(\pm)}(Q,\omega) e^{Qz} + {\mathcal{C}}^{(\pm)}_{2}(Q,\omega) e^{-Qz}  & (t_{\rm s}+t_1 < z \leq t_{\rm s}+t_1+t_2) 
      \end{array}
      \right. \ ,
\label{eq:Phi_IPP_Qhalf_fin}  
\end{multline}
\end{widetext}
for $t_{\rm s}/2 \le z < t_{\rm s}+t_1+t_2$, whereas 
$\Phi^{(\pm)}_{Q,\omega}({\bf R},z) = \pm \Phi^{(\pm)}_{Q,\omega}({\bf R},t_{\rm s}-z)$ for $-t_1-t_2 \le z < t_{\rm s}/2$ (a simple
even or odd reflection around $z=t_{\rm s}/2$).   
In this expression we have indicated simply with $\omega$ the eigen-frequency $\omega_{Q}^{(i,\pm)}$ and we have implicitly lumped the
spatial dependence of the potential into the dimensionless function $\phi^{(\alpha,i,\pm)}_{Q,\omega}(z)$.\\
Figure~\ref{fig:IPP_potential} shows the scattering potential $\Phi^{(i+)}_{Q}(z)=\sum_{\alpha}\Phi_{Q,\omega_{Q}^{(i)}}^{(\alpha,i,\pm)}({\bf 0},z)$ for
each of the symmetric hybrid modes calculated at $Q=2.37 \times 10^7$/cm (in the Landau-damping region) weighed by the Bose occupation factor 
$n_{Q}^{1/2}$. The square of this quantity is approximately proportional to the contribution of each IPP to the total scattering rate. Note that at such a large
value of $Q$ the presence of the SiO$_2$ interface-layer depresses the large contribution of the low-energy HfO$_2$ mode.\\
      
The top panel of Fig.~\ref{fig:IPP_rates} show the calculated scattering rates (as always averaged over equienergy shells)
for the symmetric IPPs. To emphasize the importance of dynamic screening (that is, the hybridization of the insulator optical phonons and the 2D plasmons,
often ignored in the literature (see, for example, Refs.~\cite{Konar_2010,Toniutti_2012,Ma_2023,Tuan_2026}), 
we show the partial scattering rates resulting separately from processes
involving transfer wave vectors $Q \le Q_{\rm LD}$ (plasmons not Landau-damped) and those in the Landau-damping region, $Q > Q_{\rm LD}$, over which 
the integration in Eq.~(\ref{eq:rate5}) is performed. 
The long-wavelength scattering rates, shown in the left panel of Fig.~\ref{fig:IPP_rates}, are significantly larger than those at short wavelength. 
The importance of accounting for dynamic screening has already been stressed by Hauber and Fahy~\cite{Hauber_2017} and by Macheda and Sohier~\cite{Macheda_2026}.\\
The fact that small-$Q$ processes dominate has profound consequences on the effect of IPP scattering on the off-equilibrium transport properties of
transistors with extremely short channels, $\lesssim$ 10~nm: At or near equilibrium, in the Ohmic regime, being small-angle scattering processes, 
they affect the mobility not so much by redirecting the electron velocity but by reducing it as a consequence of energy loss. However, at high fields,
a few IPP emissions are sufficient to reduce the electron energy, preventing them to transfer to higher-energy subbands. Indeed, we shall show below that 
IPP scattering boosts the saturation velocity. In short-channel transistors one may expect even more dramatic effects, given the streaming nature of
transport induced by these inelastic, small-angle collisions.\\
 
Returning to Fig.~\ref{fig:IPP_rates}, note how `mode 2' (mainly a low-energy HfO$_2$ phonon at small $Q$ and a high-energy HfO$_2$ phonon at large $Q$) 
dominates at long wavelengths. This is due to the fact that, at long wavelengths, the HfO$_2$ layers are `seen' close enough to matter. Since the presence 
of HfO$_2$ is felt only at long wavelength, the mobility is indeed depressed by its proximity, but not as much as it would happen if the interfacial SiO$_2$ 
layer and the 'nearby' metal gates (that screen the IPP scattering potential) were not present. Indeed, `mode 2' plays a negligible role at short wavelength, 
a regime in which `mode 3' (mainly a low/high-energy SO$_2$ phonon at small/large $Q$) controls transport.\\
  
In a similar way, the bottom panel of Fig.~\ref{fig:IPP_rates} shows the scattering rates with the anti-symmetric IPPs in the 
region of wavelength in which plasmons are not Landau-damped ($Q \le Q_{\rm LD}$) and in the Landau-damped region, $Q > Q_{\rm LD}$. The first thing to
notice is their lower values when compared to scattering with the symmetric IPPs. This is the result of the fact that the antisymmetric scattering potential
vanishes at long wavelengths (as $Q \rightarrow 0$: at large distances the electrons feel a net null total polarization charge) 
and from the symmetry of the modes: Inter-subband scattering
involves symmetric or antisymmetric initial and final electronic states and the matrix element {\em almost} vanishes (`almost', since the structure is `almost'
but not {\rm exactly} atomically symmetric). The same consideration applies to inter-subband inter-valley scattering between electronic states with similar
symmetry.\\
      
Equation~(\ref{eq:rate6}) clarifies some interesting selection rules: Odd-parity IPPs result in vanishing intra-valley/intra-subband  transition between 
states on similar parity (that is, $\zeta_{\mu}$ and $\zeta_{\mu'}$ both even or both odd). This results in weaker scattering processes mediated by
odd-parity IPP modes, as seen by comparing the rates shown in the top panel of Fig.~\ref{fig:IPP_rates} with those show in the bottom panel. 
Conversely, even-parity IPPs do not contribute to intra-valley, inter-subband processes involving states of opposite parity.      
Interestingly, regardless of the
parity of the electron states, odd-parity IPPs yield a vanishing contribution at long wavelengths (small $Q$). Indeed, the potential
$\Phi_{Q,\omega}^{(\alpha,i,-)}({\bf R},z)$ due to odd-parity IPPs vanishes in the middle of the Si layer (that is, at $z=t_{\rm s}/2$). However, at small
$Q$, the potential inside the Si layers is almost constant and must vanish as $Q \rightarrow \infty$.\\ 
\begin{table*}[tb]
\centering
\caption{Low-field phonon-limited electron mobility and saturated drift velocity calculated using various boundary conditions for the confined phonons:
         Free-standing (FSBC) and clamped (CBC) at the Si/SiO$_2$ and SiO$_2$/HfO$_2$ interfaces, using bulk (unconfined) phonons with the elastic, high-temperature
         approximation and with acoustic phonons confined at the SiO$_2$/HfO$_2$ interfaces accounting for screening by the insulators, as discussed above. 
         They have been obtained for intrinsic nanosheet, except for mobility calculated accounting for IPP scattering (last row) that has been obtained assuming 
         an electron sheet density of $1.15 \times 10^{13}$ cm$^{-2}$.}
\label{Tab:phonon-limited-mobility}
\begin{tabular}{lcc}
\hline\\
                                                                     &{\bf electron mobility}  & {\bf saturated velocity}   \\ 
                                                                     &(cm$^2$V$^{-1}$s$^{-1}$) & $(10^6$~cm/s)              \\
\hline\\
FSBC@Si/SiO$_2$                                                      &            160          &        --                  \\
CBC@Si/SiO$_2$                                                       &           1060          &        --                  \\
bulk, elastic                                                        &            700          &        --                  \\ 
CBC@SiO$_2$/HfO$_2$  undoped, unscreened                             &            260          &       $\gtrsim$ 5.6        \\ 
CBC@SiO$_2$/HfO$_2$, undoped, screened                               &            360          &       $\gtrsim$ 8.9        \\
CBC@SiO$_2$/HfO$_2$, IPP, $1.15 \times 10^{13}$~cm$^{-2}$, screened  &            200          &       $\gtrsim$ 22.0       \\           
\hline
\end{tabular}
\end{table*}
     
\section{Low-field mobility and high-field characteristics}
\label{sec:Results}
Finally, in this section we present our results regarding the low-field electron mobility and high-field characteristics.\\

The center column in Table~\ref{Tab:phonon-limited-mobility} shows the low-field electron mobility calculated using the various models described above. As anticipated, it takes the low
value of about 160~cm$^2$V$^{-1}$s$^{-1}$ when assuming acoustic phonons clamped at the 'inner' Si/SiO$_2$ interfaces. However, when assuming the phonons
to be clamped at outer SiO$_2$/HfO$_2$ interfaces but ignoring the screening effects of the insulators, this value increases significantly to about 260~cm$^2$V$^{-1}$s$^{-1}$. Dielectric screening by the environment boosts the mobility even further, to about 360~cm$^2$V$^{-1}$s$^{-1}$. The effect of a high-$\kappa$ insulator
may be expected to cause a stronger beneficial effect. However, although the gate stack does include a high-$\kappa$ insulator, HfO$_2$, the interface SiO$_2$ layer,
a low-$\kappa$ dielectric, prevents a stronger reduction of the field lines outside
the nanosheet. It is also interesting to note that assuming scattering with bulk phonons in the elastic/high-temperature approximation yields a mobility
consistent with what had been reported before~\cite{Gamiz_2001}. More interesting is the observation that the choice of boundary conditions to confine
the phonons results in huge variations of the mobility, as already noted before: Assuming phonons clamped at the Si/SiO$_2$ interfaces lifts the minimum
energy of the acoustic phonons so much as to suppress scattering so significantly as to result in a mobility exceeding 1000~cm$^2$V$^{-1}$s$^{-1}$. The opposite is
found when assuming FSBCs at the Si/SiO$_2$ interfaces. Finally, IPP scattering reduces the mobility from 360~cm$^2$V$^{-1}$s$^{-1}$ to 200~cm$^2$V$^{-1}$s$^{-1}$, 
but the effect is relatively small, since the interfacial SiO$_2$ layer keeps the high-$\kappa$ HfO$_2$ film away from the channel. Although at small $Q$ 
(long range) this effect weakens, screening by the metallic gates reduces the negative effect of the high-$\kappa$ film. Note that we are comparing results 
calculated at two different electron sheet densities, intrinsic (no IPP  scattering) and at $1.15 \times 10^{13}$~cm$^{-2}$ (IPP scattering). However,
the mobility in ultra-thin Si films has been shown to depend very weakly on the electron density (or gate field) (as long as scattering with interface
roughness remains negligible), both theoretically (see Fig. 7 of
Ref.~\cite{Gamiz_2001}) and experimentally (see Fig. 10 of Ref.~\cite{Uchida_2003} and Fig. 7 of Ref~\cite{Schmidt_2009}).\\
\begin{figure}[tb!]
\centerline 
{\hbox{
\includegraphics[width=8.60cm]{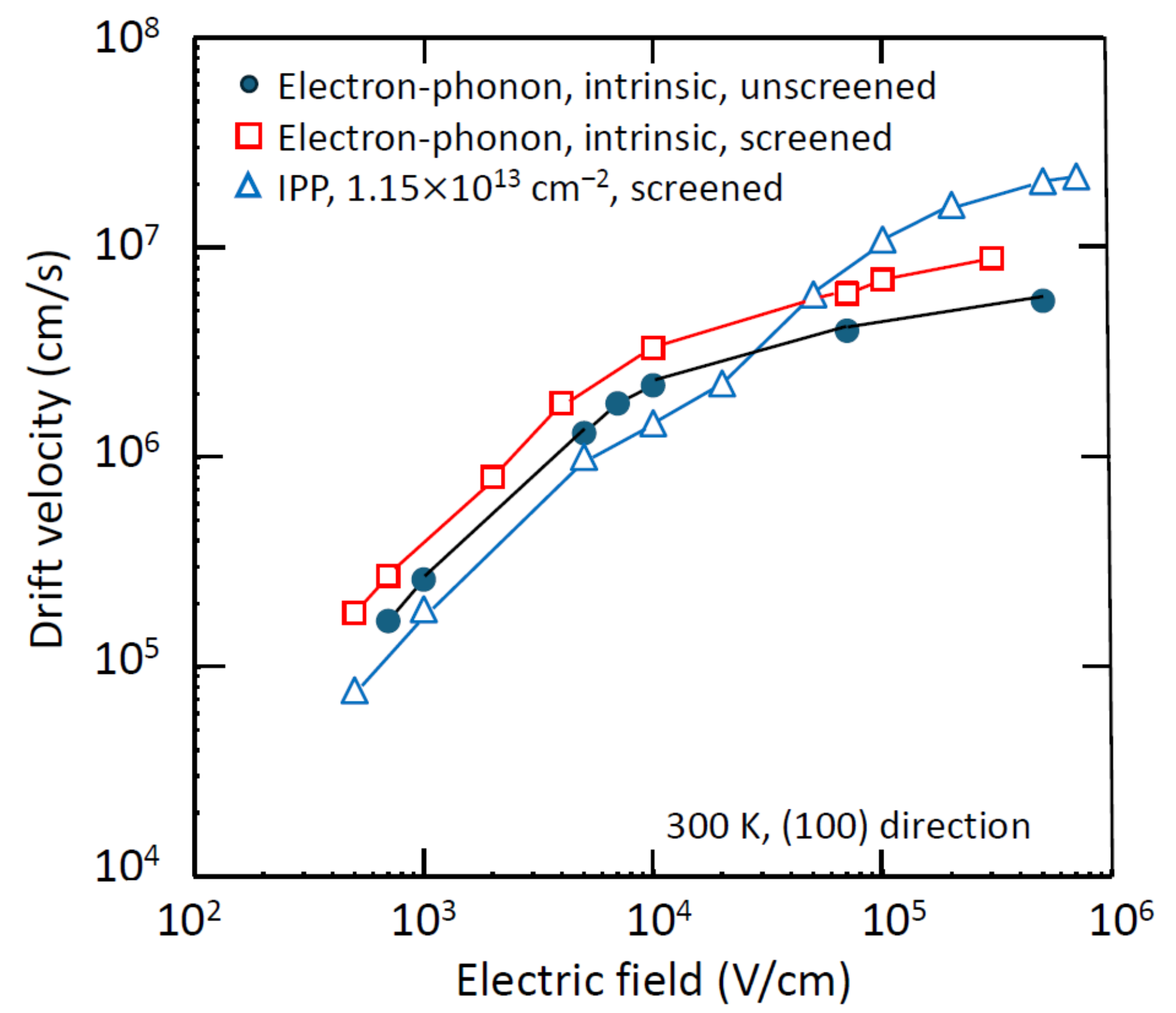}
}}
\caption{Electron drift velocity plotted as a function of a uniform electric field along the (100) direction at 300~K accounting only for unscreened scattering 
         with phonons in an intrinsic (undoped) Si nanosheet, (black dots), for scattering with phonons but accounting also for dielectric screening via
         Eq.~(\ref{eq:tau_ep_sheet_screened}) (open red square) and, finally, accounting also for scattering with the IPPs assuming a sheet electron density of 
         $1.15 \times 10^{13}$~cm$^{-2}$ (open blue triangles). The lines connecting the symbols are only a guide to the eye.}
\label{fig:vdrift_3-cell_vs_field}
\end{figure} 

Moving to off-equilibrium transport at high fields,  
the black dots in Fig.~\ref{fig:vdrift_3-cell_vs_field} shows the electron drift velocity {\it vs.} field characteristics 
assuming acoustic phonons clamped boundary conditions at the SiO$_2$/HfO$_2$ interfaces in a 3-cell-thick, undoped Si sheet but ignoring dielectric screening. 
To facilitate the comparison, we also show the velocity-field characteristics accounting for screening by the dielectrics using Eq.~(\ref{eq:tau_ep_sheet_screened})
(open red squares), and also accounting for scattering with the IPPs (open blue triangles).

\begin{figure*}[tb]
\centerline 
{\hbox{
\includegraphics[width=15.0cm]{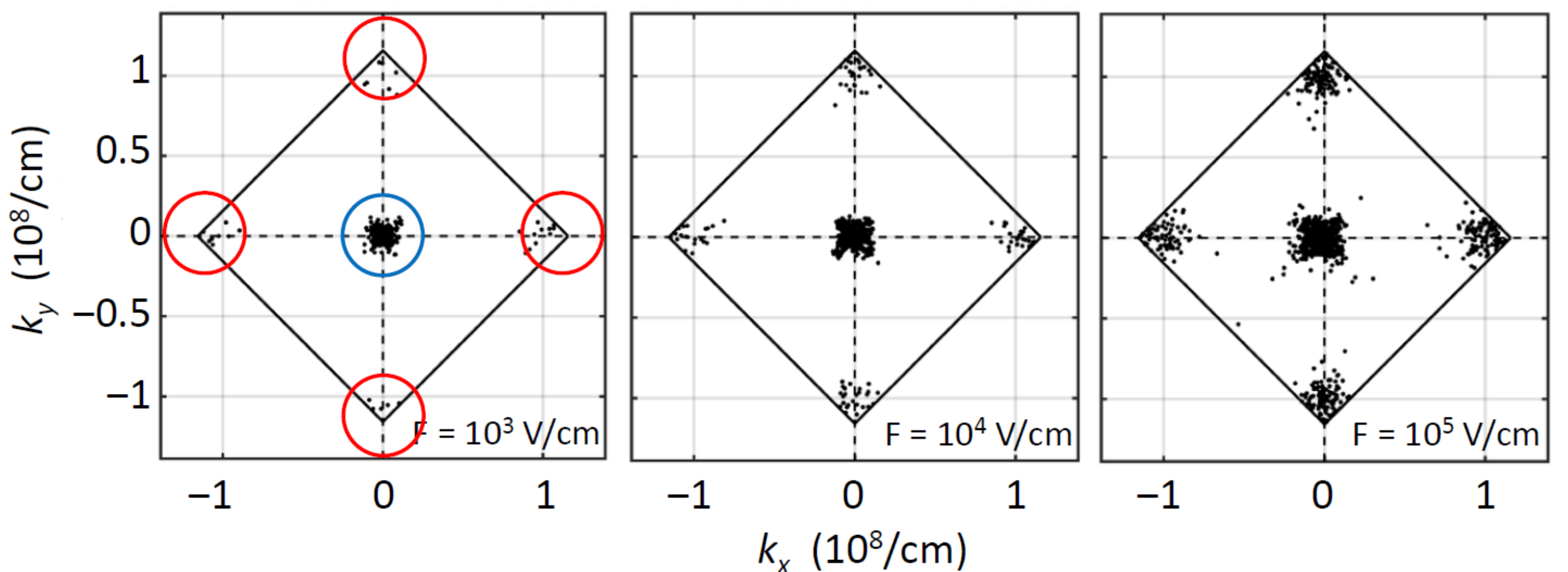}
}}
\caption{Electron distribution in the 2D Brillouin zone as they are accelerated by a uniform field with strength ranging from $10^3$ (left panel) to 
        $10^5$~V/cm (right panel). In the left panel, the blue and red circles highlight the position of the unprimed and primed subbands, respectively. 
        Note the increasing population of the primed subbands at high fields that results in a reduced saturated velocity.}  
\label{fig:El-ph_3-cell_HfO2_BZ}
\end{figure*}

Note that the electron saturation velocity approaches a value of about 5.5$\times 10^{6}$~cm/s, lower than the value in bulk Si. This is a nontrivial result 
which is in agreement with experimental results obtained in the similar situation of a Si 2DEG in inversion layer. 
References ~\cite{Fang_1970,Coen_1978,Modelli_1988,Assaderadi_1993} report values of 6.5$\times 10^6$~cm/s, 5.0$\times 10^6$~cm/s, 
6.0$\times 10^6$~cm/s (in SOIs), and 5.8$\times 10^6$~cm/s, respectively, at electron sheet densities ranging from 2.3$\times 10^{11}$~cm$^{-2}$ to 
3.6$\times 10^{12}$~cm$^{-2}$ (that is, at normal fields ranging from 3.6 to $5.6 \times 10^5$~V/cm). The fact that Nelson and Cooper~\cite{Nelson_1982} 
obtained a value close to the bulk, $10^7$~cm/s, using a rather complicated (and hard to interpret) time-of-flight technique, has created confusion. Such a confusion
has been reinforced by that fact that previous theoretical calculations~\cite{Basu_1977,Basu_1978,Zimmermann_1980,Imanaga_1991,Fischetti_1993} have been unable 
to explain the lower values of the saturation velocity, all predicting values close to $10^7$~cm/s. To our knowledge, our theoretical results are the first to 
predict the lower saturated velocity observed experimentally (with the exception of the Nelson and Cooper results that have been obtained using a technique
that does not permit an easy determination of the electric fields and flight time of the injected packet of carriers). The ultimate reason for such a
lower saturated velocity lies in the transfer of the electrons from the unprimed to the primed subbands at high fields, 
as shown in Fig.\ref{fig:El-ph_3-cell_HfO2_BZ}, and on 
the rather complicated and 'flat' dispersion of the fourfold-degenerate primed ladder (seen in Fig.~\ref{fig:Bands_3_4-cells} for a thin Si
sheet, but also predicted for inversion layers~\cite{thebook}), an effect missed by previous calculations, all based on the effective-mass approximation. 
It is interesting to note that the experimental and theoretical literature on this problems is somewhat dated: We could not find anything more recent than our 
own work from 1993~\cite{Fischetti_1993}.\\

The open red squares in Fig.~\ref{fig:vdrift_3-cell_vs_field} show the velocity-field characteristics accounting for the dielectric 
screening of the
electron-phonon interaction due to the insulators in our double-gate geometry, as described by Eq.~(\ref{eq:tau_ep_sheet_screened}). Both the mobility as well
as the saturated velocity are larger, as expected. In particular, the saturated velocity reaches a value of about $8.9 \times 10^{6}$~cm/s. Although larger
than the value obtained ignoring screening, this is still lower than the value of bulk Si.\\
The effect of IPP scattering on the electron drift velocity is shown by the
open blue triangles in Fig.~\ref{fig:vdrift_3-cell_vs_field}. As we found in the case of WS$_2$ bilayers~\cite{Mansoori_2025}, IPP scattering depresses 
the drift velocity in the Ohmic regime (that is, the mobility) but yields a higher velocity at large fields, with a saturated value exceeding  
$2 \times 10^{7}$~cm/s. In the of WS$_2$ we attributed this to the fact that the inelastic IPP interactions depress the electron energy preventing them 
from undergoing transitions to higher-mass satellite valleys. Here, the cause is similar, as lower-energy electrons are prevented from undergoing inter-subband
scattering. This is also evident from the fact that the Ohmic regime extends to fields almost as high as $10^5$~V/cm in the presence of IPP scattering, whereas
electrons become hot at fields as low as a few kV/cm in its absence.\\         

Finally, the values given in Table \ref{Tab:phonon-limited-mobility} represent the angular average of the electron mobility, 
calculated at zero field from the diffusion constant for intrinsic nanosheets (no IPP scattering) or at an electron density of $1.15 \times 10^{13}$ cm$^{-2}$
(IPP scattering) using the Einstein's relation, corrected for degeneracy~\cite{hnote8} when needed. Indeed, Pauli blocking is 
implemented in our Monte Carlo program as described in Ref.~\cite{Fischetti_1988}. 
Whereas at higher energies the dispersion of the ground-state unprimed subbands exhibits anisotropic nonparabolicity with different effective masses 
along the (100) and (110) direction, with the effective mass along the $(11)$ direction being significantly larger than along the $(100)$ direction,
this anisotropy becomes significant only at very high densities and or at high electric fields, as indeed observed experimentally by 
Mochizuki {\it et al.}~\cite{Mochizuki_2023}. However, at the thermal energy the dispersion is almost isotropic and, indeed, by looking at the angular dependence 
of the diffusion constant we found that the mobility calculated in at $1.15 \times 10^{13}$ cm$^{-2}$ is still largely isotropic.\\

Unfortunately, experimental data for these extremely thin Si films are scant. Results are available for sheets thicker than about 
3~nm~\cite{Uchida_2002,Uchida_2003,Esseni_2004,Uchida_2007,Park_2025a}, showing a sharp drop below this thickness. For the 1.63~nm-thick films considered here,
Schmidt and coworkers~\cite{Schmidt_2009} have measured a mobility ranging from $\approx$~15 to $\approx$~50~cm$^2$V$^{-1}$s$^{-1}$ for 
an electron sheet density ranging from $\approx 5 \times 10^{11}$ to $10^{13}$~cm$^{-2}$ in single gated SOIs with SiO$_2$ gate insulators. 
Similar values have been reported by Shimizu and Hiramoto~\cite{Shimizu_2007}, who found that in films so thin the electron mobility is not depressed when 
moving from single-gate to double-gated structures. These studies also found the weak dependence of the mobility on electron density, already predicted 
theoretically in Ref.~\cite{Gamiz_2001} and experimentally observed in Ref.~\cite{Uchida_2003}, A more recent experimental study on Si nanosheet transistors~\cite{Agrawal_2024} reports what we presume is an off-equilibrium field-effect mobility of about 100~cm$^2$V$^{-1}$s$^{-1}$ for a 1.5~nm-thick Si 
sheet in an 18~nm gate-length multi-sheet transistor with an unspecified gate stack. Since the field-effect mobility is an off-equilibrium quantity,
we expect it to be larger than the low-field mobility. A value of 60-to-70~~cm$^2$V$^{-1}$s$^{-1}$ has been reported by 
Schmidt {\it et al.} for even thinner films, 0.9~nm thick, a value that is comparable to what has been measured by Uchida {\it et al.}~\cite{Uchida_2003} who 
reported values in the range of 45-to-65~cm$^2$V$^{-1}$s$^{-1}$ and 80-to-130~cm$^2$V$^{-1}$s$^{-1}$ for films 0.7 and 0.9~nm thin, respectively.  
From these consideration, all we can conclude is that our calculations yield values that are in the correct range. 
However, we cannot draw any exact quantitative conclusion, considering that parasitic effects and scattering with defects and, most important, scattering
with interface roughness affect the experimental results. Moreover, different gate-insulator stacks modify the confinement of phonons and we saw that  
the electron mobility is very sensitive to this effect. 

\section{Conclusions}
\label{sec:Conclusions}
Using full-band Monte Carlo simulations, we have studied theoretically electron transport in an extremely scaled top- and bottom-gated 1.6~nm-thin silicon 
nanosheet, with the limited goal of elucidating the role played by phonon confinement and IPP (`remote phonon') scattering. The structure is taken to be 
sandwiched by top and bottom SiO$_2$/HfO$_2$ gate stacks. The electron dispersion for the H-terminated sheet has been obtained using local empirical 
pseudopotentials, whereas the elastic and dielectric continuum limits have been used to deal with phonon confinement and IPPs.\\
 
We have found that the low-field electron mobility at room temperature is extremely sensitive to the choice of boundary conditions 
for the confined phonons, ranging from 160~cm$^2$V$^{-1}$s$^{-1}$ in an undoped 1.6~nm-thick Si nanosheet, when assuming free-standing phonon boundary conditions at
the Si/SiO$_2$ interfaces, to 1060~cm$^2$V$^{-1}$s$^{-1}$ when the phonons are assumed to be clamped at these interfaces. 
The more realistic assumptions of acoustic phonons clamped sat the `outer' SiO$_2$/HfO$_2$ interfaces yields a mobility of about 260~cm$^2$V$^{-1}$s$^{-1}$ when
ignoring dielectric screening, increasing to about 360~cm$^2$V$^{-1}$s$^{-1}$ when accounting for screening effects. 
As a result of the complex dispersion of the primed fourfold-degenerate subbands, we have also found that the high-field saturation velocity is about 
$5.5 \times 10^6$~cm/s, significantly lower than its bulk value (no screening, up to $8.9 \times 10^6$~cm/s with screening), 
but consistent with experimental values reported for the similar case of electrons in Si inversion layers.\\ 

Scattering with the IPPs depresses the mobility from 360 to about 200~cm$^2$V$^{-1}$s$^{-1}$, an effect that is smaller than anticipated:
Large-$Q$, short-range processes
do not 'see' the high-$\kappa$ HfO$_2$ layers, thanks to the presence of the SiO$_2$ interfacial layers; at long distances (small $Q$), the potentially large IPP scattering potential due to the presence of the HfO$_2$ layers is screened by the nearby ideal metal gates. Moreover, the high-field drift velocity is significantly
enhanced, since the energy losses to the IPPs prevent electrons from reaching an energy large enough to permit inter-subband scattering. Ignoring parasitic effects,
 such as the quantum access resistance at the source/nanosheet 'junction', we expect that IPP-scattering, rather than depressing the performance 
of short-channel transistors, may actually be beneficial, thanks to the the higher velocity in off-equilibrium conditions.

\acknowledgments
\noindent We acknowledge the help of the Texas Advanced Computing Center (TACC) for having provided computing resources, a Texas Instruments for the Endowment and
TSMC for having provided financial support. 
\begin{appendices}
\renewcommand{\thesection}{\Alph{section}}
\numberwithin{equation}{section}
\section{Poisson Green's functions}
\label{AppendixC}
In this Appendix we derive the analytical expression for the Poisson's Green's function $G^{(2)}_{Q}(z,z')$ in the case in which the
source-charge is in the nanosheet, $0 \le z' \le t_{\rm s}$.\\ 

      We must solve the equation:
      \begin{equation}
         \nabla \cdot \nabla [ \epsilon(z) G({\bf r},{\bf r}') ] \ = \ - \ \delta ({\bf r}-{\bf r}')  \ ,
      \label{eq:AGreeneq}
      \end{equation}
      for our geometry consisting of the Si layer in the region $0 < z \le t_{\rm s}$, an interface insulator labeled as `1' (SiO$_2$) in the regions 
      $-t_{\rm b1} < z \le 0$ and $t_{\rm s} < z \le t_{\rm s}+t_{\rm t1}$, a high-$\kappa$ insulator `2' (HfO$_2$) in the regions 
      $-t_{\rm b1}-t_{\rm b2} < z \le -t_{\rm b1}$ and  $t_{\rm s}+t_{\rm b1} < z \le t_{\rm s}+t_{\rm b1}+t_{\rm b2}$ and, finally, ideal metal gates at 
      $z=-t_{\rm b1}-t_{\rm b2}$ and $z= t_{\rm s}+t_{\rm b1}+t_{\rm b2}$.
      For simplicity we have omitted the time dependence of the Green's function, so that 
      $G({\bf r},{\bf r}')$ may actually be a function also of time, $G({\bf r},{\bf r}'; t, t')$ and its Fourier-Bessel transform $G_{Q}(z,z')$ 
      defined implicitly by Eq.~(\ref{eq:AFourier_z}) below should be
      viewed as also dependent on the frequency $\omega$, so that $G_{Q}(z,z')$ should be read as $G_{Q,\omega}(z,z')$.\\
            
      The dielectric functions are assumed to be isotropic, as expected for amorphous insulators, constant on the $(x,y)$ plane, and piece-wise 
      constant along $z$. The Fourier-transform (or, more appropriately, the  Fourier-Bessel-transform) of the Green's function $G$ on the $(x,y)$-plane, 
      using its translational invariance on this plane, is:
      \begin{multline}
      G({\bf r},{\bf r}') \ = \ \frac{1}{\Omega} \ \sum_{\bf Q} \ G_{Q}(z,z') \ e^{i{\bf Q} \cdot ({\bf R}-{\bf R}')} \\ =
         \frac{1}{2 \pi} \int_{0}^{\infty} {\rm d} Q \ Q \ J_{0}(Q|{\bf R}-{\bf R}'|) \ G_{Q}(z,z') \ ,
         \vspace{-0.15cm}
      \label{eq:AFourier_z}
      \end{multline}
      where $J_0(x)$ is the Bessel function of the first kind of order 0 and we have
      omitted to write explicitly the time and frequency-dependence of $G$ and $G_Q$ to simplify the notation. 
      As usual, upper-case bold characters denote 2-vectors on the $(x,y)$ plane and $\Omega$ now is the normalization area. 
      We have also already used the fact 
      that $G_{Q}(z,z')$ depends only on the magnitude of ${\bf Q}$.\\ 
      
      The problem simplifies in our case of a symmetric double-gate structure in which the top insulators `1' and `2' 
      (SiO$_2$ and HfO$_2$, respectively) have the same thickness as the bottom insulators; 
      that is, $t_{\rm b1} = t_{\rm t1} \equiv t_1$ and $t_{\rm b2} = t_{\rm t2} \equiv t_{2}$.
      In the same spirit, in the following $\epsilon_1(\omega)$ and $\epsilon_2(\omega)$ will denote the dielectric functions of insulators `1' (SiO$_2$) 
      and `2' (HfO$_2$), respectively.\\
     
      Inserting Eq.~(\ref{eq:AFourier_z}) into Eq.~(\ref{eq:AGreeneq}), we find that $G_{Q}(z,z')$ must satisfy the equation:
      \begin{equation}
      \frac{\rm d}{{\rm d}z} \left [ \epsilon(z) \frac{{\rm d} G_{Q}(z,z')}{{\rm d}z} \right ] - Q^{2} \epsilon(z) G_{Q}(z,z') \ = \ - \ \delta(z-z') \ ,     
      \label{eq:AGreenz}
      \end{equation}
      where we have used total derivatives since $z'$ is treated as a parameter. This equation must satisfy the boundary conditions expressing the continuity 
      of $G_{Q}(z.z')$ and $\epsilon(z) {\rm d} G_{Q}(z,z')/{\rm d}z$ at four interfaces (8 conditions), the vanishing of $G_{Q}(z,z')$ at bottom and top gate/oxide 
      interfaces due to the presence of the ideal metal gates (two more conditions), the continuity of $G_{Q}(z,z')$ at $z=z'$ and, finally, the condition 
      reflecting the presence of the source-charge at $z=z'$, that is:
      \begin{multline}
            { \underset{\eta \rightarrow 0} {\rm lim} } \ \left [ \epsilon(z'-\eta) \frac{{\rm d}G_{Q}(z'-\eta,z')}{{\rm d}z} \right. - \\
                                          \left. \epsilon(z'+\eta) \frac{{\rm d}G_{Q}(z'+\eta,z')}{{\rm d}z} \right ] \ = \ 1 \ .
      \label{eq:Asource}
      \end{multline}
   
      We consider here only the case of a source-charge in the nanosheet, as needed in the main section of the text. 
      When $0 \le z' < t_{\rm s}$, the solution of Eq.~(\ref{eq:AGreenz}) with the twelve boundary conditions appropriate to our symmetric 
      geometry can be written as:
      \vspace*{-0.2cm}
      \begin{widetext}
      \begin{equation}
      G^{\rm (2)}_{Q}(z,z') \ = \left \{
      \begin{array}{ll}
      a_1 e^{Q(z+t_1+t_2)}         \ + \ a_2 e^{-Q(z+t_1+t_2)}         & (-t_2-t_1 \le z \le -t_1) \vspace*{0.0cm}  \\
      b_1 e^{Q(z+t_1)}             \ + \ b_2 e^{-Q(z+t_1)}             & (-t_1     \le z \le 0)    \vspace*{0.0cm}  \\
      c_1 e^{Qz}                   \ + \ c_2 e^{-Qz}                   & (0 < z \le z')            \vspace*{0.0cm}  \\
      d_1 e^{Q(z-z')}              \ + \ d_2 e^{-Q(z-z')}              & (z' < z \le t_{\rm s})    \vspace*{0.0cm}  \\
      f_1 e^{Q(z-t_{\rm s})}       \ + \ f_2 e^{-Q(z-t_{\rm s})}       & (t_{\rm s} < z \le t_{\rm s}+t_1) \vspace*{0.0cm}  \\
      g_1 e^{Q(z-t_{\rm s}-t_1)}   \ + \ g_2 e^{-Q(z-t_{\rm s}-t_1)}   & (t_{\rm s}+t_1 < z \le t_{\rm s}+t_1+t_2) \\
      \end{array}
      \right. \ .
      \label{eq:AGQ}  
      \end{equation}
      The system of equations resulting from the boundary conditions can now be written as:
      \begin{subequations}
      \begin{align}
      & a_1 + a_2                                                                   = 0                                 \label{eq:Asystema}  \\[-0.150cm] 
      & a_1 e^{Qt_2} + a_2 e^{-Qt_2}                                                = b_1 + b_2                         \label{eq:Asystemb}  \\[-0.150cm]
      & \epsilon_2 Q (a_1 e^{Qt_2} - a_2 e^{-Qt_2})                                 = \epsilon_1 Q (b_1 - b_2)          \label{eq:Asystemc}  \\[-0.150cm]                                  
      & b_1 e^{Qt_1} + b_2 e^{-Qt_1}                                                = c_1 +c_2                          \label{eq:Asystemd}  \\[-0.150cm]
      & \epsilon_1 Q (b_1 e^{Qt_1} - b_2 e^{-Qt_1})                                 = \epsilon_{\rm s} Q (c_1-c_2)      \label{eq:Asysteme}  \\[-0.150cm]
      & c_1 e^{Qz'} + c_2 e^{-Qz'}                                                  = d_1 +d_2                          \label{eq:Asystemf}  \\[-0.150cm]
      & \epsilon_{\rm s} Q \ (c_1 e^{Qz'} - c_2 e^{-Qz'})                           = 1+\epsilon_{\rm s} Q (d_1 - d_2)  \label{eq:Asystemg}  \\[-0.150cm]
      & d_1 e^{Q(t_{\rm s}-z')} + d_2 e^{-Q(t_{\rm s}-z')}                          = f_1 + f_2                         \label{eq:Asystemh}  \\[-0.150cm]
      & \epsilon_{\rm s} Q \ (d_1 e^{Q(t_{\rm s}-z')} - d_2 e^{-Q(t_{\rm s}-z')})   = \epsilon_1 Q \ (f_1 - f_2 )       \label{eq:Asystemi}  \\[-0.150cm]
      & f_1 e^{Qt_1} + f_2 e^{-Qt_1}                                                = g_1 + g_2                         \label{eq:Asystemj}  \\[-0.150cm]
      & \epsilon_1 Q (f_1 e^{Qt_1} - f_2 e^{-Qt_1})                                 = \epsilon_2 Q (g_1 - g_2)          \label{eq:Asystemk}  \\[-0.150cm]
      & g_1 e^{Qt_2} + g_{2} e^{-Qt_2}                                              = 0         \ .                     \label{eq:Asysteml}
      \end{align}
      \end{subequations}
      \end{widetext}
      The choice of the expressions that must be used for the dielectric functions $\epsilon_1$, $\epsilon_2$, and $\epsilon_{\rm s}$ depends on the context in which 
      the Green's function is used. For example, when dealing with dielectric screening of the electron-phonon interaction, we should
      set $\epsilon_{\rm s}$ equal to its dynamic expression $\epsilon({\bf Q},\omega)$, Eq.~(\ref{eq:scalar_0_eps}). On the contrary, when looking for the
      statically screened Coulomb potential of a charge in the 2D layer, we should set $\epsilon_{\rm s}$ equal to $\epsilon^{\rm (scalar)}(Q,\omega=0)$. 
      As discussed in the main text, we have used this static approximation for $\epsilon_{\rm s}$, as well as for $\epsilon_1$ and $\epsilon_2$, also when
      screening the electron-phonon interactions.     
      For simplicity, we have not indicated explicitly the dependence of these unknowns, $a_1$ through $g_2$, 
      on $Q$, $\omega$, and $z'$.\\

\begin{figure*}[tb]
\centerline 
{\hbox{
\includegraphics[width=14.1cm]{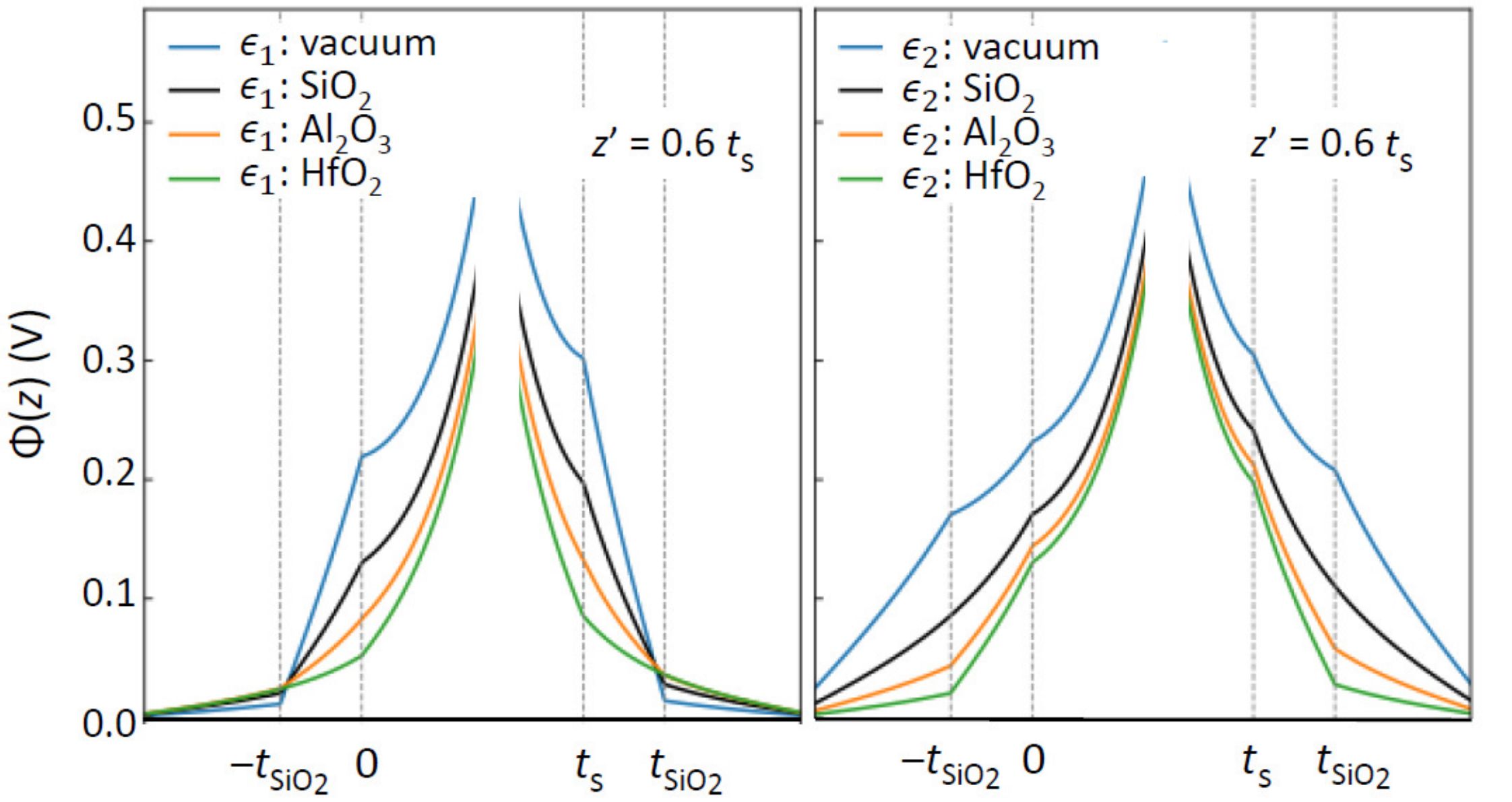}
}}
\caption{{\bf Left:} Potential $\Phi(z)=(2\pi)^{-1}\int {\rm d}Q \ Q \ G^{(2)}(z,z')$ for a point charge located at $z'=0.6~t_{\rm s}$ in a structure with
$t_{\rm s}$ = 1.63~nm, $t_{\rm SiO_{2}} = t_1$ = 0.6~nm, $t_{\rm HfO_{2}} = t_2$ = 1.6~nm. The curves are parametrized by various values of the dielectric
constant of insulator `1' and show how the potential in the Si nanosheet decreases when the dielectric constant increases. 
{\bf Right:} The same, but the curves 
are parametrized by the dielectric constant of insulator `2'. In both cases, the effect is due to the stronger dielectric screening of insulators with a 
higher static dielectric constant. All curves have been truncated at a distance $\approx 1/Q_{\rm max}$ away from the source-charge
$Q_{\rm max}$ being the maximum value of $Q$ used for the numerical integration.
}  
\label{fig:G2a}
\end{figure*}

\begin{figure*}[tb]
\centerline 
{\hbox{
\includegraphics[width=14.0cm]{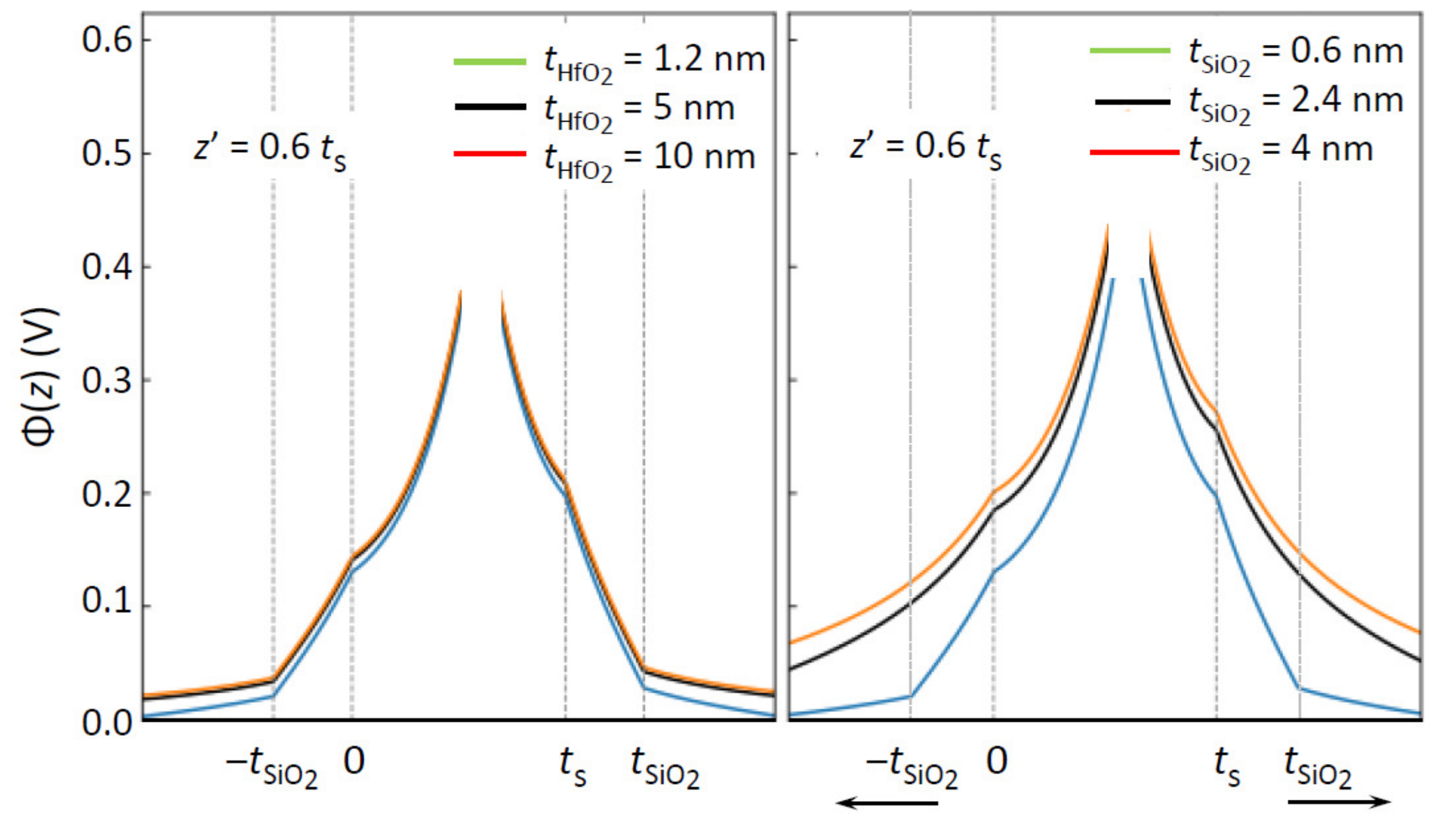}
}}
\caption{As in Fig.~\ref{fig:G2a}, but the curves are parametrized by the thickness of insulator `2' (left) and `1' (right), 
showing how the potential in the Si nanosheet decreases when the thickness insulators decreases, as the screening effects of the closer metal gates increase.
}  
\label{fig:G2}
\end{figure*}

      Equations~(\ref{eq:Asystema})-(\ref{eq:Asysteml}) constitute an in-homogeneous linear system of twelve equations in twelve unknowns. Solving this
      system yields the following form for the coefficients entering the Bessel-Fourier (Hankel) components of the 
      Poisson Green's function, Eq.~(\ref{eq:AGQ}):
      \begin{widetext}
      \begin{subequations}
      \begin{align}
      & a_1 = \frac{b_1}{2\epsilon_2}
         e^{-Qt_2} \left [ \left ( \epsilon_2 + \epsilon_1 \right ) + 
                                        \Lambda \left ( \epsilon_2 - \epsilon_1 \right ) \right ]   \label{eq:G2_a1} \\
      & a_2 = \frac{b_1}{2\epsilon_2}  
         e^{Qt_2} \left [ \left ( \epsilon_2 -  \epsilon_1 \right ) + 
                                        \Lambda \left ( \epsilon_2 + \epsilon_1 \right ) \right ]   \label{eq:G2_a2} \\
      & b_1 = \frac{1}{\epsilon_{\rm s}Q} 
               \frac{({\mathcal A} - 1 ) e^{Q z'}} 
                      { e^{Qt_1}(1 - \epsilon_1/\epsilon_{\rm s}) 
                           + \Lambda e^{-Qt_1}(1+ \epsilon_1/\epsilon_{\rm s})
                                        + {\mathcal A} \ e^{2Q z'} \left[ e^{Qt_1}(1+ \epsilon_1/\epsilon_{\rm s}) 
                           + \Lambda e^{-Qt_1}(1 - \epsilon_1/\epsilon_{\rm s}) \right ] } \label{eq:G2_b1} \\                                    
      & b_2 = b_1 \frac {\sinh(Qt_2)-(\epsilon_2/\epsilon_1)\cosh(Qt_2)}
                  {\sinh(Qt_2)+(\epsilon_2/\epsilon_1)\cosh(Qt_2)} \label{eq:G2_b2} \\ 
      & c_1 = \frac{b_1}{2\epsilon_{\rm s}} \ \left [ e^{Qt_1}  \left ( \epsilon_{\rm s} + \epsilon_1 \right ) + 
                                        \Lambda e^{-Qt_1} \left ( \epsilon_{\rm s} - \epsilon_1 \right ) \right ]  \label{eq:G2_c1} \\
      & c_2 = \frac{b_1}{2\epsilon_{\rm s}} \ \left [ e^{Qt_1}  \left ( \epsilon_{\rm s} - \epsilon_1 \right ) + 
                                        \Lambda e^{-Qt_1} \left ( \epsilon_{\rm s} + \epsilon_1 \right ) \right ]  \label{eq:G2_c2} \\  
      & d_1 = \frac{b_1}{2\epsilon_{\rm s}} \ \left [ e^{Qt_1} \left ( \epsilon_{\rm s}+\epsilon_1 \right ) + 
                 \Lambda e^{-Qt_1} \left ( \epsilon_{\rm s}-\epsilon_1 \right ) \right ] e^{Qz'}  - \frac{1}{2\epsilon_{\rm s} Q}  \label{eq:G2_d1} \\
      & d_2 = \frac{b_1}{2\epsilon_{\rm s}} \ \left [ e^{Qt_1} \left ( \epsilon_{\rm s}-\epsilon_1 \right ) + 
                 \Lambda e^{-Qt_1} \left ( \epsilon_{\rm s}+\epsilon_1 \right ) \right ] e^{-Qz'} + \frac{1}{2\epsilon_{\rm s} Q}  \label{eq:G2_d2} \\
      & f_1 = \frac{1}{2\epsilon_1} \ \left [  d_1 e^{Q(t_{\rm s}-z')}  \left ( \epsilon_1+\epsilon_{\rm s} \right ) 
                                 + d_2 e^{-Q(t_{\rm s}-z')} \left ( \epsilon_1-\epsilon_{\rm s} \right ) \right ]   \label{eq:G2_f1} \\
      & f_2 = \frac{1}{2\epsilon_1} \ \left [  d_1 e^{Q(t_{\rm s}-z')}  \left ( \epsilon_1-\epsilon_{\rm s} \right ) 
                                 + d_2 e^{-Q(t_{\rm s}-z')} \left ( \epsilon_1+\epsilon_{\rm s} \right ) \right ]   \label{eq:G2_f2} \\ 
      & g_1 = \frac{1}{2\epsilon_2} \ \left [ f_1 \ e^{Qt_1} \left ( \epsilon_2+\epsilon_1 \right ) 
                                 + f_2 e^{-Qt_1} \left ( \epsilon_2-\epsilon_1 \right ) \right ]   \label{eq:G2_g1} \\
      & g_2 = \frac{1}{2\epsilon_2} \ \left [ f_1 \ e^{Qt_1} \left ( \epsilon_2-\epsilon_1 \right ) 
                                 + f_2 e^{-Qt_1} \left ( \epsilon_2+\epsilon_1 \right )  \right ]  \label{eq:G2_g2}
      \end{align}
      \end{subequations}
      \end{widetext}
      where:
      \begin{subequations}
      \begin{align}
      & \Lambda = \frac {\sinh(Qt_2)-(\epsilon_2/\epsilon_1)\cosh(Qt_2)}
                  {\sinh(Qt_2)+(\epsilon_2/\epsilon_1)\cosh(Qt_2)}  \label{eq:G2_Lambda} \\  
      & {\mathcal A}  = e^{2Q(h - z')}
         \frac{{\mathcal B} (\epsilon_1 + \epsilon_{\rm s}) + (\epsilon_1 - \epsilon_{\rm s})} 
               {{\mathcal B} (\epsilon_1 - \epsilon_{\rm s}) + (\epsilon_1 + \epsilon_{\rm s})} \label{eq:G2_bigA} \\
      & {\mathcal B} = \frac{e^{Qt_1}(\epsilon_2 - \epsilon_1) + e^{Q(2t_2+t_1)} (\epsilon_2 + \epsilon_1)}
                          {e^{-Qt_1}(\epsilon_2 + \epsilon_1) + e^{Q(2t_2-t_1)} (\epsilon_2 - \epsilon_1)}  
                                                                                                \label{eq:G2_bigB}
      \end{align}
      \end{subequations}
      Figure~\ref{fig:G2a} shows the static potential $\Phi(z)=(2\pi)^{-1} \int {\rm d}Q \ Q \ G^{(2)}(z,z')$ for a point charge located at $z'=0.6~t_{\rm s}$ in 
      a structure with $t_{\rm s}$ = 1.63~nm, $t_{\rm SiO_{2}} = t_1$ = 0.6~nm, $t_{\rm HfO_{2}} = t_2$ = 1.6~nm. The left frame 
      shows that the potential in the Si nanosheet is reduced as the dielectric constant of insulator `1' (nominally SiO$_2$ with 
      $\epsilon_1$ = 3.9~$\epsilon_{0}$) increases to 12~$\epsilon_{0}$ (Al$_2$O$_3$) and 22~$\epsilon_{0}$ (HfO$_2$). The right frame shows 
      the same effect when changing the  dielectric constant of insulator `2' (nominally HfO$_2$), $\epsilon_2$. 
      Figure~\ref{fig:G2} shows the screening effects of the metal gates: As the thickness of insulators increases, so does the potential in the nanosheet 
      as the screening gate is removed farther away.

\end{appendices}

\clearpage
\bibliography{Si_nanosheets_ref_titles}

\end{document}